# *Best Bulk Bennu ($B^3$): Elemental compositions of aggregate samples from asteroid Bennu*

Wang, K.[1]*, Lodders, K.[1], Koefoed, P.[1], Broussard, M.[1], Wimpenny, J[2], Render, J.[2], Shollenberger, Q.R.[2], Kruijer, T.S.[2], Brennecka, G.A.[2], Amini, M.[3], Lai, V.[3], Weis, D.[3], Welten, K.C.[4], Nishiizumi, K.[4], Glavin, D.P.[5], Dworkin, J.P.[5], Alexander, C.M.O'D.[6], Barrat, J.-A.[7], Haenecour, P.[8], Barnes, J.J.[8], Nguyen, A.N.[9], Connolly, H.C. Jr.[8,10,11], and Lauretta, D.S.[8]

[1] McDonnell Center for the Space Sciences and Department of Earth, Environment, and Planetary Sciences, Washington University in St. Louis, MO, USA.
[2] Lawrence Livermore National Laboratory, Livermore, CA, USA.
[3] Pacific Centre for Isotopic and Geochemical Research, Department of Earth, Ocean and Atmospheric Sciences, The University of British Columbia, Vancouver, British Columbia, Canada.
[4] Space Sciences Laboratory, University of California, Berkeley, Berkeley, CA, USA.
[5] Solar System Exploration Division, NASA Goddard Space Flight Center, Greenbelt, MD, USA.
[6] Earth and Planets Laboratory, Carnegie Institution for Science, Washington, DC, USA.
[7] Institut Universitaire Européen de la Mer, Université de Bretagne Occidentale, CNRS UMS 3113, F-29280 Plouzané, France.
[8] Lunar and Planetary Laboratory, University of Arizona, Tucson, AZ, USA.
[9] Astromaterials Research and Exploration Science Division, NASA Johnson Space Center, Houston, TX, USA.
[10] Department of Geology, Rowan University, Glassboro, NJ, USA.
[11] Department of Earth and Planetary Sciences, American Museum of Natural History, New York, NY, USA.

* Corresponding authors: wangkun@wustl.edu

# Abstract

**Ivuna-type (CI) carbonaceous chondrites have long served as proxies for the bulk Solar System composition due to their close agreement with solar photospheric abundances and enrichment in volatile elements relative to other chondrite classes. However, CI chondrites have been subject to terrestrial alteration, which may have modified their original mineralogies and redistributed aqueously mobile elements. NASA's OSIRIS-REx mission delivered pristine material from the carbonaceous asteroid Bennu, and initial analyses revealed close compositional similarities with CI chondrites. This provides a valuable opportunity to refine estimates of the bulk Solar System composition. We report concentrations of 63 elements in aggregate (bulk, unsorted) Bennu material (representing a combined mass of ~8 g) analyzed across four laboratories. Within analytical uncertainties, we find no statistically significant differences among aliquots of two aggregate samples that were homogenized into powders, or among subsamples (including two previously reported) of a non-homogenized aggregate sample. Nor do we find measurable variations among sample masses ranging from tens of milligrams to several grams. These findings indicate that the material returned from Bennu is well mixed at the aggregate scale. Combining all of the data, we derive an average bulk elemental composition that we term Best Bulk Bennu ($B^3$). Comparison of $B^3$ with an average CI chondrite composition shows excellent agreement within analytical uncertainties, confirming that terrestrial alteration of CI chondrites is effectively isochemical for non-highly volatile elements. $B^3$ and the CI chondrite average composition compare similarly to the solar photospheric composition. Our findings suggest that both CI meteorites and returned samples of carbonaceous asteroids are the best available materials for estimating the bulk Solar System composition of non-highly volatile elements.**

# 1. Introduction

Knowledge of the bulk chemical composition of our Solar System is fundamental to astrophysics, cosmochemistry, and planetary science. It provides essential constraints for the origin and formation of the Solar System and terrestrial planets, serves as a baseline composition for studying other stars and exoplanetary systems, and is crucial for testing models of stellar evolution, element nucleosynthesis, and galactic chemical evolution (Lodders et al., 2025a). Starting in 1925, astronomers began using absorption lines in the spectra of stars (Payne, 1925) and the solar photosphere (Russell, 1929) to determine the Sun's bulk composition and, by extension, the elemental abundances of the Solar System, given that the Sun contains more than 99% of its mass. To date, the abundances of 68 of the 83 naturally occurring stable and long-lived elements have been determined through spectral analysis of the solar photosphere, sunspots, and corona (e.g., Asplund et al., 2021; Truong et al., 2024; Lodders et al., 2025a). However, significant challenges remain in deriving solar abundances from spectral data: (1) the abundances are not directly measured but inferred through one-dimensional or three-dimensional modeling of the solar atmosphere and radiative transfer; (2) the abundances of some elements, including As, Se, Cd, Sb, Te, Cs, Ta, Pt, and U, cannot be determined by this method due to their weak spectral lines or line blending; and (3) the associated uncertainties are considerable, with only 10–20 elements determined to within 15%, while uncertainties for most other elements are around 30–50% (see review in Lodders et al., 2025a).

Contemporaneously with the early efforts to estimate solar abundances from photospheric spectra, an alternative approach was developed: determining the bulk solar composition through the analysis of meteorite samples (see Lodders, 2020, for a detailed historical account). The early approach used weighted averages of meteoritic silicate, metal, and sulfide samples (e.g., Goldschmidt, 1922, 1938; Brown, 1949; Suess, 1949). Chondrite averages were employed by Merrill (1909, 1915, 1916), Harkins (1917), Saslawsky (1933), and Suess and Urey (1956). Urey (1952) thought that the carbonaceous chondrites were probably the best choice but briefly changed his mind to favor ordinary chondrites (Urey and Edwards, 1956). The Ivuna-type (CI) carbonaceous chondrites, only identified as a separate group by Wiik in 1956, were found to be the best match to abundances expected from nuclear systematics by Reed, Kigoshi, and Turekevitch (1960), and later were also found to be a close match to the elemental abundances derived from solar photospheric spectra (e.g., Anders, 1971; Cameron, 1968, 1973). Compared to spectroscopy, the determination of the solar composition from meteorites provides several advantages: (1) it is based on direct measurements referenced to standards of known composition, rather than model-dependent inferences; (2) it enables the determination of all 83 primordial, naturally occurring elements, with the caveat that abundances of highly volatile elements (H, C, N, O, and the noble gases) are not representative of those of the Solar System; and (3) it achieves higher analytical precision, often better than 5–10% for most elements. Consequently, CI chondrites have become the standard reference for establishing the most accurate and precise estimates of the bulk Solar System composition.

Only 10 CI chondrites have been identified: five are observed falls (Alais in 1806, Orgueil in 1864, Tonk in 1911, Ivuna in 1938, and Revelstoke in 1965), four are finds from Antarctica (Yamato 86029, Yamato 86737, Yamato 980115, and Yamato 980134), and one is a find in a hot desert environment (Oued Chebeika 002; Gattacceca et al., 2025; Zhu et al., 2025; Broussard et al., 2026). Although CI chondrites have long served as key references for the bulk Solar System composition, concerns remain about sample heterogeneity, as well as terrestrial alteration during their residence on Earth's surface, recovery, and long-term curation in museum environments (e.g., Gounelle and Zolensky, 2001; Bland et al., 2006; Barrat et al., 2012; King et al., 2020; Greenwood et al., 2023). Due to their porous and friable nature, CI chondrites are especially susceptible to modification by atmospheric O and humidity (e.g., Daubree, 1864; King et al., 2020; Greenwood et al., 2023).

In addition to terrestrial alteration (mainly from sulfide oxidation and hygroscopic salt formation), terrestrial contamination is a long-standing issue for all meteorites. For CI chondrites, contamination is known to affect H and O elemental and isotopic abundances through oxidation and water uptake (Boato, 1954). Many samples of the Orgueil CI chondrite are heavily contaminated with Hg (e.g., Ehmann and Lovering, 1967; Hintenberger et al., 1973; Palme and Beer, 1993) or with pollen and organics from museum drawers (Anders et al., 1964). However, for other elements, no clear indications of widespread contamination have been reported.

Two sample-return missions to near Earth carbonaceous asteroids — JAXA's Hayabusa2 to Ryugu and NASA's OSIRIS-REx to Bennu — have delivered materials directly from space that were immediately placed into ultra-clean, dry and $O_2$-free curation environments (e.g., under dry $N_2$). This approach minimized the complications associated with the long and challenging handling histories of meteorites on Earth (Righter et al., 2023; Yokoyama et al., 2023; Lauretta and Connolly et al., 2024).

In the case of the OSIRIS-REx mission, 121.6 g of regolith was brought back from Bennu, including loose aggregate material (unsorted particles ~<0.5 cm) and individual stones (>0.5 cm) (Lauretta and Connolly et al., 2024). These samples consist of at least three different particle types that may relate to the different types of boulders observed on Bennu by the spacecraft (Lauretta and Connolly et al., 2024; Connolly and Lauretta et al., 2025; Ryan et al., 2026). Similar to CI chondrites and Ryugu samples, the Bennu samples are composed of phyllosilicates (>80 vol.%), along with varying amounts of sulfide, magnetite, carbonate, phosphate, other rarer evaporite minerals, and minor to trace amounts of olivine and pyroxene (McCoy and Russell et al., 2025; Zega and McCoy et al., 2025). A 20.66 mg aggregate sample (OREX-803015-0), analyzed by quadrupole inductively coupled plasma mass spectrometry (Q-ICP-MS) at Washington University in St. Louis (WUSTL) and by high-resolution (HR)-ICP-MS at Lawrence Livermore National Laboratory (LLNL), closely matched the composition of CI chondrites (Lauretta and Connolly et al., 2024; Barnes and Nguyen et al., 2025). No terrestrial alteration products, such as Ca and/or Mg sulfates, which are frequently found in CI chondrites, have been detected in Bennu samples. Thus, the CI-like yet pristine samples returned from Bennu (and Ryugu) offer an unprecedented opportunity to refine estimates of the bulk Solar System composition.

Here, we present concentrations of 63 elements in aliquots of large, homogenized Bennu aggregate samples (representing ~8 g total), as well as subsamples of a smaller non-homogenized aggregate Bennu sample, analyzed across multiple laboratories using different methods. In the case of the non-homogenized aggregate sample, we include data from subsamples previously reported in the literature (Lauretta and Connolly et al., 2024; Barnes and Ngyuen et al., 2025) in addition to data acquired in this study. Using the new and previously published elemental data, we define a Best Bulk Bennu ($B^3$) composition. We then compare $B^3$ to the average compositions of CI chondrites, Ryugu samples, and the Sun as inferred from spectroscopic observations. Finally, we discuss implications for refining estimates of the bulk Solar System composition.

# 2. Methods

## 2.1 Samples

All Bennu samples used in this study, along with their respective masses, are listed in **Table 1**. The relationships between parent samples and subsamples are shown in **Supplementary Figures S1–S3**.

Two large aggregate samples retrieved from the sample chamber (or "head") of the spacecraft's Touch-and-Go Sample Acquisition Mechanism (TAGSAM) (Bierhaus et al., 2018) were allocated to separate working groups within the OSIRIS-REx Sample Analysis Team (Lauretta et al., 2023) for the preparation of homogenized powders. The Sample Elements & Isotopes Working Group (SEIWG) received aggregate sample OREX-800117-0, weighing 1288 mg; four aliquots ranging from 14 mg to 464 mg of its homogenized powder were analyzed at four different laboratories (**Table 1**). The Sample Organic Analysis Working Group (SOAWG) received aggregate sample OREX-800107-0, weighing 6425 mg; a 10.56 mg aliquot of the resulting homogenized powder was analyzed at one laboratory (**Table 1**).

These two large aggregate samples were independently homogenized by their respective working groups. The SOAWG aggregate sample was powdered and homogenized following the procedure described in Ryan et al. (2026). Briefly, the sample was mechanically ground to reduce compositional heterogeneity and better represent the bulk composition. Sample processing took place under a laminar flow bench in a clean room using quartz mortar and pestle sets. All glassware and tools in direct contact with the samples were baked at 500 °C overnight in air. As a processing control, ~7 g of granular fused silica chips (Isomass Analytics, 0.85–1.7 mm), preheated at 500 °C overnight, underwent the same preparation steps as the Bennu aggregate. Approximately 2 g of the Bennu aggregate was powdered at a time by crushing it in cycles to minimize frictional heating. The mortar was rotated in the opposite direction of the pestle. The resulting powder was transferred to the original glass container using a quartz funnel, with the mortar interior scraped clean using the end of a glass pipette. This procedure was designed to minimize organic contamination. After homogenization, the resulting powder was scanned by XCT, revealing that the majority (>95%) consisted of particles smaller than 300 µm (Glavin et al., 2025). In contrast, the median particle size of the non-homogenized aggregate is 1.2 mm (Lauretta and Connolly et al., 2024).

The SEIWG aggregate sample was processed at LLNL. It was homogenized using a fused quartz mortar and pestle set dedicated for use with Bennu samples. The mortar and pestle were successively cleaned with fused silica, MQ water, 2N hydrochloric acid (HCl), and methanol to minimize inorganic contamination, with a final rinse with MQ water. The grinding method followed the same procedure as for the SOAWG aggregate sample (Ryan et al., 2026). The homogenized sample was then divided into multiple aliquots for distribution (**Table 1** and **Supplementary Figure S2)**. A fused silica sample (180 mg) was crushed in parallel as a control. No baking of glassware or tools at 500 °C was conducted during the SEIWG sample preparation as organic contamination is not a concern for this working group. Their homogenization procedure was designed to minimize inorganic contamination.

In addition to these two large, homogenized aggregates, three small sub-samples from non-homogenized aggregate parent sample OREX-800033-0, also retrieved from the TAGSAM head, were analyzed (**Table 1**). The analytical data for two of these sub-samples were previously reported (Lauretta and Connolly et al., 2024; Barnes and Ngyuen et al., 2025), and one is newly reported herein (**Table 1**).

At each laboratory, at least one CI chondrite sample (Orgueil or Ivuna) was analyzed for quality control. The fused silica chips used as blank controls for the SEIWG and SOAWG powdering procedures were also analyzed to monitor potential contamination during preparation. Two original fused silica chips, unbaked and baked (ie., baking is one step during SOAWG fused silica processing, see above), are also measured for comparison. All these samples were analyzed together with the Bennu samples using the same methods, described below.

## 2.2 Analytical Techniques

Aliquots of the large, homogenized powders and small aggregates were analyzed at four laboratories: WUSTL, LLNL, the University of California, Berkeley (Berkeley), and the University of British Columbia (Pacific Centre for Isotopic and Geochemical Research, UBC). A summary of the laboratories, instruments, calibration standards, and analytical methods is provided in **Table 2** and **Supplementary Table 1**.

**WUSTL**: Samples were dissolved using a mixture of concentrated hydrofluoric acid (HF) and nitric acid ($HNO_3$) in a 3:1 ratio and heated for 48 hours at 170 °C in closed beakers. Following dissolution, the samples were fluxed in concentrated $HNO_3$ and HCl. During the $HNO_3$ flux step, 1 mL of hydrogen peroxide ($H_2O_2$) was added slowly to remove any remaining organics. Once the dissolution process was complete, the solutions were brought up in 5 mL of 0.5 M $HNO_3$. From these, aliquots were prepared by weight at total dilution factors of approximately 5,000 and 50,000 in a final 2 vol.% $HNO_3$ matrix for analysis. All elemental analyses were carried out using a Thermo Fisher iCAP Q-ICP-MS. Detector cross calibration and instrument tuning were performed prior to each analytical session. Analyses were conducted in kinetic energy discrimination (KED) mode. A 5 $\mu g\,g^{-1}$ internal standard solution containing Rh and Re was used throughout the analytical run: Rh was applied for internal correction of elements Ba and lighter, while Re was used for elements La and heavier. Each sample was analyzed three times at both ~5,000 and ~50,000 dilution factors. Linear calibration curves were constructed using a synthetic chondrite reference solution (prepared at dilution factors ranging from ~3,000 to ~70,000) and two certified rock standards (US Geological Survey (USGS) BHVO-2 and BIR-1, analyzed at dilution factors of ~5,000 and ~50,000). Calibration curves were evaluated for each element and adjusted as needed. Reference materials with concentrations significantly outside the range observed in the samples were excluded from the relevant calibrations. Final elemental concentrations for each sample were determined by averaging the results of three replicate measurements: trace elements were quantified using the ~5,000 dilution data and major elements using the ~50,000 dilution data. If the averages from both dilution factors agreed, the results from all six replicates were combined to calculate the final concentration. The analytical uncertainties are the standard deviation (SD) of the three or six replicate measurements.

**LLNL**: The homogenized powder was digested in pre-cleaned Savillex® PFA vials using a 2:1 mixture of concentrated HF and $HNO_3$ acids. The sample was dried and re-dissolved in a 3:1 mixture of concentrated $HNO_3$ and HCl. The remaining residue was placed in a Parr bomb with 2:1 HF and $HNO_3$ acids in an oven at ~220 °C for 72 hrs. Major and trace element concentrations were measured via high-resolution sector field ICP-MS (SF-ICP-MS) Thermo Fisher Element XR. An aliquot of the bulk digestion representing approximately 0.5 mg was removed, dried down, and redissolved in 5 ml of internal standard solution. The internal standard solution used at LLNL consists of 2% $HNO_3$ + 0.005 M HF, that is spiked with 1 ng $g^{-1}$ of In, Re, and Bi. Indium was used as the internal standard element at all resolutions to correct for instrument drift and sample matrix effects during measurement. To ensure measurement quality, a series of solution standards and certified rock standards (USGS) were prepared in parallel and diluted using the same internal standard solution. The instrument was fitted with standard 'H' sample and skimmer cones, and solutions were aspirated using a 100 ml $min^{-1}$ nebulizer (Glass Expansion). The Element was tuned for a mix of maximum sensitivity and reduced oxide formation, resulting in typical count rates between $1.2 \times 10^6$ and $1.5 \times 10^6$ cps for 1 ng $g^{-1}$ of In and oxide formation (UO/U) at ~5% or less. Most elements were measured using low-resolution mode; however, elements that are commonly subject to interferences (e.g., transition metals), were measured at medium or high resolution (where low resolution is R = 300, medium resolution is R = 4,000 and high resolution is R = 10,000, where $R = m/\Delta(m)$ and $m$ is mass). Measured intensities (in counts per second) were quantified using a combination of reference solutions (Inorganic Ventures) and USGS rock standards. Each sample was measured using multiple dilutions of the sample stock solution, and elements were individually quantified using the most appropriate dilution based on the count rates. The accuracy of these analyses was assessed using the USGS basalt standard BHVO-2, with most concentrations falling within 10% of reference values. Hafnium and W concentrations were determined by the isotope dilution method (Kleine et al., 2004; Kruijer et al., 2014). For this purpose, a separate ~10% solution aliquot taken from the bulk digestion was spiked with a mixed $^{180}$Hf-$^{183}$W tracer. Following purification of Hf and W by ion chromatography, Hf and W isotope dilution measurements were performed using the ThermoScientific Neptune Plus MC-ICP-MS. The total masses of Hf and W blanks associated with this procedure were 3 and 16 pg, respectively, which is negligible given the analyte amounts.

**Berkeley**: Samples of Bennu and the Orgueil CI chondrite (Muséum national d'Histoire naturelle, MNHN-Paris) were weighed in an Al weighing boat on a Mettler analytical balance. Each sample, along with a carrier solution containing Be and Cl, was dissolved in a 3:1 mixture of concentrated HF/$HNO_3$ in Parr Teflon digestion vessels placed in an oven at 120–130 °C for 20–24 hours. After dissolution, the chloride fraction was separated as AgCl for analysis of cosmogenic $^{36}$Cl (not reported here), and the remaining sample was fumed three times to dryness with 0.5 ml concentrated $HClO_4$. The residue was dissolved in dilute HCl, and a small aliquot (~5%) of the dissolved sample was taken for chemical analysis. From these solutions, two aliquots were prepared with dilution factors of approximately 12,000x (dilution-1) and 60,000x (dilution-2) in a matrix of 0.5 M HCl. The concentrations of Al, K, Ca, Ti, Mn, Co, and Ni in dilution-1 and of Be, Mg, S, and Fe in dilution-2 were measured by ICP–optical emission spectrometry (OES), using a Thermo Fisher iCAP 6300 Duo instrument. The instrument was calibrated with the standard solutions used for analysis of all extraterrestrial samples in this laboratory. Each dilution was measured three times. A single elemental line was used for Al, S, K, and Co; the average of two lines was used for Be, Mg, Ca, and Mn; and the average of three elemental lines was used for Ti, Fe, and Ni. The performance of the instrument was verified using the measured Be concentration, which agrees within 1–3% with the added amount of Be carrier (average yield = 100.4 ± 1.3%). Typical uncertainties (2SD) of the reported values are 2–3% for Be, Mn, Fe, and Ni; 3–5% for Mg, Al, Ca; and 5–10% for S, K, Ti, and Co.

**UBC**: Three splits of 121 mg of the homogenized powder were dissolved in an Anton Paar Multiwave 5000 microwave system using an acid mixture of 6 ml 15 M $HNO_3$, 2 ml 28 M HF, and 4 ml 12 M HCl at a temperature and pressure of up to 240 °C and 60 bar for 1 hour. All digested samples, quality control materials, and blanks were re-dissolved and stored in 3 M $HNO_3$ as a "stock solution" from which aliquots were further processed in preparation for ICP-MS analysis. Analyses of 52 major and trace elements were carried on high-resolution SF-ICP-MS Thermo Scientific Element 2 in two separate sessions, one each for major and trace elements, and in three resolution modes — low ($R$ = 300, LR), medium ($R$ = 4,000, MR), and high ($R$ = 8,000, HR) — to account for isobaric and molecular interferences on specific masses of interest. The instrument was optimized for the highest sensitivity, signal stability across the whole mass range, and low oxide formation using a PFA Scott-type spray chamber, self-aspirating PFA ESI ST nebulizer (100 ul min$^{-1}$), and Pt cones. Two different dilutions were prepared from the stock solutions, one for the trace elements at ~2,000 dilution factor and from this another at ~100 dilution factor for major elements. Efforts were made to keep all analytes well above the detection limits but below 100 ng g$^{-1}$ to prevent strong matrix effects in plasma, except for Ca, which was run at higher levels up to 1 µg g$^{-1}$. Samples, standard solutions, blanks, and reference materials were all re-dissolved in the same batch of 2% $HNO_3$ + 0.05% HF containing 1 ng $g^{-1}$ In for internal standardization. For Ca and Sr external calibration, single-element calibration standard solutions were prepared to monitor their mutual interference due to dimer and doubly charged ion formation. For all other elements, a five-point calibration from 0 to 100 ng $g^{-1}$ was established using mixed synthetic standard solutions. Matrix effects were corrected using measurements of an aliquot of the Orgueil-T sample. Orgueil-T is a powdered aliquot of a >3 g sample whose trace-element homogeneity has been demonstrated previously (Beer et al., 1984; Rocholl and Jochum, 1993; Jochum et al., 2000). The meteorites Orgueil and Ivuna, as well as the geological reference material BCR-2 (USGS Basalt Columbia River), were included as quality control materials. Every sample run was followed by a clean acid solution to prevent and minimize carry-over and memory effects. Results reported are averages of at least two duplicates, each with at least three replicate analyses, from different digestions and sessions. Overall external precision of all aliquots measured, including duplicate digests analyzed in different sessions, varies within 10% for the major elements and most trace elements, with some trace elements varying up to 15%.

# 3. Results

## 3.1 Blank fused silica samples and laboratory total procedural blanks

Two types of blank samples were analyzed in this study. The first type consists of fused silica chips used during both the SEIWG and SOAWG homogenization processes as contamination controls. The second type comprises subsequent analytical laboratory total procedural blanks, which include the acid digestion, dilution, labware, and instrumental analysis workflow. Results for the fused silica blanks and laboratory procedural blanks are provided in **Supplementary Tables 2** and **3**, respectively. **Supplementary Figure S4 and S5** show the elemental compositions of the two fused silica blanks (SEIWG and SOAWG), normalized to the original fused silica and to the average Bennu composition (Discussion Section 4.2), respectively. For most elements, the fused silica blanks exhibit negligible contributions relative to Bennu (well below analytical uncertainty). However, a few elements, such as U, the light rare earth elements (LREE: La and Ce), W, and Th, display notable enrichments (~3-5 times relative to the average Bennu composition), especially in the SOAWG fused silica blank. The magnitude and inconsistency of these enrichments (i.e., La, Ce, and W) between the SEIWG and SOAWG blanks indicate contamination introduced during the SOAWG powdering process. This is supported by similar spikes (La, Ce, and W) observed in the SOAWG Bennu sample. In contrast, elements such as U and Th show comparable enrichments in both SEIWG and SOAWG blanks, suggesting that they are inherent to the commercial fused silica chips rather than introduced during processing.

Beyond contamination associated with homogenization, contributions from analytical laboratory procedural blanks (acids, labware, and instrument backgrounds) are negligible and not subtracted (**Supplementary Table 3**). Although a few elements (e.g., Na, Al, P, K, and Ca) in some laboratory procedural blanks were above detection limits, the corresponding Bennu results from those laboratories do not show noticeable offsets relative to results from other laboratories. We therefore conclude that no measurable laboratory-derived contamination was introduced during sample digestion and analysis stages.

## 3.2 CI chondrites (Orgueil and Ivuna)

At each laboratory, at least one CI chondrite sample (Orgueil or Ivuna) was analyzed alongside the Bennu samples for quality control. The results are summarized in **Table 3** and **Figure 1**. Geological reference materials (BCR-2) for quality control are also analyzed and presented in **Supplementary Table 4**.

Overall, the CI chondrite measurements in all laboratories agree with the average CI values reported in the literature (Lodders et al., 2025a). Minor discrepancies can be attributed to sample heterogeneity, as the masses of CI chondrites analyzed at different laboratories varied from only 1 mg to 510.2 mg (**Table 3**), and the chemical heterogeneity of CI chondrites is known to vary as a function of sample mass (e.g., Barrat et al., 2012; Palme and Zipfel, 2022). Such "nugget effects" likely explain the small deviations observed in **Figure 1**.

In addition to these small deviations, a few elements show large (>20 %) departures from the average CI value (Lodders et al., 2025a) at individual laboratories (see labeled elements in **Figure 1**). For example, Sn abundances in two Orgueil samples (from two separate collections) measured at WUSTL are more than 20% lower than the CI average, which we attribute to an analytical artifact. This may be related to the use of HCl during sample digestion, where Sn can be lost during evaporation as $SnCl_4$ or during precipitation as insoluble $SnO_2$. The extent of Sn loss during evaporation in HCl does not appear to be reproducible, according to the literature (Bragagni et al., 2018; Kirchenbaur et al., 2018). We therefore paid close attention to elements with large deviations from CI values at different laboratories. When calculating $B^3$ (Discussion Section 4.2), we exclude data for any element-laboratory pair where the Bennu results show a similarly clear offset (>20%) from the inter-laboratory consensus.

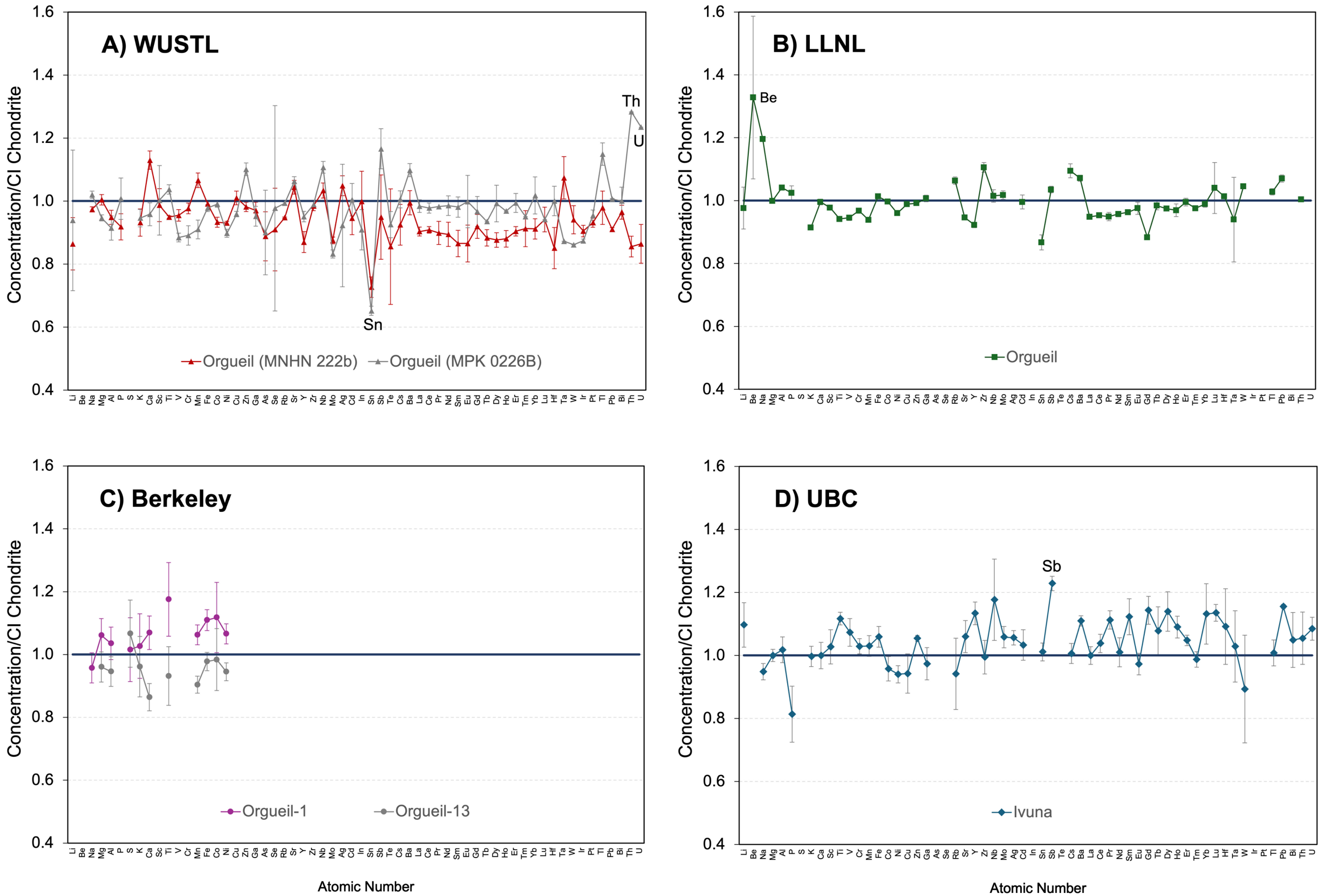
A) WUSTL
B) LLNL
C) Berkeley
D) UBC
Concentration/CI Chondrite
Atomic Number
1.6
1.4
1.2
1.0
0.8
0.6
0.4
Th
U
Sn
Be
Sb
Orgueil (MNHN 222b)
Orgueil (MPK 0226B)
Orgueil
Orgueil-1
Orgueil-13
Ivuna

***Figure 1. Bulk elemental compositions of CI meteorites (Orgueil and Ivuna) measured alongside Bennu samples in the same analytical sessions at different laboratories.*** *Concentrations are normalized to the literature averages for CI chondrites (Lodders et al., 2025a). Elements that deviate by more than 20% from the CI average are labeled. Data for these elements from individual laboratories are excluded from the final Best Bulk Bennu ($B^3$) calculation if these elements also differ from the results of other laboratories for Bennu samples.*

## 3.3 Small sub-samples of the non-homogenized aggregate sample (OREX-800033-0)

**Table 1** and **Supplementary Figure S1** show the split of the non-homogenized parent sample OREX-800033-0 into small subsamples (9.85–20.66 mg) analyzed in this study and reported in the literature (Lauretta and Connolly et al., 2024; Barnes and Ngyuen et al., 2025). These subsamples consist of hundreds of millimeter-sized and smaller particles. Analytical data are compiled in **Table 4** and **Figure 2** for side-by-side comparison and to assess any intrinsic differences among subsamples derived from the same non-homogenized parent sample.

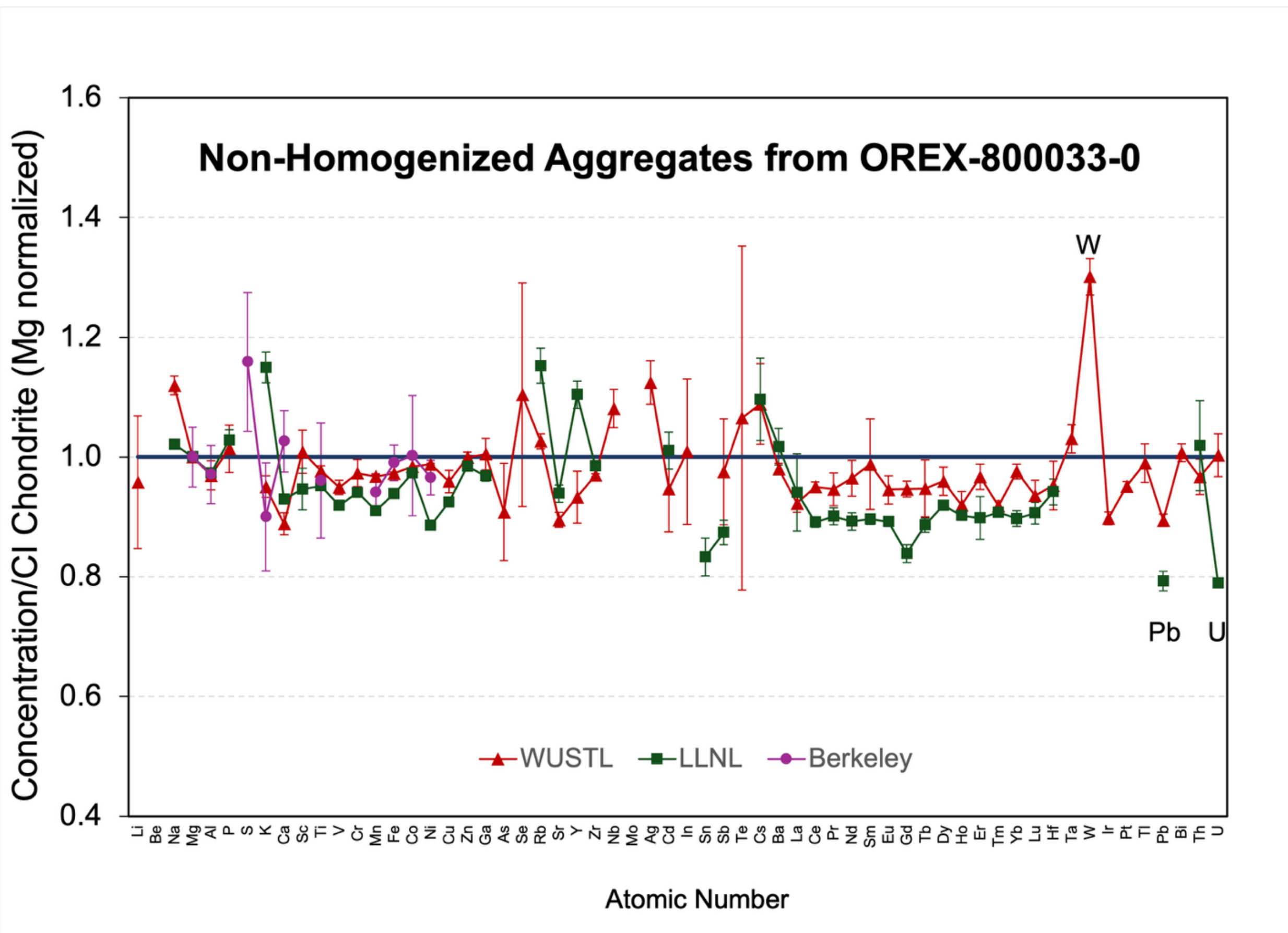

***Figure 2: Bulk elemental compositions of small sub-samples of the non-homogenized aggregate OREX-800033-0.*** *Concentrations are double-normalized to Mg and the literature average values for CI chondrites (Lodders et al., 2025a). The uncertainties shown are* ±1 SD. *The data from WUSTL and LLNL were previously reported (Lauretta and Connolly et al., 2024;*

*Barnes and Ngyuen et al., 2025). The WUSTL (OREX-803015-0) and LLNL (OREX-803015-101) samples are splits from the same dissolved solution. The Berkeley sample (OREX-803047-0) is a different subsample from the same parent sample OREX-800033-0. Elements that deviate by more than 20% from the CI average are labeled.*

In **Figure 2**, elemental concentrations were double-normalized to Mg and the average CI values determined by Lodders et al. (2025a) to eliminate variations in absolute concentrations caused by differing amounts of absorbed $H_2O$ and phyllosilicate $H_2O$/OH. Only three elements — W, Pb, and U — deviate by more than 20% from the CI average. These three elements exhibit notable inter-laboratory discrepancies between LLNL and WUSTL. However, this difference cannot be attributed to sample heterogeneity because the aliquots analyzed at LLNL and WUSTL were even splits of the same sample digestion (**Table 1** and **Supplementary Figure S1**). We consider the discrepancies in W, Pb, and U to reflect analytical challenges and standardization issues rather than true compositional differences. Tungsten is known to be difficult to analyze using single-collector ICP-MS, and there can be additional W blank contributions from Teflon and metal tools (Kleine et al., 2004; Kruijer et al., 2014). Lead is also well known to contaminate samples/standards easily in the laboratory.

As illustrated in **Figure 2**, the elemental data obtained by all three laboratories are consistent with each other within analytical uncertainties. Overall, there are no measurable differences among small (~10–20 mg), non-homogenized subsamples derived from the same parent sample (OREX-800033-0).

## 3.4 Large, homogenized aggregate powder for SEIWG (OREX-800117-0)

A large (1288 mg) aggregate, OREX-800117-0, was homogenized into a fine powder to reduce potential compositional heterogeneity for better representing the bulk composition. Four aliquots of this homogenized powder were analyzed at all four laboratories in this study. The results are summarized in **Table 5** and **Figure 3**. The comparison between the compositions of the homogenized SEIWG sample (OREX-800117-0) and the non-homogenized aggregate sample OREX-800033-0 analyzed in the same laboratory is shown in **Supplementary Figure S6**.

To first order, data from the four laboratories that analyzed aliquots of the homogenized SEIWG sample (OREX-800117-0) show similar elemental patterns relative to average CI chondrites and are broadly consistent with each other (**Figure 3** and **Table 5**). Nonetheless, interlaboratory discrepancies and outliers are evident. For example, Li, Na, Ca, Ba, W, and U measured at UBC, Tl measured at WUSTL, and S measured at Berkeley deviate from CI values by more than 20%. We attribute these deviations to analytical outliers rather than true sample heterogeneity, based on the following:

1. Overall interlaboratory agreement: Despite different digestion and calibration methods, for most elements, at least three laboratories agree within analytical uncertainty, indicating homogeneity of the powder.

2. UBC data show slightly more deviations (often depletions) from the CI average than data from other laboratories. UBC is the only laboratory that used microwave-assisted digestion and synthetic calibration standards with a subsequent matrix correction derived from Orgueil data. This may explain the slight discrepancies.

3. Contamination-prone elements: W, Tl, and U are known to be susceptible to laboratory contamination, especially at low levels. The apparent Tl anomaly may also reflect uncertainties associated with the calibration standards.

4. Lack of mineralogical/geochemical control: These outlier elements do not relate to any specific mineral phases (e.g., carbonates or phosphates) or share common geochemical affinities (e.g., lithophile, siderophile, chalcophile) or aqueous mobilities (e.g., alkali elements associated with brines; McCoy and Russell et al., 2025), and no known natural process would selectively enrich or deplete them.

In summary, we interpret the observed differences as laboratory-related rather than reflecting real compositional heterogeneity. OREX-800117-0 is homogeneous to the level of our analytical uncertainties, and its composition is consistent with CI values.

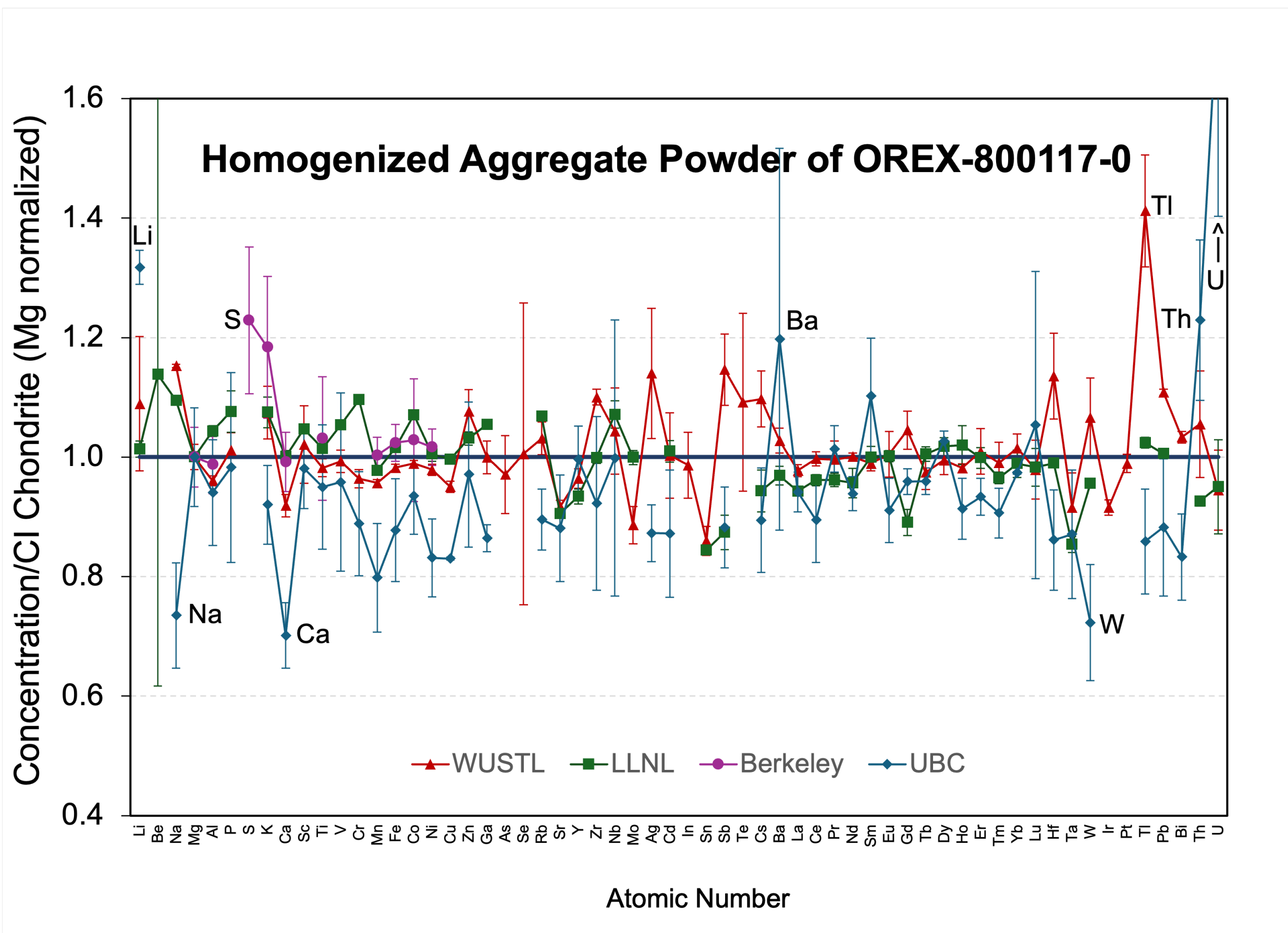

***Figure 3: Bulk elemental compositions of the large homogenized aggregate powder for SEIWG (OREX-800117-0).*** *Concentrations are double-normalized to Mg and the literature average values for CI chondrites (Lodders et al., 2025a). The uncertainties shown are ±1 SD. All samples are aliquots from the same homogenized aggregate powder: WUSTL(OREX-800117-*

*107); LLNL (OREX-800117-111); Berkeley (OREX-800117-103); UBC (OREX-800117-109). Elements that deviate by more than 20% from the CI average are labeled.*

## 3.5 Large, homogenized aggregate powder for SOAWG (OREX-800107-0)

Another large aggregate sample, OREX-800107-0 (6425 mg), was homogenized into a fine powder for SOAWG. A 10.56 mg aliquot of this powder was analyzed at WUSTL in this study, and the results are presented in **Table 6** and **Figure 4**. The comparison between the compositions of the homogenized SOAWG sample (OREX-800107-0) and the homogenized SEIWG sample (OREX-800117-0), both analyzed in the same laboratory, is shown in **Supplementary Figure S6**.

As shown in **Figure 4**, the elemental pattern of this sample generally agrees with the average CI values (**Table 6** and **Figure 4**). Four elements (Nb, La, Ce, and W) deviate by more than 20% from CI. Lanthanum, Ce, and W are extreme outliers, enriched by 4.87, 5.02, and 2.60 times the CI values, respectively. No known mineral phases can produce such selective enrichment of these elements while other REEs remain at 1 × CI levels. We therefore conclude that these anomalies are the result of contamination introduced during the homogenization process.

Our conclusion is supported by the analyses of blank fused silica chips used as contamination controls (**Supplementary Table 2** and **Supplementary Figures S4** and **S5**). The SOAWG blank fused silica is also highly enriched in La, Ce, and W, with no enrichment of other REEs. In contrast, the SEIWG blank fused silica shows no enrichment in La, Ce, or W (**Supplementary Figures S4** and **S5**). The SOAWG homogenization process differed from the SEIWG process in its focus on organic cleanliness, requiring all glassware and tools to be heated at 500 °C overnight, a step not included in the SEIWG procedure. This suggests that the SOAWG sample may have been contaminated during homogenization, although the specific sources of La, Ce, and W contamination remains unknown.

After removing La, Ce, and W, the only remaining element with more than 20% deviation from CI is Nb. Because this anomaly does not correlate with other elements, we attribute it to analytical issues rather than a true compositional difference. Overall, the composition of OREX-800107-0 is consistent with the average CI composition.

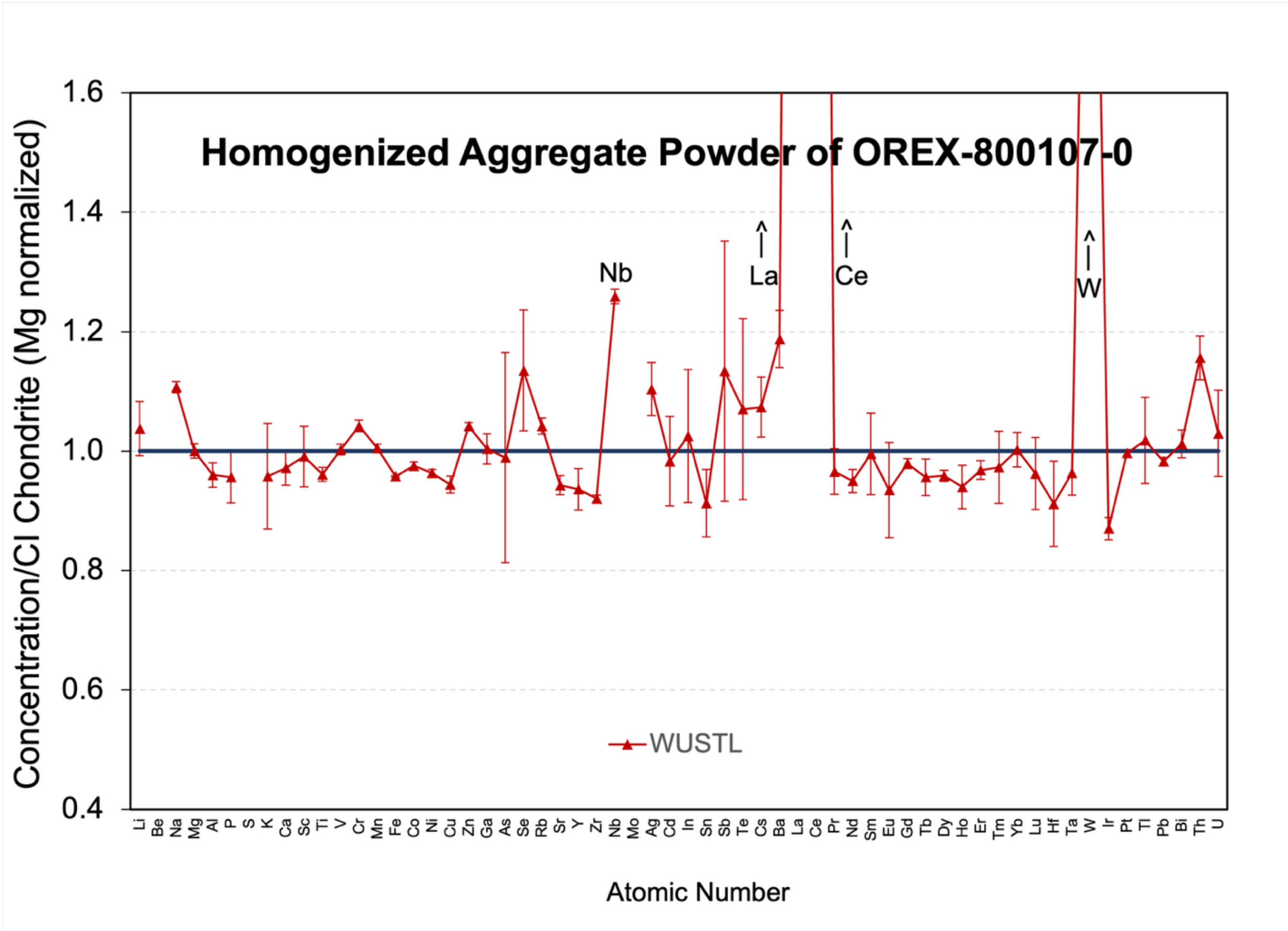

***Figure 4: Bulk elemental compositions of the large, homogenized aggregate powder for SOAWG (OREX-800107-0).*** *Concentrations are double-normalized to Mg and the literature average values for CI chondrites (Lodders et al., 2025a). The uncertainties shown are ±1 SD. Elements that deviate by more than 20% from the CI average are labeled.*

# 4. Discussion

## 4.1 Sample heterogeneity among Bennu aggregates

As summarized in **Table 1**, multiple Bennu aggregate samples have been analyzed for their bulk elemental compositions in this and previous studies (Lauretta and Connolly et al., 2024; Barnes and Ngyuen et al., 2025). These combined datasets allow us to assess the extent of compositional heterogeneity among Bennu aggregates.

The three non-homogenized subsamples of the parent aggregate sample OREX-800033-0 range in mass from 9.85 mg to 20.66 mg and represent random scoops from the OSIRIS-REx collection. As shown in **Figure 2**, no systematic variations in elemental abundances are observed among these aggregates. Minor discrepancies between these samples can be attributed to analytical variability rather than true compositional differences. For example, K concentrations measured at LLNL are significantly higher than those reported by WUSTL and Berkeley (**Figure 2**); however, LLNL and WUSTL analyzed aliquots (split 50/50) from the same dissolved solution of the same subsample, whereas Berkeley analyzed a different subsample. Collectively, these observations indicate that there are no significant or measurable elemental heterogeneities among small (~10–20 mg) non-homogenized aggregates, and the observed discrepancies likely arise from inter-laboratory analytical issues due to differences in instruments, calibration standards, etc. (**Table 2**).

As shown in **Figure 3**, the four aliquots from the homogenized powder of OREX-800117-0 show no systematic differences, aside from minor discrepancies that can be attributed to analytical variability among laboratories. This outcome is unsurprising: given that no measurable elemental heterogeneity was detected among the small, non-homogenized aggregates from parent OREX-800033-0, aliquots of homogenized fine powders should likewise exhibit no detectable heterogeneity.

Furthermore, we compared the average elemental compositions of the three Bennu samples considered in this work (OREX-800033-0, OREX-800117-0, and OREX-800107-0), using all available data from this study and the literature (**Table 1**). The results are presented in **Figure 5**. With the exception of Nb and W, the three samples agree with each other for all elements within ±1 SD analytical uncertainties. The apparent offsets in Nb and W are likely analytical artifact or contamination, as each element was measured by only a single laboratory for the respective samples, and both are known to be analytically challenging or prone to contaminate, as discussed in the Results section. Therefore, we conclude that there are no systematic or measurable differences in bulk elemental compositions among these three Bennu aggregate samples.

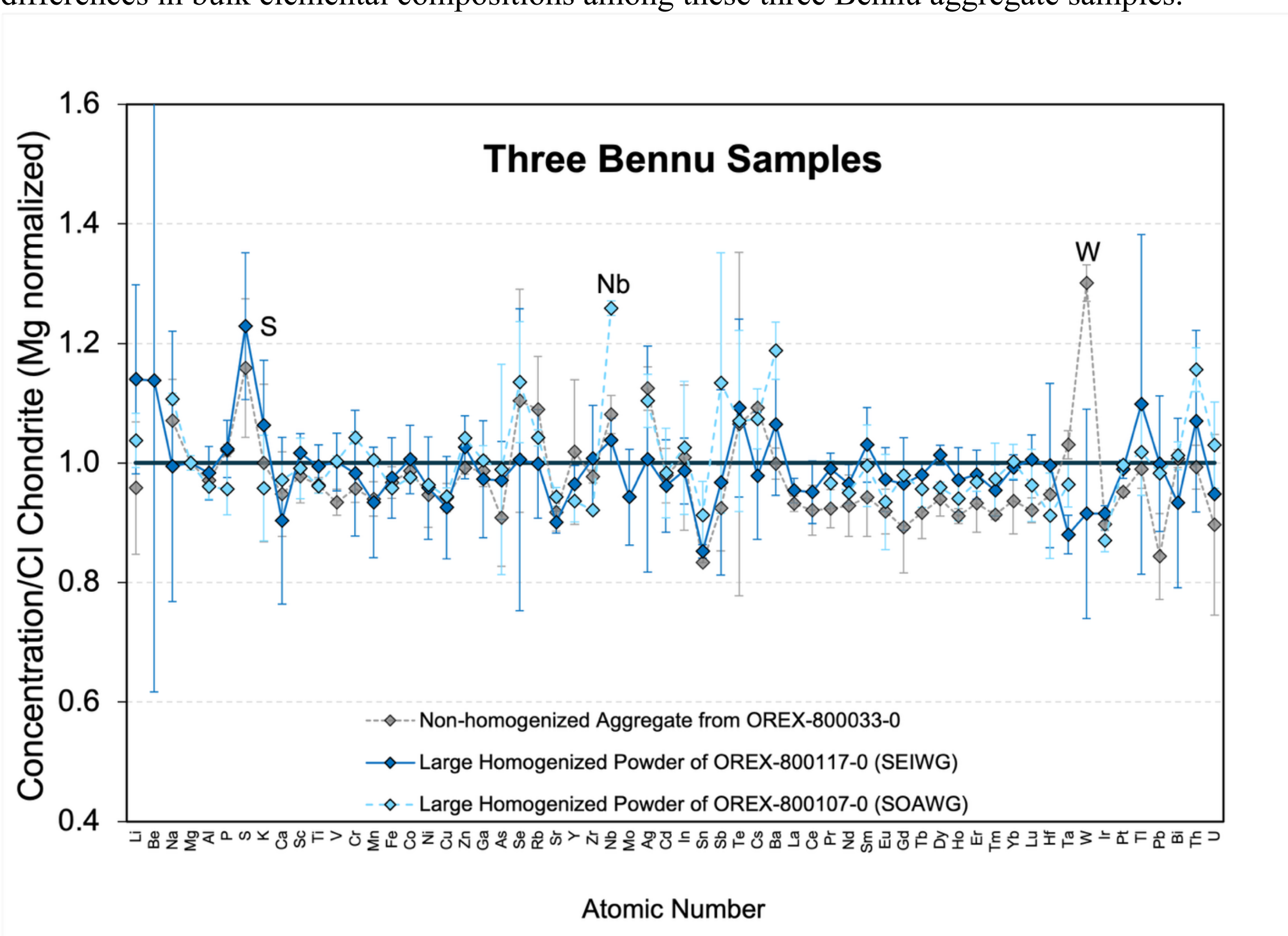

***Figure 5: Comparison of the averages of the Bennu aggregate samples (OREX-800033-0, OREX-800117-0, OREX-800107-0) from four different laboratories.*** *Concentrations are*

*double-normalized to Mg and the literature average values for CI chondrites (Lodders et al., 2025a). The uncertainties shown are ±1 SD. Elements that deviate by more than 20% from the CI average are labeled. Four extreme outliers from individual analyses have been excluded (see Table 7) and are not shown in this figure.*

To further evaluate sample heterogeneity, we investigated whether the small variations among different subsamples could be explained by nugget effects, as observed in Ryugu samples (Yokoyama et al., 2025), associated with secondary minerals such as dolomite, apatite, magnetite, and pyrrhotite. **Supplementary Figure S7** shows correlations between CI-normalized elemental ratios involving elements (e.g., Mn, Ca, Sr, P, Fe, and Ti) associated with these minerals. We further modeled the addition and removal of these secondary minerals relative to the bulk CI chondrite composition following the approach of Yokoyama et al. (2025). Compared with the Ryugu bulk samples reported by Yokoyama et al. (2025), the Bennu bulk samples analyzed in this study exhibit much smaller variations in these elemental ratios. Therefore, unlike the Ryugu samples (Yokoyama et al., 2025; López García et al., 2026), we do not observe significant compositional variations among the Bennu subsamples that can be attributed to nugget effects resulting from heterogeneous distributions of secondary minerals such as dolomite, apatite, magnetite, and pyrrhotite.

Another approach to evaluating nugget effects is to examine the relationship between sample mass and elemental deviation from CI values (Yokoyama et al., 2025). If nugget effects are responsible for the observed elemental ratio variations, the data dispersion should decrease with increasing sample mass (Dauphas and Pourmand, 2015). Yokoyama et al. (2025) demonstrated this relationship for Ryugu samples. Following their method, we plotted sample mass against elemental deviation in **Supplementary Figures S8 and S9**. For sample mass, we considered both the dissolved mass and the homogenized powder mass. We also calculated the expected data dispersion based on the variability observed in smaller samples (see methodology in Yokoyama et al., 2025). Again, unlike the Ryugu samples, we do not observe a clear decrease in elemental variability with increasing sample mass in the Bennu samples.

We therefore conclude the small variations among different Bennu subsamples to be primarily attributable to analytical uncertainties among different laboratories, rather than true sample heterogeneity caused by nugget effects. The difference between Bennu and Ryugu samples is further discussed in Section 4.4.

In summary, our results suggest that aggregates of varying masses — even small aliquots — are representative of the bulk composition of the Bennu material collected by OSIRIS-REx. The overall consistency among the three Bennu aggregates (OREX-800033-0, OREX-800117-0, and OREX-800107-0; **Figure 5**) and between their respective subsamples/aliquots (**Figures 2–4**) indicates effective mixing of material within the TAGSAM head (Bierhaus et al., 2018). This outcome is expected given that the TAGSAM head consists of a single chamber and that materials likely mixed at the aggregate level during sample collection, the two-year return to Earth, atmospheric entry, and subsequent curation handling processes (Lauretta and Connolly et al., 2024).

## 4.2 Best Bulk Bennu ($B^3$) composition

Given that the three parent samples considered in this work have been shown to be homogeneous at the aggregate level, we calculated the average composition of Bennu aggregate material using all the data in **Table 1** from this study and the literature (Lauretta and Connolly et al., 2024; Barnes and Ngyuen et al., 2025). We refer to this average as the Best Bulk Bennu ($B^3$) composition, presented in **Table 7** and **Figure 6**.

The $B^3$ value is a simple (unweighted) mean. We considered several alternative weighting schemes, including (1) weighting by analyzed mass, (2) weighting by total sample mass, (3) weighting by analytical uncertainty, and (4) first averaging aliquots from the same parent sample and then averaging across parents. However, we selected the simple mean for the following reasons:

1. Large differences in analyzed mass introduce bias for mass-based weighting. For example, LLNL digested 464 mg of OREX-800117-0, whereas WUSTL and Berkeley analyzed only 22.22 and 14.42 mg, respectively. Yet, as shown in **Figure 3**, the results from all three laboratories are highly consistent. Weighting by analyzed mass would only be justified if samples were compositionally distinct, which is not observed. Given the grain size (e.g., the median particle size is 1.2 mm; Lauretta et al., 2024), even small aliquots (10-20 mg) consist of hundreds of mm-sized (and smaller) particles and are expected to represent the bulk. Measurements from multiple subsamples show minimal differences, supporting equal treatment of all analyses.
2. Weighting by total sample mass also excessively biases the result. This approach would disproportionately favor the largest sample (OREX-800107-0 for SOAWG; 6425 mg total), relative to small subsamples of OREX-800033-0 (9.85–20.66 mg), resulting in >80–90% overrepresentation. In addition, the largest sample was measured by only one laboratory, whereas other samples were analyzed by three to four laboratories (**Table 1**).
3. Analytical uncertainties reflect precision, not accuracy. Reported errors primarily represent within-laboratory reproducibility. Because laboratories use different methods, instruments, and reference materials, higher precision does not imply higher accuracy, as demonstrated by inter-laboratory comparisons (**Figures 1–4**). Therefore, weighting by analytical error is not justified.
4. Testing of weighted approaches confirmed that poorer results were obtained using such approaches. Weighted averages produced more variable and noisier patterns relative to the simple mean because they became dominated by a small number of analyses or samples.

For these reasons, we adopt the simple mean to compute the $B^3$ values, treating each analysis as an independent yet *representative* measurement of Bennu's bulk composition (however, this may not be applicable if very small (e.g., 1 mg) samples are considered). This approach also facilitates future updates, as additional Bennu analyses can be incorporated directly into the $B^3$ dataset without recalibrating a weighting scheme. A similar approach has been applied in the current compilations of the CI chondrite average composition (e.g., Lodders et al., 2025a).

The only two exceptions are Hf and W, for which the $B^3$ values were chosen solely for one large sample (OREX-800117-111) using the isotope dilution method at LLNL. This method was chosen because it more reliably resolves blank issues, particularly for W, where contamination from Teflon labware is a known concern, and therefore this method yields more accurate concentration data for these elements (Kleine et al., 2004; Kruijer et al., 2014; see also Lodders et al., 2025a for discussion on refractory lithophiles).

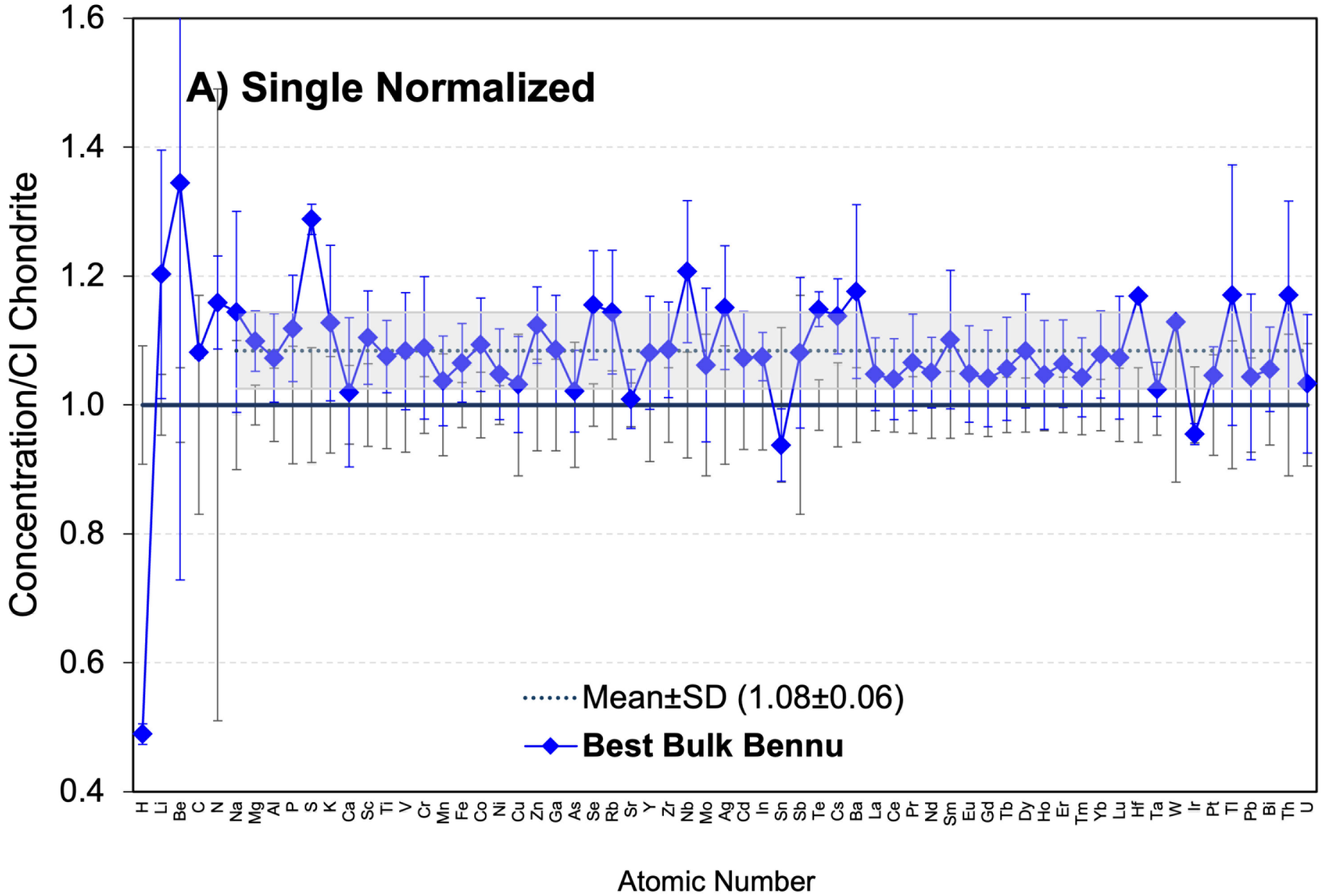

A) Single Normalized
Concentration/CI Chondrite
Mean±SD (1.08±0.06)
Best Bulk Bennu
Atomic Number
H Li Be C N Na Mg Al P S K Ca Sc Ti V Cr Mn Fe Co Ni Cu Zn Ga As Se Rb Sr Y Zr Nb Mo Ag Cd In Sn Sb Te Cs Ba La Ce Pr Nd Sm Eu Gd Tb Dy Ho Er Tm Yb Lu Hf Ta W Ir Pt Tl Pb Bi Th U

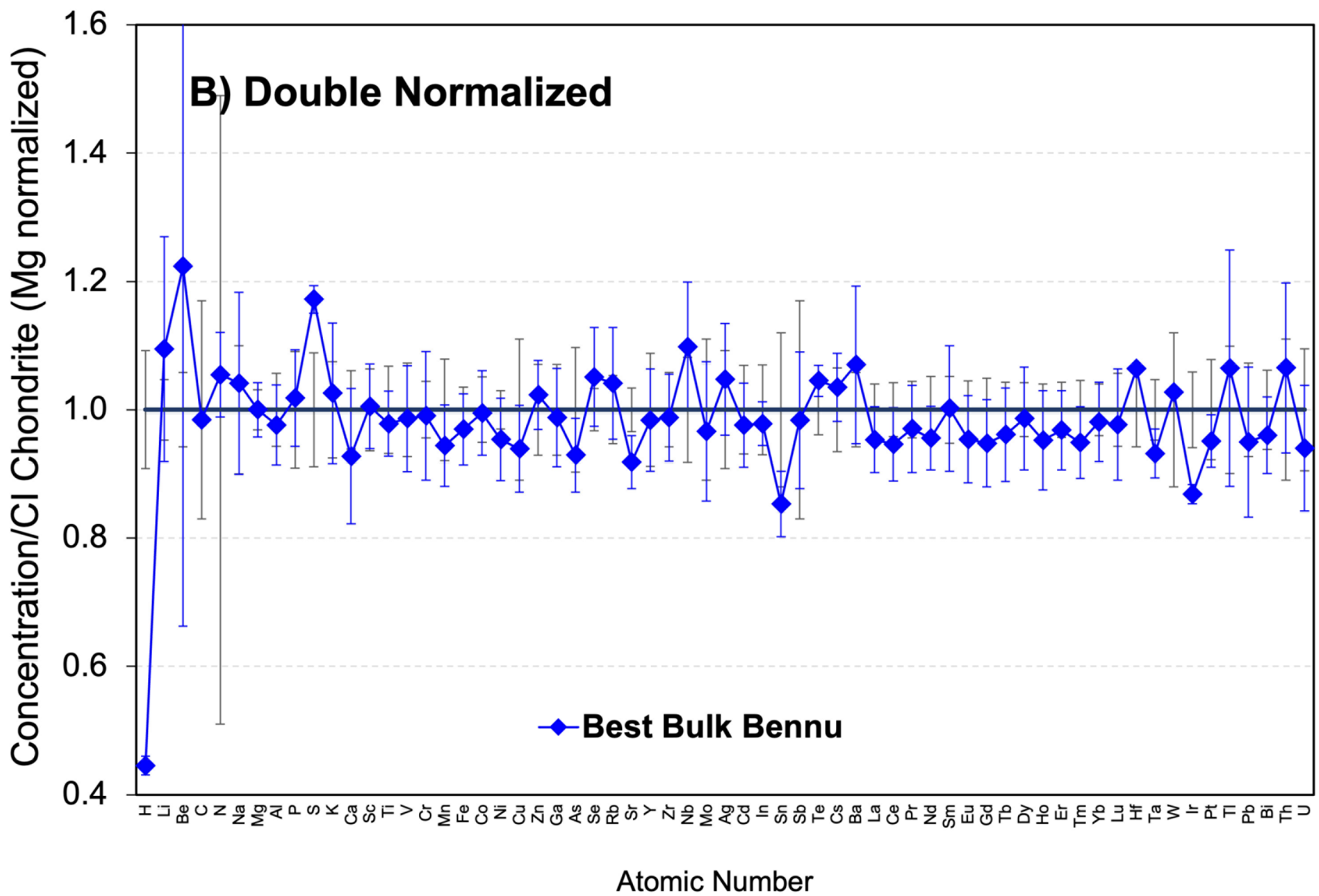

B) Double Normalized
Concentration/CI Chondrite (Mg normalized)
Best Bulk Bennu
Atomic Number
H Li Be C N Na Mg Al P S K Ca Sc Ti V Cr Mn Fe Co Ni Cu Zn Ga As Se Rb Sr Y Zr Nb Mo Ag Cd In Sn Sb Te Cs Ba La Ce Pr Nd Sm Eu Gd Tb Dy Ho Er Tm Yb Lu Hf Ta W Ir Pt Tl Pb Bi Th U

***Figure 6: B³ elemental composition.*** *(A) Singlely-normalized to the average CI composition (Lodders et al., 2025a). (B) Double-normalized to Mg and the average CI composition (Lodders et al., 2025a). Errors shown are ±1 SD. The mean value from Na to U is 1.08 ± 0.06 (1 SD).*

## 4.3 Comparisons to the average CI chondrite composition

As shown in **Figure 6**, and consistent with findings reported by Lauretta and Connolly et al. (2024) and Barnes and Nyugen et al. (2025), the $B^3$ composition proposed here closely matches the average CI composition (Lodders et al., 2025a). This agreement reinforces previous interpretations from telescopic and spacecraft observations that Bennu originates from an extensively aqueously altered parent body (Clark et al., 2011; Lauretta and DellaGiustina et al., 2019).

Closer inspection reveals resolvable differences between the $B^3$ and CI compositions. In the straight ratio plot (**Figure 6A**), most elements in $B^3$ (except for H) are slightly enriched relative to CI, with values generally above the 1.0 ratio line. For example, the average $B^3$/CI ratio for non-highly volatile elements (Na to U) is 1.08 ± 0.06 (1 SD) (**Figure 6A**). This enrichment reflects the tendency of CI meteorites to gain terrestrial water and atmospheric O during their residence time on Earth (e.g., Bland et al., 2006), resulting in a slight dilution of their absolute elemental concentrations other than H and O. This interpretation is consistent with the lower H concentration measured in Bennu relative to CIs (Lauretta and Connolly et al., 2024).

Like Bennu, the recently recovered CI meteorite Oued Chebeika 002, which is not as terrestrially altered as other CIs, shows uniform enrichment in all elements (except H) relative to the CI average samples (Gattacceca et al., 2025; Zhu et al., 2025; Broussard et al., 2025). Together, these findings support the hypothesis that the bulk compositions of many CI meteorites have been diluted through the addition of terrestrial water and O over time (e.g., Bland et al. 2006, Gounelle and Zolensky, 2014; Greenwood et al., 2023). Alternatively, the difference in H between Bennu and CI meteorites may reflect other processes, such as surface exposure of Bennu samples to vacuum and solar radiation, or variations in the abundances of OH-bearing clay minerals between Bennu and CI meteorites.

After correcting for potential dilution effects by double-normalizing to a more refractory major element (e.g., Mg), all but three elements in $B^3$ agree with the CI values within ±1 SD uncertainty (**Figure 6B**). Only three elements (H, S, and Ir) fall outside of the ±1 SD range.

The difference in H can be understood from terrestrial water gain by CI chondrites. The higher S and lower Ir of $B^3$ relative to CI chondrites, however, may be attributable to analytical issues and require further confirmation by future studies. In particular, S and Ir abundances were each measured by only one laboratory, Berkeley and WUSTL, respectively. The S abundance of $B^3$ (6.67 ± 0.12 wt.%) is higher than the CI average (5.18 ± 0.46 wt.%; Lodders et al., 2025a) and exceeds the range reported for individual CI meteorites (4.31–5.52 wt.%) by Braukmüller et al. (2019, 2018) and Alexander et al. (2022).

Overall, the $B^3$ composition matches the average CI composition for 60 out of 63 elements within 1 SD uncertainties. The discrepancies observed in the remaining three elements are likely due to terrestrial alteration of the meteorites (i.e., for H) or analytical issues in the laboratory (e.g., S, and Ir). Any potential differences will require additional confirmation by future studies.

## 4.4 Comparisons to the average Ryugu sample composition

Overall, the $B^3$ composition aligns well with that of Ryugu samples returned by the Hayabusa2 mission (Nakamura et al., 2022; Ito et al., 2022;Yokoyama et al., 2023, 2025; Shollenberger et al., 2025), which are also broadly comparable to CI chondrites (**Figure 7**).

Because Hayabusa2 returned substantially less material than OSIRIS-REx (5.4 g vs. 121.6 g; Yada et al., 2022; Tachibana et al., 2022; Lauretta and Connolly et al., 2024), only small masses (0.65–33 mg) of Ryugu samples have been analyzed to date (Nakamura et al., 2022; Ito et al., 2022;Yokoyama et al., 2023, 2025;Shollenberger et al., 2025) and there is no large homogenized Ryugu sample like those from Bennu in this study. Because of the small sample masses, the average Ryugu composition is based on a sample-mass weighted average.

Some elements, however, show significant differences between $B^3$ and the average Ryugu composition (see labeled elements in **Figure 7**). For example, the Ryugu material is highly enriched in Ta, due to contamination from the Ta projectile used during sample collection (Yokoyama et al., 2025). Several other elements are also elevated in the average Ryugu sample composition relative to $B^3$. For instance, Ca, Mn, and Sr are enriched perhaps due to a high abundance of carbonates in the samples, and heavy REEs and P are enriched, perhaps owing to phosphate phases in the analyzed sample (Yokoyama et al., 2025; López García et al., 2026). These carbonates and phosphates are interpreted to be secondary minerals formed during aqueous alteration (Nakamura et al., 2023; Yamaguchi et al., 2023; Tkalcec et al., 2024; Iizuka et al., 2025). The sample mass-weighted averaging should take care of the nugget effect to some extent (Yokoyama et al., 2025), unless the majority of analyzed samples were enriched in carbonates and phosphates. Currently, however, it remains uncertain whether the bulk Ryugu material is truly more enriched in carbonates and phosphates than bulk Bennu and CI chondrites. Future analyses of larger, homogenized Ryugu samples will be critical for better constraining Ryugu's bulk composition and evaluating the origin of these differences (Yokoyama et al., 2026).

The Bennu samples have also experienced multiple episodes of aqueous alteration, ranging from early global closed-system alteration to later localized open-system alteration (Lauretta and Connolly et al., 2026). Evidence for these alteration events is clearly reflected in the mineralogical and petrological characteristics of the samples, as well as in the elemental variations among different particle types and their isotopic systematics (Lauretta and Connolly et al., 2026). In contrast, the present study focuses on the bulk compositions of the Bennu aggregates. More detailed discussions of fluid–rock interactions in the parent bodies of Bennu and Ryugu can be found in recent publications (e.g., Nakamura et al., 2022; Yokoyama et al., 2023; Iizuka et al., 2025; Lauretta and Connolly et al., 2026).

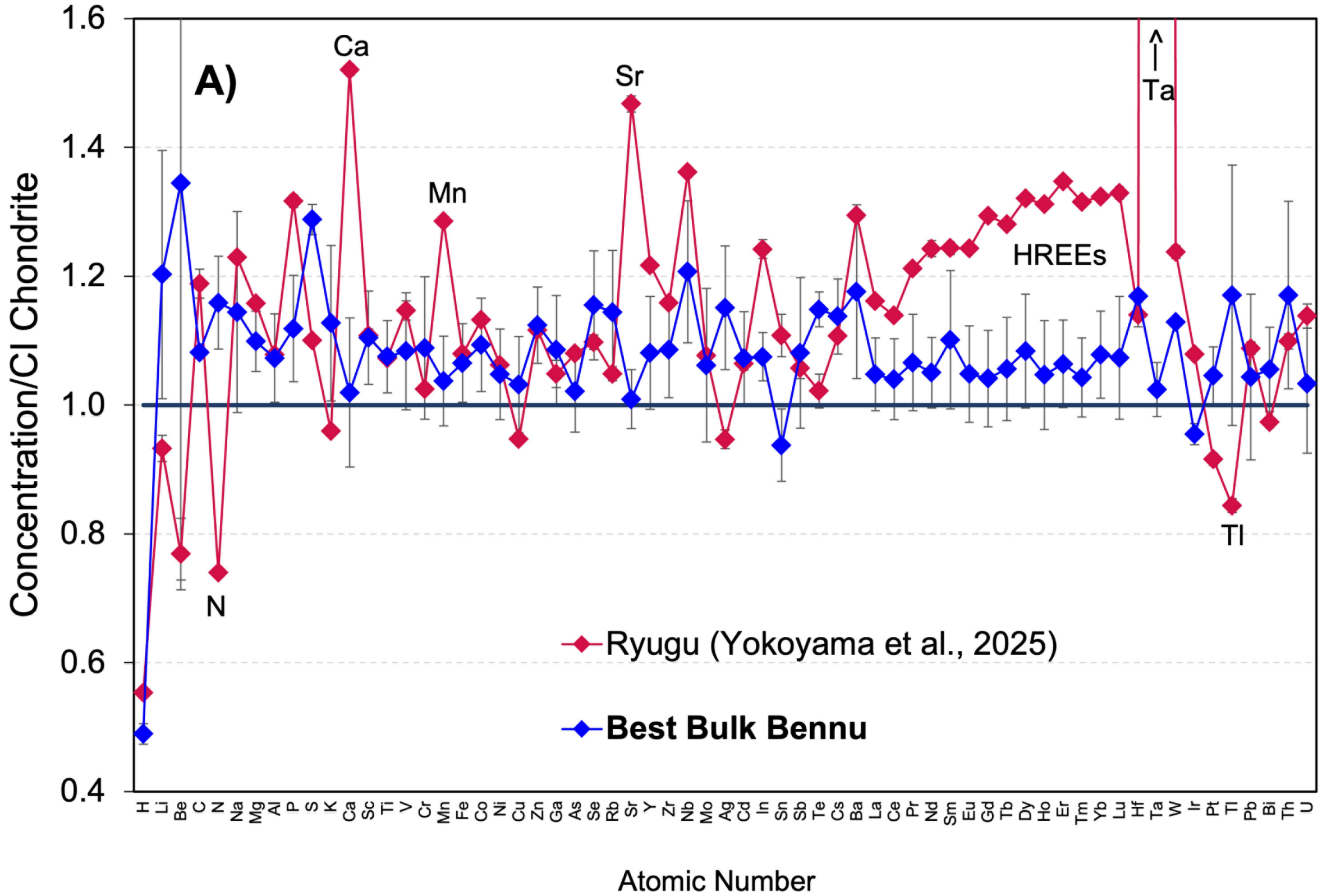
A)
1.6
1.4
1.2
1.0
0.8
0.6
0.4
Concentration/CI Chondrite
Ca
Sr
Mn
N
HREEs
Ta
Tl
Ryugu (Yokoyama et al., 2025)
Best Bulk Bennu
H Li Be C N Na Mg Al P S K Ca Sc Ti V Cr Mn Fe Co Ni Cu Zn Ga As Se Rb Sr Y Zr Nb Mo Ag Cd In Sn Sb Te Cs Ba La Ce Pr Nd Sm Eu Gd Tb Dy Ho Er Tm Yb Lu Hf Ta W Ir Pt Tl Pb Bi Th U
Atomic Number

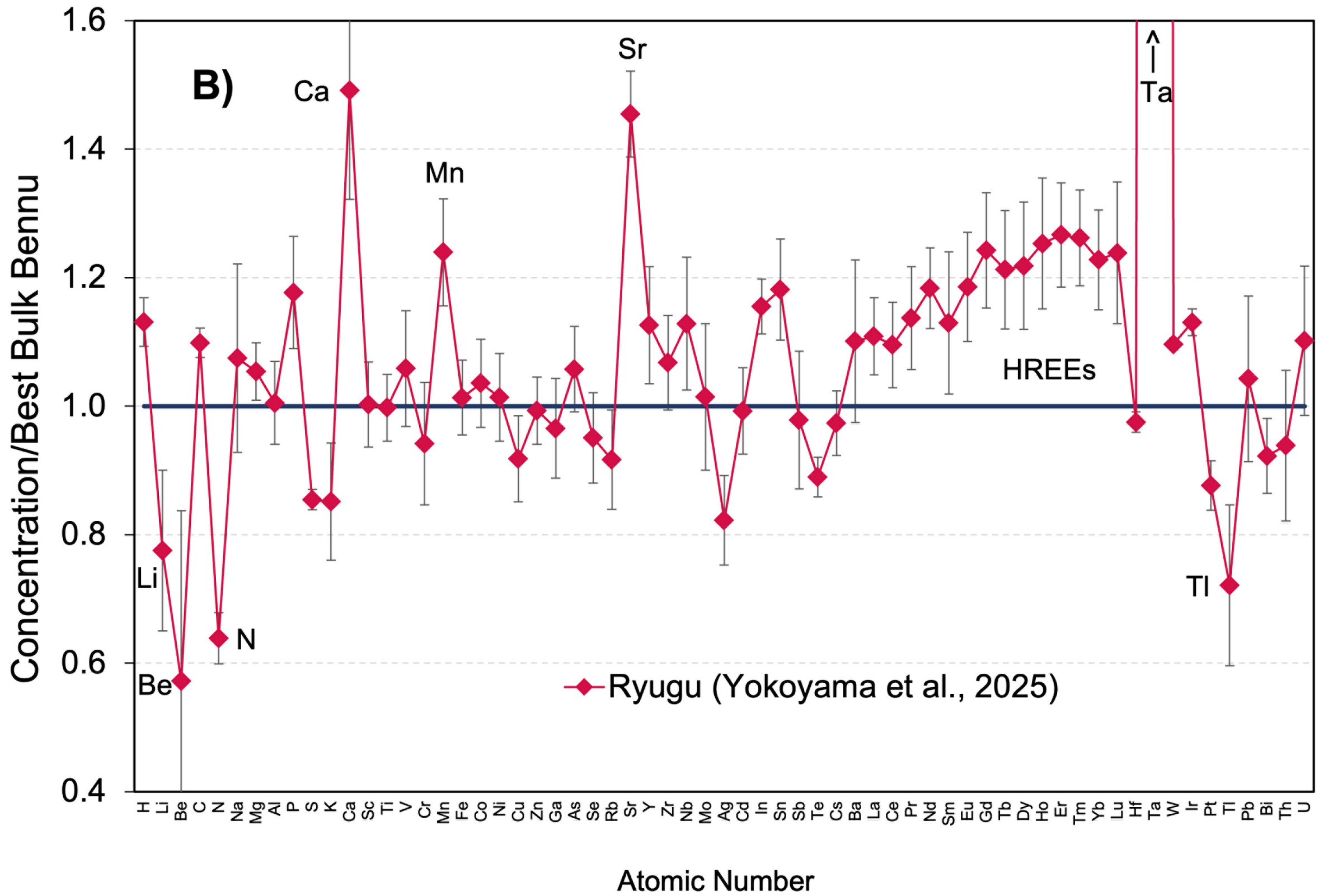
B)
1.6
1.4
1.2
1.0
0.8
0.6
0.4
Concentration/Best Bulk Bennu
Ca
Sr
Mn
Ta
HREEs
Li
N
Be
Tl
Ryugu (Yokoyama et al., 2025)
H Li Be C N Na Mg Al P S K Ca Sc Ti V Cr Mn Fe Co Ni Cu Zn Ga As Se Rb Sr Y Zr Nb Mo Ag Cd In Sn Sb Te Cs Ba La Ce Pr Nd Sm Eu Gd Tb Dy Ho Er Tm Yb Lu Hf Ta W Ir Pt Tl Pb Bi Th U
Atomic Number

***Figure 7. Bulk elemental compositions of Ryugu and $B^3$. (A)*** *Concentrations are normalized to the literature average values for CI chondrites (Lodders et al., 2025a). The Ryugu values are from Yokoyama et al. (2025). The error bars are ±1 SD as reported in each study.* ***(B)*** *Concentrations are normalized to the $B^3$ values in this study. The error bars are the combined ±1 SD of both Bennu and Ryugu data. Only elements that do not agree within 20% between Ryugu and Bennu are labeled.*

## 4.5 Comparisons to other carbonaceous chondrites

Among carbonaceous chondrites, it has long been recognized that volatile element abundances show a progressive depletion from CI to CM, CV, CO, CK, and CR chondrites (e.g., Kallemeyn and Wasson, 1981; Lodders et al., 2025b). This trend has been interpreted as reflecting either nebular condensation processes (Wasson and Chou, 1974) or varying proportions of volatile-rich (e.g., matrix) and volatile-poor (e.g., chondrule) components derived from distinct early Solar System reservoirs (e.g., Anders, 1964; Alexander et al., 2001; Braukmüller et al., 2018, Alexander, 2019).

In this study, we have shown that CI chondrites and $B^3$ are indistinguishable for nearly all elemental abundances except H (**Figure 6**). Because $B^3$ represents a pristine returned sample, we normalize all carbonaceous chondrites, including CI chondrites, to $B^3$ to reassess the classical volatile depletion trends. As shown in **Figure 8**, elemental abundances for the major carbonaceous chondrite groups are plotted in order (from left to right) of increasing volatility based on 50% condensation temperatures (Lodders et al., 2025b). Similar to conventional CI-normalized plots, the chondrite groups show progressively lower abundances of more volatile elements when normalized to $B^3$.

Unexpectedly, CI chondrites themselves appear to display a slight depletion in volatile elements relative to $B^3$, although this is within the combined uncertainties of the Bennu and CI chondritic averages. This difference is not statistically significant (**Section 4.3**), yet future higher-precision analyses may further test the hypothesis that CI chondrites are subtly depleted in volatile elements compared to $B^3$. If the depletion in volatiles in CI chondrites is confirmed, $B^3$ would currently represent the most volatile-rich chondritic material analyzed to date and could serve as a better reference for the bulk Solar System composition (Wang et al. 2025).

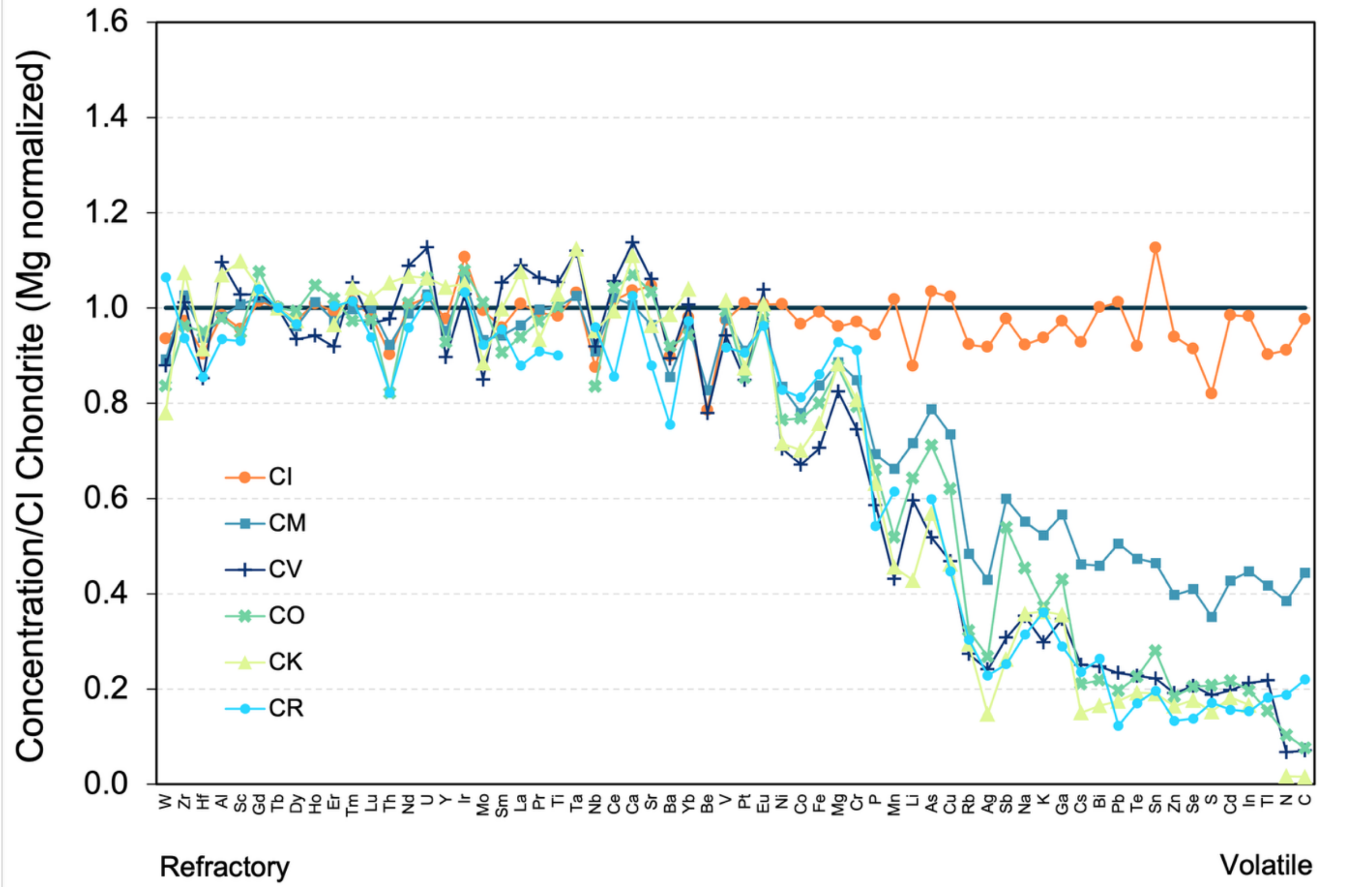

***Figure 8: Bulk elemental compositions of carbonaceous chondrites (CI, CM, CV, CO, CK, and CR) double-normalized to Mg and the B³ composition.*** *Elemental concentrations are shown in order of increasing volatility (Lodders et al., 2025b). The average CI composition is from Lodders et al. (2025a). The averages of other carbonaceous chondrites are from Lodders (2021).*

## 4.6 Comparisons to the solar photospheric composition and implications for refining the bulk Solar System composition

CI chondrites have traditionally been regarded as the most volatile-rich astromaterials and have thus served as the reference for bulk Solar System composition for more than 50 years (Anders, 1971; Cameron, 1973), as not all elements can be quantitatively measured in the solar photosphere (Lodders et al., 2025a). This study confirms that the bulk composition of Bennu ($B^3$) closely resembles that of CI chondrites (Lauretta and Connolly et al., 2024; Barnes and Nguyen et al., 2025). This raises a critical question: should both $B^3$ and CI be used as the proxy for bulk solar composition? To address this, we compare the $B^3$ composition with the most recent compilation of solar photospheric abundances derived from spectroscopic data (Lodders et al., 2025a).

In **Figures 9** and **10**, element abundances in the solar photosphere generally agree with those of $B^3$ within the combined analytical uncertainties (1 SD), except for six elements: H, Li, Be, C, N, and Ag. The highly volatile elements H, C, and N are depleted in rocky materials (e.g.,

Bennu and CI chondrites), as they remain largely in the gas phase and can be easily lost during nebular and parent-body processes. Lithium is depleted in the solar photosphere (hence enriched in $B^3$ and CI samples relative to solar) due to its destruction during the pre–main-sequence stage and at the hot base of the solar convection zone (Lodders et al., 2025a). The apparent discrepancy in Be and Ag is likely attributable to the large uncertainties in solar photospheric abundance measurements; the photospheric Be and Ag abundances agree with those of Bennu and CI chondrites within 2 SD.

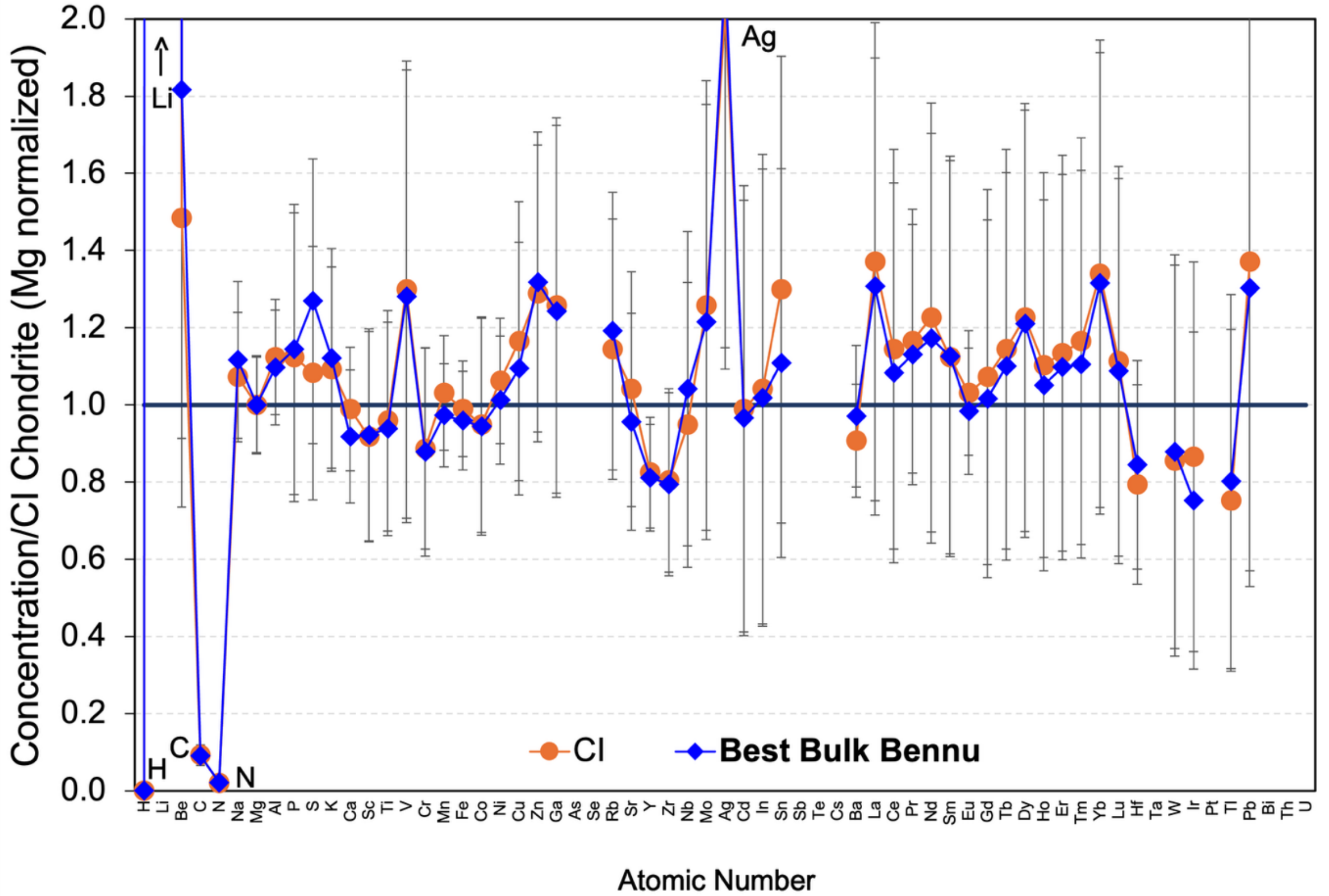

***Figure 9: Bulk elemental compositions of $B^3$ and average CI chondrites double-normalized to the solar photosphere and Mg.*** *The compositions of average CI chondrites and the solar photosphere is from Lodders et al. (2025a). The 1 SD error bars shown are combined errors from both solar photosphere and meteorite/Bennu analyses but are largely dominated by the uncertainties in solar values (Lodders et al., 2025a). The H, C, and N abundances are from Foustoukos et al. (2026). Elements that do not agree between solar photosphere and meteorite/Bennu analyses are labeled (H, Li, C, N, and Ag).*

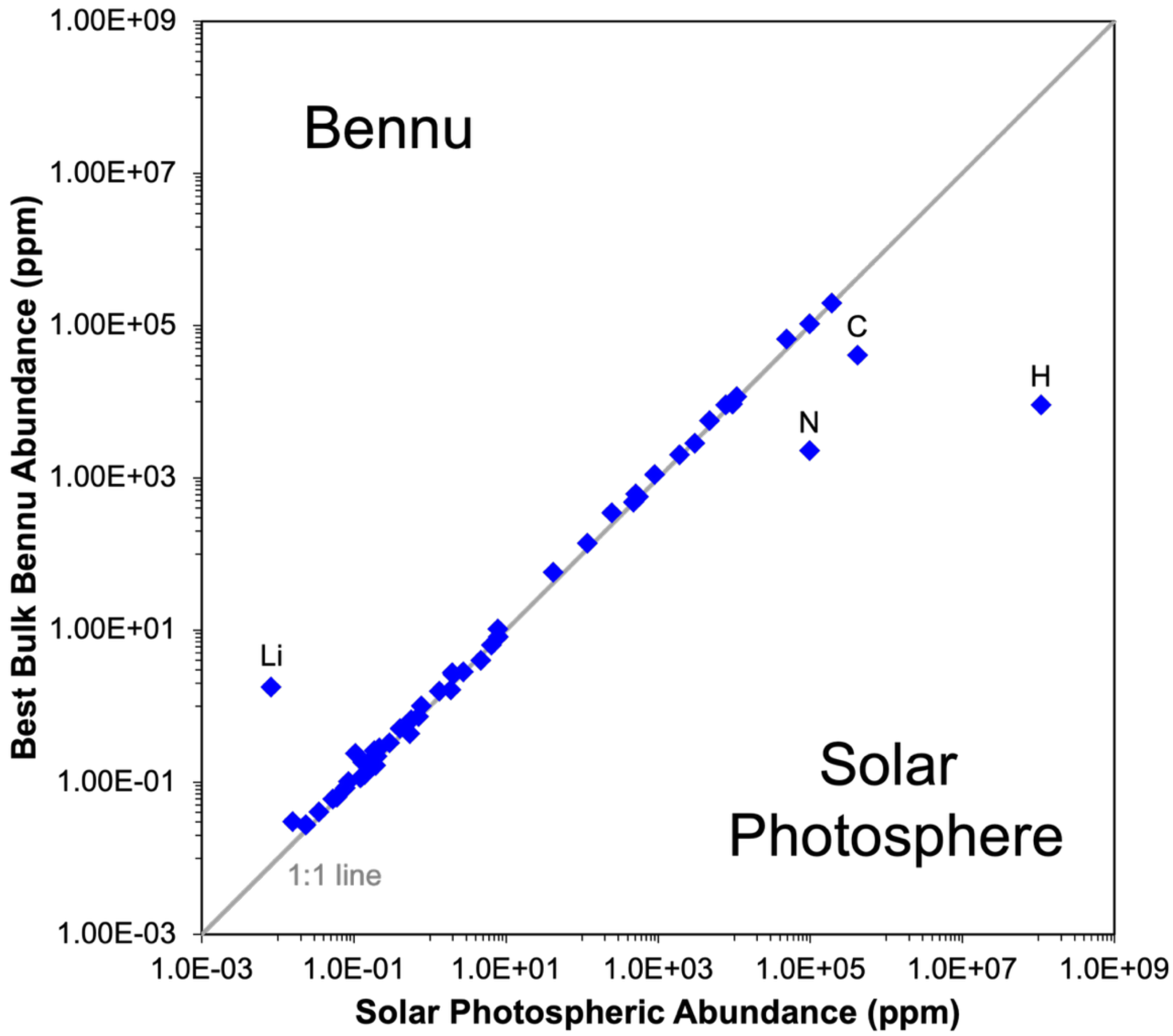

***Figure 10. Comparison of elemental abundances by mass in the solar photosphere and $B^3$ on a logarithmic scale.*** *Solar photosphere data are from Lodders et al. (2025a). Not every element measured in $B^3$ is available in the solar photospheric dataset. The 1:1 reference line is also plotted for comparison.*

Overall, considering the substantial uncertainties, primarily from the solar photospheric measurements, both $B^3$ and CI chondrites display good agreement with the solar abundances derived from spectroscopic data where the comparison can be made. Neither $B^3$ nor CI has a clear advantage in terms of matching the solar photospheric composition more precisely. Thus, we propose that both $B^3$ and CI chondrites are suitable to represent the bulk Solar System composition.

Currently, CI meteorites remain the best choice for estimating the bulk Solar System abundances of non-highly volatile elements, as extensive analytical data has been collected from these chondrites, large masses are available for future analysis, and all non-volatile elements have been measured — whereas, for instance, Si and the halogens (F, Cl, Br, I) have not yet been measured in Bennu samples. The $B^3$ composition determined in this study represents an initial contribution toward refining estimates of the bulk Solar System composition. Additional analyses of representative Ryugu and Bennu samples, particularly for elements not yet quantified, are essential to place any refinement of Solar System abundances on a robust statistical foundation.

# 5. Conclusions

In this study, we report new bulk elemental compositions of aggregate samples from the Bennu material returned by OSIRIS-REx. Combining these new data with published results, we derived the Best Bulk Bennu ($B^3$) composition for 63 elements. Our main conclusions are as follows:

1. Bennu aggregate samples, even those as small as ~10 mg, are overall homogeneous in bulk elemental composition, with no significant differences observed among subsamples, whether from homogenized or non-homogenized parent samples.
2. The $B^3$ composition agrees with the average CI chondrite composition, except for hydrogen, whose higher abundance in CI meteorites is attributed to terrestrial water absorption. Minor discrepancies in other elements are likely due to analytical outliers in this study. Further analyses are required to assess any real compositional differences between Bennu and CI material.
3. The $B^3$ composition also agrees well with the average bulk Ryugu composition, except for some elements likely associated with carbonates and phosphates. These differences probably capture small-sample heterogeneity arising from aqueous alteration and the precipitation of these minerals, rather than being scalable to intrinsic compositional differences between the two asteroids. Overall, both Bennu and Ryugu are compositionally similar to CI chondrites.
4. When compared to the bulk Solar System composition inferred from the solar photosphere, both the $B^3$ and CI average compositions show consistent patterns. There is currently no clear advantage to using one over the other for constraining the bulk Solar System composition.
5. Both CI chondrites and samples delivered by spacecraft from asteroids, such as Bennu and Ryugu, should be used to refine estimates of bulk Solar System composition, especially as more elemental data from these returned asteroid samples become available. This study represents an initial step in that direction.

Overall, since the first stellar abundance determinations by Payne (1925) a century ago, we now have come to an age where pristine asteroid materials delivered to Earth by the Hayabusa2 and OSIRIS-REx missions can be added to the inventory for determining the Solar System elemental abundances.

## Acknowledgements

This material is based upon work supported by NASA under Contract NNM10AA11C issued through the New Frontiers Program. We acknowledge the NASA Centers task award (NNH09ZDA007O). Several authors are supported by NASA's Participating Scientist Program: K.W. and P.K. under grant #80NSSC22K1689; K.C.W. and K.N. under grant #80NSSC22K1693; J.W., J.R., Q.R.S., T.S.K., and G.A.B. under grant #NNH230B64A. Work by K.L. is supported by NASA (grant #80NSSC24K1284) and the McDonnell Center for the Space Sciences. Part of this work was performed under the auspices of the U.S. Department of Energy by Lawrence Livermore National Laboratory under contract DE-AC52-07NA27344 with release number LLNL-JRNL-2013386.

We acknowledge the entire OSIRIS-REx team for enabling the return and analysis of samples from Bennu. This research was also made possible by the careful work of Curation staff at NASA's Johnson Space Center.

We thank Dr. Jutta Zipfel, curator at the Naturmuseum Senckenberg, and Dr. Mathieu Roskosz, curator at Muséum National d'Histoire Naturelle, for providing Orgueil samples. The LLNL team greatly appreciates Orgueil sample USNM 2980 from the Division of Meteorites, Department of Mineral Sciences, Smithsonian Institution.

**Author contributions**

Conceptualization of the study, K.W., K.L., J.J.B., A.N.N., H.C.C., D.S.L; Methodology and Investigation, P.K., J. R., Q.R.S., J.W., T.S.K., G.A.B., M.A., V.L., K.C.W., D.P.G., J.P.D., C.M.O'D.A., J-A.B., P.H.; Writing – Original Draft, K.W., K.L.; Writing – Review & Editing, all co-authors.

**Appendix A. Supplementary Material**

The supplementary material includes five supplementary tables and nine supplementary figures:
**Supplementary Table S1**. Extended summary of laboratories, instruments and methods in this study.
**Supplementary Table S2**. Bulk elemental compositions of blank fused silica samples.
**Supplementary Table S3**. Bulk elemental compositions of total process blanks.
**Supplementary Table S4**. Bulk elemental compositions of geological reference materials.
**Supplementary Table S5**. The DOIs associated with the data detailed in this paper.
**Supplementary Figure S1**. Flow chart illustrating the division history of the small non-homogenized aggregate sample (OREX-800033-0).
**Supplementary Figure S2**. Flow chart illustrating the division history of the large homogenized aggregate powder for SEIWG (OREX-800117-0).
**Supplementary Figure S3**. Flow chart illustrating the division history of the large homogenized aggregate powder for SOAWG (OREX-800107-0).
**Supplementary Figure S4**. Bulk elemental compositions of SEIWG, SOAWG, and baked original blank fused silica samples, normalized to the original blank fused silica sample.
**Supplementary Figure S5**. Bulk elemental compositions of SEIWG, SOAWG, baked and unbaked original blank fused silica samples, normalized to the $B^3$ composition.
**Supplementary Figure S6**. Compositions of Bennu samples normalized to another sample analyzed in the same laboratory.
**Supplementary Figure S7**. Correlations between CI-normalized elemental ratios.
**Supplementary Figure S8**. CI-normalized elemental ratios plotted against dissolved sample mass.
**Supplementary Figure S9**. CI-normalized elemental ratios plotted against homogenized sample mass.

**Data availability**

Instrument data products that underpin the findings of this study are available via astromat.org at the DOIs listed in Supplementary Table S5.

***Table 1. Samples included in the calculation of the Best Bulk Bennu ($B^3$) values, either analyzed in this study or reported in the literature***

| Sample | Number of Analyses | WUSTL | LLNL | Berkeley | UBC |
|---|---|---|---|---|---|
| **Small Non-Homogenized Aggregate: OREX-800033-0 (30.51 mg)** | 3 | OREX-803015-0 (20.66 mg shared with LLNL)* | OREX-803015-101 (20.66 mg shared with WUSTL)† | OREX-803047-0 (9.85 mg) | |
| **Large Homogenized Aggregate Powder for SEIWG‡: OREX-800117-0 (1288 mg)** | 4 | OREX-800117-107 (22.22 mg) | OREX-800117-111 (464 mg) | OREX-800117-103 (14.42 mg) | OREX-800117-109 (121 mg) |
| **Large Homogenized Aggregate Powder for SOAWG‡: OREX-800107-0 (6425 mg)** | 1 | OREX-800107-101 (10.56 mg) | | | |

* Data published in Lauretta and Connolly et al. (2024).
† Data published in Barnes and Ngyuen et al. (2025).
‡ SEIWG: Sample Elements & Isotopes Working Group; SOAWG: Sample Organic Analysis Working Group.

*Table 2. Summary of laboratories, instruments and methods in this study*

| Laboratory | WUSTL | LLNL | Berkeley | UBC |
|---|---|---|---|---|
| **Instruments** | Q-ICP-MS (Quadrupole Inductively Coupled Plasma Mass Spectrometry) | SF-ICP-MS (High-Resolution Inductively Coupled Plasma Mass Spectrometry) * | ICP-OES (Inductively Coupled Plasma Optical Emission Spectrometry) | SF-ICP-MS (High-Resolution Inductively Coupled Plasma Mass Spectrometry) |
| **Calibrating Standards** | Synthetic CM chondrite reference solutions | Geological reference materials | Synthetic chondrite reference solutions | Multi element solutions |
| **Additional Normalization** | No | No | No | Orgueil in the same run for matrix-correction |
| **Elements Analyzed** | 58 | 51 | 10 (only major elements) | 52 |

*The Hf and W concentrations of OREX-800117-111 are measured using isotope dilution method at LLNL with a MC-ICP-MS (see Methods).

### *Table 3. Bulk elemental compositions of CI meteorites (Orgueil and Ivuna) measured alongside Bennu samples in the same analytical sessions at different laboratories*

| | **Orgueil (MNHN 222b)** | | **Orgueil (MPK 0226B)** | | **Orgueil (USNM 2980)** | | **Orgueil-1** | | **Orgueil-13** | | **Ivuna** | |
|---|---|---|---|---|---|---|---|---|---|---|---|---|
| **Lab** | **WUSTL** | | **WUSTL** | | **LLNL** | | **Berkeley** | | **Berkeley** | | **UBC** | |
| **Method** | **Q-ICP-MS** | | **Q-ICP-MS** | | **SF-ICP-MS** | | **ICP-OES** | | **ICP-OES** | | **SF-ICP-MS** | |
| **Mass (mg)** | **36.98** | | **13.73** | | **510.2** | | **1** | | **13.12** | | **13.80** | |
| | **µg/g** | ±SD | **µg/g** | ±SD | **µg/g** | ±SD | **µg/g** | ±SD | **µg/g** | ±SD | **µg/g** | ±SD |
| **Li** | **1.28** | 0.12 | **1.39** | 0.33 | **1.44** | 0.10 | | | | | **1.62** | 0.10 |
| **Be** | | | | | **0.030** | 0.006 | | | | | | |
| **Na** | **4828** | 41 | **5061** | 56 | **5933** | 41 | **4748** | 237 | | | **4704** | 129 |
| **Mg** | **95992** | 1600 | **90467** | 918 | **95460** | 573 | **101449** | 5072 | **91826** | 4591 | **95583** | 1851 |
| **Al** | **8032** | 193 | **7743** | 326 | **8821** | 52 | **8771** | 439 | **8011** | 401 | **8619** | 345 |
| **P** | **908** | 41 | **997** | 65 | **1013** | 23 | | | | | **804** | 88 |
| **S** | | | | | | | **52607** | 5261 | **55250** | 5525 | | |
| **K** | **507** | 24 | **515** | 15 | **497** | 4 | **559** | 56 | **523** | 52 | **542** | 18 |
| **Ca** | **10335** | 267 | **8765** | 333 | **9108** | 82 | **9784** | 489 | **7907** | 395 | **9144** | 382 |
| **Sc** | **5.69** | 0.30 | **5.77** | 0.64 | **5.63** | 0.03 | | | | | **5.91** | 0.31 |
| **Ti** | **419** | 2 | **458** | 7 | **416** | 4 | **520** | 52 | **412** | 41 | **493** | 9 |
| **V** | **50.7** | 1.0 | **46.9** | 0.7 | **50.2** | 0.5 | | | | | **57.0** | 2.3 |
| **Cr** | **2554** | 45 | **2331** | 81 | **2531** | 19 | | | | | **2691** | 63 |
| **Mn** | **2063** | 45 | **1762** | 57 | **1817** | 8 | **2057** | 62 | **1750** | 53 | **1994** | 64 |
| **Fe** | **183410** | 1465 | **180920** | 2060 | **187436** | 704 | **205324** | 6160 | **180883** | 5426 | **195910** | 6056 |
| **Co** | **479** | 8 | **509** | 5 | **512** | 3 | **575** | 57 | **506** | 51 | **492** | 20 |
| **Ni** | **10399** | 62 | **10036** | 145 | **10732** | 59 | **11918** | 358 | **10567** | 317 | **10510** | 304 |
| **Cu** | **134** | 3 | **127** | 1 | **132** | 1 | | | | | **125** | 8 |
| **Zn** | **304** | 5 | **341** | 6 | **307** | 1 | | | | | **327** | 3 |
| **Ga** | **9.25** | 0.25 | **9.08** | 0.30 | **9.61** | 0.10 | | | | | **9.29** | 0.49 |

| | | | | | | | | | | |
|---|---|---|---|---|---|---|---|---|---|---|
| **As** | **1.55** | 0.14 | **1.58** | 0.24 | | | | | | |
| **Se** | **19.6** | 2.8 | **21.0** | 7.0 | | | | | | |
| **Rb** | **2.14** | 0.01 | **2.24** | 0.01 | **2.40** | 0.03 | | | **2.13** | 0.26 |
| **Sr** | **8.39** | 0.18 | **8.54** | 0.12 | **7.61** | 0.06 | | | **8.52** | 0.41 |
| **Y** | **1.32** | 0.05 | **1.45** | 0.03 | **1.40** | 0.01 | | | **1.72** | 0.06 |
| **Zr** | **3.59** | 0.05 | **3.61** | 0.05 | **4.04** | 0.06 | | | **3.63** | 0.19 |
| **Nb** | **0.280** | 0.007 | **0.300** | 0.005 | **0.275** | 0.005 | | | **0.319** | 0.035 |
| **Mo** | **0.828** | 0.010 | **0.788** | 0.012 | **0.964** | 0.013 | | | **1.002** | 0.032 |
| **Ag** | **0.216** | 0.007 | **0.190** | 0.040 | | | | | **0.218** | 0.005 |
| **Cd** | **0.645** | 0.036 | **0.685** | 0.035 | **0.679** | 0.015 | | | **0.704** | 0.033 |
| **In** | **0.0780** | 0.0075 | **0.0710** | 0.0050 | | | | | | |
| **Sn** | **1.21** | 0.05 | **1.09** | 0.02 | **1.45** | 0.04 | | | **1.69** | 0.05 |
| **Sb** | **0.149** | 0.021 | **0.183** | 0.010 | **0.163** | 0.002 | | | **0.193** | 0.004 |
| **Te** | **1.97** | 0.42 | **2.13** | 0.10 | | | | | | |
| **Cs** | **0.172** | 0.012 | **0.187** | 0.005 | **0.204** | 0.004 | | | **0.187** | 0.006 |
| **Ba** | **2.40** | 0.09 | **2.65** | 0.05 | **2.58** | 0.02 | | | **2.67** | 0.04 |
| **La** | **0.224** | 0.004 | **0.244** | 0.005 | **0.235** | 0.002 | | | **0.248** | 0.007 |
| **Ce** | **0.575** | 0.007 | **0.619** | 0.010 | **0.603** | 0.005 | | | **0.657** | 0.018 |
| **Pr** | **0.0860** | 0.0035 | **0.0940** | 0.0000 | **0.0908** | 0.0012 | | | **0.1065** | 0.0028 |
| **Nd** | **0.431** | 0.019 | **0.475** | 0.015 | **0.461** | 0.004 | | | **0.487** | 0.022 |
| **Sm** | **0.135** | 0.007 | **0.153** | 0.005 | **0.150** | 0.001 | | | **0.175** | 0.009 |
| **Eu** | **0.0520** | 0.0035 | **0.0600** | 0.0050 | **0.0586** | 0.0012 | | | **0.0584** | 0.0021 |
| **Gd** | **0.194** | 0.008 | **0.204** | 0.010 | **0.186** | 0.001 | | | **0.241** | 0.009 |
| **Tb** | **0.0340** | 0.0010 | **0.0360** | 0.0000 | **0.0379** | 0.0005 | | | **0.0415** | 0.0029 |
| **Dy** | **0.227** | 0.006 | **0.257** | 0.015 | **0.252** | 0.002 | | | **0.295** | 0.016 |
| **Ho** | **0.0500** | 0.0015 | **0.0550** | 0.0000 | **0.0550** | 0.0011 | | | **0.0619** | 0.0020 |
| **Er** | **0.151** | 0.003 | **0.166** | 0.005 | **0.166** | 0.002 | | | **0.175** | 0.003 |
| **Tm** | **0.0240** | 0.0015 | **0.0250** | 0.0000 | **0.0257** | 0.0001 | | | **0.0260** | 0.0006 |
| **Yb** | **0.155** | 0.006 | **0.173** | 0.010 | **0.168** | 0.002 | | | **0.192** | 0.016 |
| **Lu** | **0.0240** | 0.0010 | **0.0240** | 0.0000 | **0.0265** | 0.0021 | | | **0.0289** | 0.0007 |
| **Hf** | **0.091** | 0.007 | **0.107** | 0.005 | **0.1084***| 0.0003 | | | **0.117** | 0.013 |
| **Ta** | **0.0160** | 0.0010 | **0.0130** | 0.0000 | **0.0140** | 0.0020 | | | **0.0153** | 0.0017 |
| **W** | **0.095** | 0.005 | **0.087** | 0.000 | **0.1056***| 0.0002 | | | **0.090** | 0.017 |
| **Ir** | **0.413** | 0.008 | **0.398** | 0.002 | | | | | | |
| **Pt** | **0.842** | 0.013 | **0.863** | 0.015 | | | | | | |
| **Tl** | **0.138** | 0.008 | **0.162** | 0.005 | **0.145** | 0.001 | | | **0.142** | 0.006 |
| **Pb** | **2.38** | 0.01 | **2.63** | 0.01 | **2.79** | 0.03 | | | **3.01** | 0.02 |
| **Bi** | **0.109** | 0.003 | **0.113** | 0.005 | | | | | **0.119** | 0.010 |
| **Th** | **0.0260** | 0.0010 | **0.0390** | 0.0000 | **0.0305** | 0.0002 | | | **0.0321** | 0.0025 |
| **U** | **0.00700** | 0.00050 | **0.01000** | 0.00000 | | | | | **0.00878** | 0.00029 |

* These Hf and W concentrations were measured using the isotope dilution method at LLNL (see Methods).

*Table 4. Bulk elemental compositions of subsamples of the non-homogenized aggregate sample (OREX-800033-0)*

| | OREX-803015-0* | | OREX-803015-101† | | OREX-803047-0 | |
|---|---|---|---|---|---|---|
| **Lab** | **WUSTL** | | **LLNL** | | **Berkeley** | |
| **Method** | **Q-ICP-MS** | | **SF-ICP-MS** | | **ICP-OES** | |
| **Mass (mg)** | **20.66** | | **20.66** | | **9.85** | |
| | **µg/g** | ±SD | **µg/g** | ±SD | **µg/g** | ±SD |
| **Li** | **1.49** | 0.17 | | | | |
| **Be** | | | | | | |
| **Na** | **5826** | 82 | **5537** | 41 | | |
| **Mg** | **100303** | 332 | **104482** | 515 | **104888** | 5244 |
| **Al** | **8614** | 217 | **9007** | 67 | **9023** | 451 |
| **P** | **1052** | 41 | **1111** | 19 | | |
| **S** | | | | | **65860** | 6586 |
| **K** | **542** | 10 | **684** | 15 | **537** | 54 |
| **Ca** | **8527** | 175 | **9292** | 50 | **10302** | 515 |
| **Sc** | **6.10** | 0.22 | **5.96** | 0.22 | | |
| **Ti** | **453** | 4 | **460** | 3 | **466** | 47 |
| **V** | **52.9** | 0.7 | **53.3** | 0.3 | | |
| **Cr** | **2671** | 63 | **2689** | 19 | | |
| **Mn** | **1965** | 9 | **1926** | 9 | **2000** | 60 |
| **Fe** | **188831** | 353 | **189805** | 703 | **200990** | 6030 |
| **Co** | **531** | 9 | **547** | 5 | **565** | 57 |
| **Ni** | **11588** | 16 | **10822** | 63 | **11845** | 355 |
| **Cu** | **134** | 3 | **134** | 1 | | |
| **Zn** | **325** | 3 | **334** | 2 | | |
| **Ga** | **10.06** | 0.26 | **10.10** | 0.09 | | |
| **As** | **1.67** | 0.15 | | | | |

| | | | | |
|---|---|---|---|---|
| **Se** | **24.9** | 4.2 | | |
| **Rb** | **2.43** | 0.03 | **2.85** | 0.07 |
| **Sr** | **7.55** | 0.11 | **8.25** | 0.13 |
| **Y** | **1.49** | 0.07 | **1.83** | 0.04 |
| **Zr** | **3.71** | 0.01 | **3.93** | 0.02 |
| **Nb** | **0.307** | 0.009 | | |
| **Mo** | | | | |
| **Ag** | **0.243** | 0.008 | | |
| **Cd** | **0.677** | 0.051 | **0.754** | 0.023 |
| **In** | **0.0827** | 0.0100 | | |
| **Sn** | | | **1.52** | 0.06 |
| **Sb** | **0.161** | 0.015 | **0.150** | 0.004 |
| **Te** | **2.57** | 0.69 | | |
| **Cs** | **0.212** | 0.013 | **0.223** | 0.014 |
| **Ba** | **2.48** | 0.02 | **2.68** | 0.08 |
| **La** | **0.240** | 0.004 | **0.255** | 0.018 |
| **Ce** | **0.631** | 0.006 | **0.617** | 0.006 |
| **Pr** | **0.0950** | 0.0028 | **0.0942** | 0.0015 |
| **Nd** | **0.488** | 0.015 | **0.470** | 0.008 |
| **Sm** | **0.162** | 0.012 | **0.153** | 0.002 |
| **Eu** | **0.0596** | 0.0015 | **0.0586** | 0.0005 |
| **Gd** | **0.210** | 0.003 | **0.193** | 0.004 |
| **Tb** | **0.0383** | 0.0019 | **0.0373** | 0.0005 |
| **Dy** | **0.261** | 0.006 | **0.260** | 0.001 |
| **Ho** | **0.0548** | 0.0014 | **0.0560** | 0.0005 |
| **Er** | **0.169** | 0.004 | **0.164** | 0.007 |
| **Tm** | **0.0253** | 0.0002 | **0.0261** | 0.0000 |
| **Yb** | **0.174** | 0.002 | **0.167** | 0.003 |
| **Lu** | **0.0250** | 0.0007 | **0.0253** | 0.0005 |
| **Hf** | **0.107** | 0.005 | **0.110** | 0.003 |
| **Ta** | **0.0161** | 0.0004 | | |
| **W** | **0.138** | 0.003 | | |
| **Ir** | **0.429** | 0.005 | | |
| **Pt** | **0.902** | 0.008 | | |
| **Tl** | **0.146** | 0.005 | | |
| **Pb** | **2.45** | 0.03 | **2.26** | 0.05 |
| **Bi** | **0.119** | 0.002 | | |
| **Th** | **0.0308** | 0.0010 | **0.0338** | 0.0025 |
| **U** | **0.00853** | 0.00030 | **0.00699** | 0.00000 |

[*] Data published in Lauretta and Connolly et al. (2024).
[†] Data published in Barnes and Ngyuen et al. (2025).

*Table 5. Bulk elemental compositions of the large homogenized aggregate powder for SEIWG (OREX-800117-0)*

| | OREX-800117-107 | | OREX-800117-111 | | OREX-800117-103 | | OREX-800117-109 | |
|---|---|---|---|---|---|---|---|---|
| **Lab** | **WUSTL** | | **LLNL** | | **Berkeley** | | **UBC** | |
| **Method** | **Q-ICP-MS** | | **SF-ICP-MS** | | **ICP-OES** | | **SF-ICP-MS** | |
| **Mass (mg)** | **22.22** | | **464** | | **14.42** | | **81.8** | |
| | **µg/g** | ±SD | **µg/g** | ±SD | **µg/g** | ±SD | **µg/g** | ±SD |
| **Li** | **1.71** | 0.18 | **1.77** | 0.02 | | | **2.25** | 0.05 |
| **Be** | | | **0.030** | 0.014 | | | | |
| **Na** | **6076** | 15 | **6415** | 31 | | | **4212** | 504 |
| **Mg** | **101608** | 2153 | **112922** | 664 | **101508** | 5075 | **110443** | 9098 |
| **Al** | **8646** | 75 | **10440** | 89 | **8885** | 444 | **9203** | 868 |
| **P** | **1063** | 31 | **1257** | 41 | | | **1123** | 182 |
| **S** | | | | | **67590** | 6759 | | |
| **K** | **621** | 26 | **691** | 16 | **684** | 68 | **578** | 41 |
| **Ca** | **8933** | 181 | **10815** | 114 | **9634** | 482 | **7412** | 580 |
| **Sc** | **6.25** | 0.40 | **7.12** | 0.05 | | | **6.53** | 0.45 |
| **Ti** | **461** | 7 | **530** | 2 | **484** | 48 | **485** | 53 |
| **V** | **56.1** | 1.1 | **66.1** | 0.4 | | | **58.8** | 9.2 |
| **Cr** | **2680** | 42 | **3387** | 11 | | | **2686** | 264 |
| **Mn** | **1968** | 13 | **2235** | 14 | **2062** | 62 | **1785** | 203 |
| **Fe** | **193134** | 1183 | **221947** | 814 | **201126** | 6034 | **187594** | 18430 |
| **Co** | **540** | 3 | **650** | 3 | **561** | 56 | **555** | 38 |
| **Ni** | **11632** | 112 | **13272** | 159 | **12069** | 362 | **10738** | 842 |
| **Cu** | **134** | 1 | **156** | 1 | | | **128** | 0 |
| **Zn** | **355** | 12 | **378** | 4 | | | **348** | 44 |
| **Ga** | **10.14** | 0.28 | **11.88** | 0.06 | | | **9.53** | 0.25 |
| **As** | **1.81** | 0.12 | | | | | | |
| **Se** | **23.0** | 5.8 | | | | | | |
| **Rb** | **2.48** | 0.07 | **2.85** | 0.02 | | | **2.34** | 0.13 |
| **Sr** | **7.84** | 0.09 | **8.60** | 0.04 | | | **8.18** | 0.83 |
| **Y** | **1.56** | 0.03 | **1.68** | 0.02 | | | **1.75** | 0.10 |
| **Zr** | **4.27** | 0.05 | **4.31** | 0.04 | | | **3.89** | 0.61 |
| **Nb** | **0.301** | 0.021 | **0.343** | 0.007 | | | **0.313** | 0.072 |
| **Mo** | **0.892** | 0.031 | **1.118** | 0.013 | | | **1.007** | 0.157 |
| **Ag** | **0.250** | 0.024 | | | | | **0.208** | 0.011 |
| **Cd** | **0.727** | 0.052 | **0.813** | 0.015 | | | **0.687** | 0.084 |
| **In** | **0.0819** | 0.0046 | | | | | | |
| **Sn** | **1.53** | 0.04 | **1.67** | 0.01 | | | **1.46** | 0.14 |
| **Sb** | **0.191** | 0.010 | **0.162** | 0.005 | | | **0.160** | 0.012 |
| **Te** | **2.67** | 0.36 | | | | | | |
| **Cs** | **0.217** | 0.009 | **0.207** | 0.008 | | | **0.192** | 0.019 |
| **Ba** | **2.63** | 0.05 | **2.76** | 0.04 | | | **3.33** | 0.89 |
| **La** | **0.258** | 0.003 | **0.276** | 0.002 | | | **0.270** | 0.010 |
| **Ce** | **0.671** | 0.008 | **0.719** | 0.007 | | | **0.654** | 0.052 |
| **Pr** | **0.1013** | 0.0032 | **0.1087** | 0.0012 | | | **0.1121** | 0.0043 |
| **Nd** | **0.513** | 0.003 | **0.545** | 0.014 | | | **0.523** | 0.016 |
| **Sm** | **0.164** | 0.002 | **0.184** | 0.003 | | | **0.199** | 0.017 |
| **Eu** | **0.0642** | 0.0024 | **0.0711** | 0.0007 | | | **0.0632** | 0.0037 |
| **Gd** | **0.234** | 0.007 | **0.222** | 0.005 | | | **0.234** | 0.005 |
| **Tb** | **0.0399** | 0.0012 | **0.0457** | 0.0005 | | | **0.0427** | 0.0010 |
| **Dy** | **0.274** | 0.007 | **0.311** | 0.005 | | | **0.307** | 0.005 |
| **Ho** | **0.0593** | 0.0004 | **0.0684** | 0.0022 | | | **0.0599** | 0.0033 |
| **Er** | **0.179** | 0.007 | **0.197** | 0.003 | | | **0.180** | 0.006 |
| **Tm** | **0.0277** | 0.0010 | **0.0300** | 0.0003 | | | **0.0275** | 0.0013 |
| **Yb** | **0.183** | 0.004 | **0.199** | 0.005 | | | **0.191** | 0.000 |
| **Lu** | **0.0265** | 0.0013 | **0.0296** | 0.0009 | | | **0.0310** | 0.0076 |
| **Hf** | **0.129** | 0.008 | **0.125*** | 0.0006 | | | **0.106** | 0.010 |
| **Ta** | **0.0145** | 0.0009 | **0.0150** | 0.0002 | | | **0.0150** | 0.0019 |
| **W** | **0.114** | 0.007 | **0.114*** | 0.0006 | | | **0.084** | 0.011 |
| **Ir** | **0.444** | 0.006 | | | | | | |
| **Pt** | **0.951** | 0.015 | | | | | | |
| **Tl** | **0.212** | 0.014 | **0.171** | 0.002 | | | **0.140** | 0.014 |
| **Pb** | **3.08** | 0.01 | **3.10** | 0.02 | | | **2.66** | 0.35 |
| **Bi** | **0.124** | 0.001 | | | | | **0.109** | 0.009 |
| **Th** | **0.0341** | 0.0029 | **0.0333** | 0.0003 | | | **0.0432** | 0.0047 |
| **U** | **0.00813** | 0.00058 | **0.00909** | 0.00075 | | | **0.01662** | 0.00349 |

[*] These Hf and W concentrations were measured using the isotope dilution method at LLNL (see Methods).

***Table 6. Bulk elemental compositions of the large homogenized aggregate powder for SOAWG (OREX-800107-0)***

| | **OREX-800107-101** | |
|---|---|---|
| **Lab** | **WUSTL** | |
| **Method** | **Q-ICP-MS** | |
| **Mass (mg)** | **10.56** | |
| | | |
| | **μg/g** | ±SD |
| **Li** | **1.67** | 0.07 |
| **Be** | | |
| **Na** | **5988** | 52 |
| **Mg** | **104241** | 1266 |
| **Al** | **8867** | 187 |
| **P** | **1031** | 47 |
| **S** | | |
| **K** | **568** | 52 |
| **Ca** | **9690** | 283 |
| **Sc** | **6.22** | 0.32 |
| **Ti** | **463** | 6 |
| **V** | **58.1** | 0.5 |
| **Cr** | **2972** | 28 |
| **Mn** | **2121** | 15 |
| **Fe** | **193197** | 319 |
| **Co** | **547** | 4 |
| **Ni** | **11744** | 70 |
| **Cu** | **137** | 2 |
| **Zn** | **352** | 2 |
| **Ga** | **10.44** | 0.26 |
| **As** | **1.89** | 0.34 |
| **Se** | **26.6** | 2.4 |
| **Rb** | **2.57** | 0.03 |
| **Sr** | **8.27** | 0.14 |
| **Y** | **1.55** | 0.06 |
| **Zr** | **3.66** | 0.02 |
| **Nb** | **0.372** | 0.004 |
| **Mo** | | |
| **Ag** | **0.248** | 0.010 |
| **Cd** | **0.731** | 0.056 |
| **In** | **0.0873** | 0.0095 |
| **Sn** | **1.66** | 0.10 |
| **Sb** | **0.194** | 0.037 |
| **Te** | **2.68** | 0.38 |
| **Cs** | **0.218** | 0.010 |
| **Ba** | **3.12** | 0.13 |
| **La** | **1.316** | 0.016 |
| **Ce** | **3.466** | 0.032 |
| **Pr** | **0.1008** | 0.0040 |
| **Nd** | **0.499** | 0.010 |
| **Sm** | **0.169** | 0.012 |
| **Eu** | **0.0613** | 0.0052 |
| **Gd** | **0.225** | 0.002 |
| **Tb** | **0.0401** | 0.0013 |
| **Dy** | **0.271** | 0.003 |
| **Ho** | **0.0582** | 0.0022 |
| **Er** | **0.176** | 0.003 |
| **Tm** | **0.0279** | 0.0017 |
| **Yb** | **0.186** | 0.005 |
| **Lu** | **0.0268** | 0.0017 |
| **Hf** | **0.106** | 0.008 |
| **Ta** | **0.0157** | 0.0006 |
| **W** | **0.287** | 0.006 |
| **Ir** | **0.433** | 0.009 |
| **Pt** | **0.982** | 0.004 |
| **Tl** | **0.157** | 0.011 |
| **Pb** | **2.80** | 0.02 |
| **Bi** | **0.125** | 0.003 |
| **Th** | **0.0383** | 0.0012 |
| **U** | **0.00909** | 0.00064 |

*Table 7. Elemental compositions of the Best Bulk Bennu ($B^3$)*

| Element | Concentration (μg/g) | ±SD | RSD | N | Excluded |
|---|---|---|---|---|---|
| H* | **9100** | **300** | 3% | 3 | |
| Li | **1.78** | **0.28** | 16% | 5 | |
| Be | **0.030** | **0.014** | 46% | 1 | |
| C* | **40900** | **361** | 1% | 3 | |
| N* | **2277** | **142** | 6% | 3 | |
| Na | **5676** | **773** | 14% | 6 | |
| Mg | **105049** | **4456** | 4% | 8 | |
| Al | **9085** | **581** | 6% | 8 | |
| P | **1106** | **82** | 7% | 6 | |
| S | **66725** | **1223** | 2% | 2 | |
| K | **613** | **66** | 11% | 8 | |
| Ca | **9326** | **1059** | 11% | 8 | |
| Sc | **6.36** | **0.42** | 7% | 6 | |
| Ti | **475** | **25** | 5% | 8 | |
| V | **57.5** | **4.8** | 8% | 6 | |
| Cr | **2847** | **289** | 10% | 6 | |
| Mn | **2008** | **135** | 7% | 8 | |
| Fe | **197078** | **11286** | 6% | 8 | |
| Co | **562** | **37** | 7% | 8 | |
| Ni | **11714** | **788** | 7% | 8 | |
| Cu | **137** | **10** | 7% | 6 | |
| Zn | **348** | **18** | 5% | 6 | |
| Ga | **10.36** | **0.80** | 8% | 6 | |
| As | **1.79** | **0.11** | 6% | 3 | |
| Se | **24.8** | **1.8** | 7% | 3 | |
| Rb | **2.59** | **0.22** | 8% | 6 | |
| Sr | **8.11** | **0.37** | 5% | 6 | |
| Y | **1.64** | **0.13** | 8% | 6 | |
| Zr | **3.96** | **0.27** | 7% | 6 | |
| Nb | **0.327** | **0.030** | 9% | 5 | |
| Mo | **1.006** | **0.113** | 11% | 3 | |
| Ag | **0.237** | **0.020** | 8% | 4 | |
| Cd | **0.732** | **0.049** | 7% | 6 | |
| In | **0.0840** | **0.0029** | 3% | 3 | |
| Sn | **1.57** | **0.09** | 6% | 5 | |
| Sb | **0.170** | **0.018** | 11% | 6 | |
| Te | **2.64** | **0.06** | 2% | 3 | |
| Cs | **0.212** | **0.011** | 5% | 6 | |
| Ba | **2.83** | **0.33** | 11% | 6 | |
| La | **0.260** | **0.014** | 5% | 5 | WUSTL (SOAWG) † |
| Ce | **0.658** | **0.040** | 6% | 5 | WUSTL (SOAWG) † |
| Pr | **0.1020** | **0.0072** | 7% | 6 | |
| Nd | **0.506** | **0.026** | 5% | 6 | |
| Sm | **0.172** | **0.017** | 10% | 6 | |
| Eu | **0.0630** | **0.0045** | 7% | 6 | |
| Gd | **0.220** | **0.016** | 7% | 6 | |
| Tb | **0.0407** | **0.0031** | 8% | 6 | |
| Dy | **0.281** | **0.023** | 8% | 6 | |
| Ho | **0.0594** | **0.0048** | 8% | 6 | |
| Er | **0.178** | **0.011** | 6% | 6 | |
| Tm | **0.0274** | **0.0016** | 6% | 6 | |
| Yb | **0.183** | **0.012** | 6% | 6 | |
| Lu | **0.0274** | **0.0024** | 9% | 6 | |
| Hf‡ | **0.125** | **0.0006** | 0.5% | 1 | |
| Ta | **0.0153** | **0.0006** | 4% | 5 | |
| W‡ | **0.114** | **0.0006** | 0.5% | 1 | WUSTL (SOAWG) † |
| Ir | **0.435** | **0.007** | 2% | 3 | |
| Pt | **0.945** | **0.040** | 4% | 3 | |
| Tl | **0.165** | **0.029** | 17% | 5 | |
| Pb | **2.72** | **0.34** | 12% | 6 | |
| Bi | **0.119** | **0.007** | 6% | 4 | |
| Th | **0.0356** | **0.0044** | 12% | 6 | |
| U | **0.00837** | **0.00087** | 10% | 5 | UBC (SEIWG) † |

*The H, C and N abundances are the averages of OREX-800117-105, OREX-800117-106, and OREX-800107-110 reported in Foustoukos et al. (2026).
†Excluded because of potential contamination (see Figures 3 and 4).
‡For the Hf and W concentrations of $B^3$, we included only the value measured at LLNL using the isotope dilution method. It is important to note that the Lu concentration of $B^3$ represents the group average (N=6), whereas the Hf concentration is derived solely from the LLNL isotope dilution measurement (N=1). Consequently, the Lu/Hf ratio is neither method-matched nor laboratory-matched, and therefore should not be regarded as an internally self-consistent pair. The group-average Hf concentration, including and excluding the LLNL isotope dilution data, is 0.114±0.010 (1SD; N=6) and 0.112±0.010 (1SD; N=5), respectively. The average laboratory-matched Lu/Hf ratio, including and excluding the LLNL isotope dilution data, is 0.241±0.029 (1SD; N=6) and 0.242±0.032 (1SD; N=5), respectively. Given the negligible difference, **we recommend using 0.241±0.029 (1SD; N=6) as the representative**

**Lu/Hf ratio for B[3]**. Nevertheless, a Lu/Hf ratio determined by isotope dilution using a mixed Lu–Hf tracer remains necessary to establish a robust benchmark for Lu–Hf chronology.

***Supplementary Table S1. Extended summary of laboratories, instruments and methods in this study***

| **Laboratory** | **WUSTL** | **LLNL** | **Berkeley** | **UBC** |
|---|---|---|---|---|
| **Instruments** | Q-ICP-MS (Quadrupole Inductively Coupled Plasma Mass Spectrometry) | SF-ICP-MS (High-Resolution Inductively Coupled Plasma Mass Spectrometry) * | ICP-OES (Inductively Coupled Plasma Optical Emission Spectrometry) | SF-ICP-MS (High-Resolution Inductively Coupled Plasma Mass Spectrometry) |
| **Calibrating Standards** | Synthetic CM chondrite reference solutions | Geological reference materials (BHVO-2, BCR-2, AGV-2 and RGM-2) | Synthetic chondrite reference solutions | Multi element solutions (mixture of single and multi-element solutions from Inorganic Ventures and Specpure) |
| **Additional Normalization** | No | No | No | Orgueil in the same run for matrix-correction |
| **Elements Analyzed** | 58 | 51 | 10 (only major elements) | 52 |
| **Digestion** | Hot plate ~3mL:1mL concentrated $HF:HNO_3$ | Combined hot plate and Parr bombs | Oven ~4.5mL:1.5mL concentrated $HF:HNO_3$ | Microwave 6 ml 15 M $HNO_3$, 2 ml 28 M HF, and 4 ml 12 M HCl |
| **Number of runs and passes** | 2 Runs with 6 Passes (3 for 5k dilution factor and 3 for 50k dilution factor) | 3 runs with 10 passes | 1 run - 3 replicates for each dilution factor | 3 Runs with 6 Passes for each resolution |
| **Dwell times** | Li, Be, P~0.1s Others~0.03s | | N/A | 10ms per mass for Low and Medium Res and 20ms for High Res, |
| **Rinse time** | | | 20s | ~300 s |
| **Sensitivity** | For 1ppb iCapQ tune solution:<br>$^{7}$Li~220 cps<br>$^{59}$Co~62305 cps<br>$^{115}$In~106330 cps<br>$^{140}$Ce~258444 cps<br>$^{238}$U~408177 cps | 1ppb tune solution: $^{115}$In ~1.5e6 cps<br>$^{238}$U ~2.5e6 cps | N/A | For 1ppb multi-element tune solution, typically:<br>Li: ~2*1e05 cps<br>In: 1.5 - 3*1e06 cps<br>U: ~3*1e06 cps |
| **Oxide rates** | $^{140}Ce^{16}O/^{140}Ce$ ~0.0085 | ~ 5%, based on UO/U | N/A | UO/U [%] ~5%<br>BaO/Ba ~0.2 |
| **Doubly charged rates** | | Unknown | N/A | $Ba^{++}/Ba^{+}$: ~3.5%<br>$Sr^{++}$ over Ca by those single element calibration: ~1.5% |
| **Uptake rates** | | 100ul/min nebulizer | ~1.5 ml/min | |
| **Isotopes chosen for analyses** | 7Li, 9Be, 23Na, 24Mg, 27Al, 31P, 39K, 43Ca, 44Ca, 45Sc, 47Ti, 48Ti, 51V, 52Cr, 55Mn, 56Fe, 57Fe, 59Co, 60Ni, 63Cu, 66Zn, 71Ga, 75As, 77Se, 82Se, 85Rb, 88Sr, 89Y, 90Zr, 93Nb, 95Mo, 107Ag, 111Cd, 115In, 118Sn, 121Sb, 125Te, 133Cs, 137Ba, 139La, 140Ce, 141Pr, 146Nd, 147Sm, 153Eu, 157Gd, 159Tb, 163Dy, 165Ho, 166Er, 169Tm, 172Yb, 175Lu, 178Hf, 181Ta, | Low resolution:<br>7Li, 9Be, 23Na, 85Rb, 88Sr, 89Y, 90Zr, 93Nb, 95Mo, 111Cd, 121Sb, 133Cs, 137Ba, 139La, 140Ce, 141Pr, 146Nd, 147Sm, 153Eu, 157Gd, 159Tb, 163Dy, 165Ho, 167Er, | N/A | Major element sequence:<br><br>23Na (MR), 25Mg (MR), 27Al (MR), 31P (MR), 39K (HR), 43Ca (MR), 49Ti (MR), 53Cr (MR), 55Mn (MR), 56Fe (MR), 61Ni (MR).<br><br>Trace element sequence:<br>7Li(LR), 45Sc(MR), 51V(MR), 59Co(MR), 63Cu(MR), 66Zn(MR), |
| | 182W, 193Ir, 195Pt, 205Tl, 208Pb, 209Bi, 232Th, 238U | 169Tb, 172Yb, 175Lu, 178Hf, 181Ta, 182W, 205Tl, 208Pb, 232Th, 238U<br><br>Medium resolution: 24Mg, 27Al, 31P, 32S, 43Ca, 45Sc, 47Ti, 51V, 53Cr, 55Mn, 57Fe, 59Co, 60Ni, 63Cu, 66Zn, 69Ga, 118Sn,<br><br>High resolution: 39K | | 69Ga(MR), 85Rb(MR), 88Sr(MR), 89Y(LR), 90Zr(MR), 93Nb(LR), 95Mo(LR), 107Ag(LR), 111Cd(LR), 118Sn(LR), 121Sb(LR), 133Cs(LR), 137Ba(MR), 139La(HR), 140Ce(HR), 141Pr(HR), 146Nd(HR), 147Sm(HR), 153Eu(HR), 160Gd(HR), 159Tb(HR), 163Dy(HR), 165Ho(HR), 166Er(HR), 169Tm(HR), 172Yb(HR), 175Lu(HR), 178Hf(LR), 181Ta(LR), 182W(LR), 203Tl(LR), 208Pb(LR), 209Bi(LR), 232Th(LR), 238U(LR). |
| **Internal Standard** | | 115In used as internal standard in all resolutions. | | |
| **Correlation coefficients for the calibrations** | Average:0.96 Mostly >0.99 | Generally >0.999 | Two-point calibrations | |
| **Interference corrections** | 57Fe=-0.0926*43Ca 59Co=-0.002187*43Ca 115In=-0.0148637*118Sn 48Ti=-0.0896453*44Ca 208Pb=+1*206Pb+1*207Pb | 115In = -0.0149*118Sn | N/A | Cd114 = -0.0268*Sn118 In115 = -0.0149*Sn118 Sb123 = -0.1272*Te125 Gd160 = -0.1238*Dy161 Dy162 = -0.0042*Er166 Er168 = -0.0059*Yb172 Er170 = -0.1393*Yb172 Yb174 = -0.0086*Hf177 Pb204 = -0.2962*Hg200 |

*The Hf and W concentrations of OREX-800117-111 are measured using isotope dilution method at LLNL with a MC-ICP-MS (see Methods).

*Supplementary Table S2. Bulk elemental compositions of blank fused silica samples*

| | Original fused silica | | Baked original fused silica | | SEIWG blank fused silica | | SOAWG blank fused silica | |
|---|---|---|---|---|---|---|---|---|
| **Lab** | **WUSTL** | | **WUSTL** | | **WUSTL** | | **WUSTL** | |
| **Method** | **Q-ICP-MS** | | **Q-ICP-MS** | | **Q-ICP-MS** | | **Q-ICP-MS** | |
| **Mass (mg)** | **99.88** | | **124.53** | | **192.44** | | **11.23** | |
| | **µg/g** | ±SD | **µg/g** | ±SD | **µg/g** | ±SD | **µg/g** | ±SD |
| **Li** | **0.380** | 0.081 | **0.382** | 0.073 | **0.760** | 0.080 | **0.379** | 0.039 |
| **Be** | | | | | | | | |
| **Na** | **1.792** | 0.297 | **2.001** | 0.061 | **135.459** | 0.598 | **22.321** | 0.263 |
| **Mg** | **0.291** | 0.044 | **0.304** | 0.030 | **0.665** | 0.069 | **7.674** | 0.693 |
| **Al** | **13.062** | 0.565 | **13.746** | 0.315 | **71.538** | 1.436 | **14.491** | 0.166 |
| **P** | **0.326** | 0.687 | **0.468** | 0.514 | **0.759** | 0.536 | **0.504** | 1.087 |
| **S** | | | | | | | | |
| **K** | **0.517** | 0.592 | **0.592** | 0.429 | **34.169** | 1.207 | **1.018** | 0.640 |
| **Ca** | **3.876** | 1.310 | **3.629** | 0.562 | **3.533** | 1.312 | **448.038** | 26.757 |
| **Sc** | **0.003** | 0.002 | **0.002** | 0.003 | **0.009** | 0.006 | **0.003** | 0.008 |
| **Ti** | **1.377** | 0.082 | **1.278** | 0.037 | **1.920** | 0.023 | **1.835** | 0.041 |
| **V** | **0.007** | 0.002 | **0.005** | 0.001 | **0.006** | 0.001 | **0.016** | 0.001 |
| **Cr** | **0.214** | 0.079 | **0.137** | 0.003 | **0.300** | 0.008 | **2.608** | 0.020 |
| **Mn** | **0.022** | 0.008 | **0.029** | 0.007 | **0.056** | 0.008 | **0.214** | 0.016 |
| **Fe** | **0.164** | 0.024 | **1.114** | 0.041 | **2.963** | 0.041 | **12.289** | 1.581 |
| **Co** | **0.009** | 0.003 | **0.004** | 0.002 | **0.012** | 0.002 | **0.025** | 0.004 |
| **Ni** | **0.119** | 0.021 | **0.061** | 0.005 | **0.163** | 0.016 | **1.221** | 0.062 |
| **Cu** | **0.116** | 0.011 | **0.100** | 0.003 | **0.101** | 0.003 | **0.898** | 0.003 |
| **Zn** | **0.231** | 0.068 | **0.341** | 0.005 | **0.129** | 0.034 | **0.515** | 0.030 |
| **Ga** | **0.004** | 0.001 | **0.003** | 0.002 | **0.011** | 0.003 | **0.130** | 0.014 |
| **As** | **0.002** | 0.004 | **0.000** | 0.003 | **0.006** | 0.004 | **0.003** | 0.005 |
| **Se** | **0.879** | 1.479 | **0.398** | 0.810 | **0.056** | 0.120 | **0.073** | 0.244 |
| **Rb** | **0.002** | 0.001 | **0.002** | 0.001 | **0.011** | 0.001 | **0.003** | 0.001 |
| **Sr** | **0.039** | 0.002 | **0.035** | 0.003 | **0.041** | 0.005 | **0.586** | 0.015 |
| **Y** | **0.191** | 0.024 | **0.220** | 0.015 | **0.167** | 0.006 | **0.158** | 0.005 |
| **Zr** | **1.004** | 0.032 | **0.927** | 0.008 | **1.797** | 0.013 | **0.921** | 0.014 |
| **Nb** | **0.001** | 0.000 | **0.001** | 0.000 | **0.003** | 0.001 | **0.004** | 0.001 |
| **Mo** | **0.018** | 0.008 | **0.011** | 0.001 | **0.017** | 0.002 | **0.054** | 0.003 |
| **Ag** | **0.039** | 0.014 | **0.065** | 0.025 | **0.010** | 0.001 | **0.055** | 0.007 |
| **Cd** | **0.000** | 0.000 | **0.000** | 0.000 | **0.000** | 0.001 | **0.001** | 0.001 |
| **In** | **0.000** | 0.000 | **0.000** | 0.000 | **0.000** | 0.000 | **0.000** | 0.000 |
| **Sn** | **0.003** | 0.001 | **0.003** | 0.001 | **0.003** | 0.001 | **0.015** | 0.000 |
| **Sb** | **0.002** | 0.000 | **0.001** | 0.000 | **0.001** | 0.001 | **0.004** | 0.001 |
| **Te** | **0.000** | 0.000 | **0.000** | 0.000 | **0.000** | 0.000 | **0.000** | 0.000 |
| **Cs** | **0.001** | 0.000 | **0.001** | 0.000 | **0.007** | 0.001 | **0.001** | 0.001 |
| **Ba** | **0.027** | 0.002 | **0.024** | 0.002 | **0.116** | 0.008 | **0.349** | 0.017 |
| **La** | **0.028** | 0.001 | **0.030** | 0.001 | **0.034** | 0.001 | **1.227** | 0.021 |
| **Ce** | **0.051** | 0.001 | **0.059** | 0.000 | **0.067** | 0.001 | **2.293** | 0.017 |
| **Pr** | **0.006** | 0.000 | **0.007** | 0.000 | **0.007** | 0.001 | **0.007** | 0.001 |
| **Nd** | **0.027** | 0.001 | **0.031** | 0.001 | **0.030** | 0.002 | **0.029** | 0.001 |
| **Sm** | **0.012** | 0.001 | **0.014** | 0.000 | **0.011** | 0.001 | **0.013** | 0.002 |
| **Eu** | **0.001** | 0.000 | **0.001** | 0.000 | **0.002** | 0.001 | **0.001** | 0.000 |
| **Gd** | **0.024** | 0.002 | **0.030** | 0.001 | **0.021** | 0.002 | **0.026** | 0.002 |
| **Tb** | **0.005** | 0.001 | **0.007** | 0.000 | **0.004** | 0.000 | **0.005** | 0.000 |
| **Dy** | **0.031** | 0.003 | **0.039** | 0.000 | **0.027** | 0.001 | **0.030** | 0.001 |
| **Ho** | **0.005** | 0.000 | **0.006** | 0.000 | **0.005** | 0.000 | **0.005** | 0.000 |
| **Er** | **0.012** | 0.001 | **0.014** | 0.001 | **0.011** | 0.000 | **0.010** | 0.002 |
| **Tm** | **0.002** | 0.000 | **0.002** | 0.000 | **0.001** | 0.000 | **0.001** | 0.000 |
| **Yb** | **0.009** | 0.001 | **0.009** | 0.000 | **0.009** | 0.001 | **0.008** | 0.001 |
| **Lu** | **0.001** | 0.000 | **0.001** | 0.000 | **0.001** | 0.000 | **0.001** | 0.000 |
| **Hf** | **0.041** | 0.003 | **0.039** | 0.001 | **0.058** | 0.001 | **0.039** | 0.001 |
| **Ta** | **0.000** | 0.000 | **0.000** | 0.000 | **0.000** | 0.000 | **0.000** | 0.000 |
| **W** | **0.005** | 0.001 | **0.009** | 0.002 | **0.089** | 0.001 | **0.197** | 0.098 |
| **Ir** | **0.000** | 0.000 | **0.000** | 0.000 | **0.000** | 0.000 | **0.000** | 0.000 |
| **Pt** | **0.001** | 0.001 | **0.001** | 0.000 | **0.000** | 0.000 | **0.001** | 0.001 |
| **Tl** | **0.000** | 0.000 | **0.000** | 0.000 | **0.012** | 0.023 | **0.000** | 0.000 |
| **Pb** | **0.030** | 0.003 | **0.030** | 0.001 | **0.028** | 0.003 | **0.052** | 0.001 |
| **Bi** | **0.000** | 0.000 | **0.000** | 0.000 | **0.000** | 0.000 | **0.000** | 0.000 |
| **Th** | **0.024** | 0.001 | **0.027** | 0.000 | **0.028** | 0.001 | **0.024** | 0.001 |
| **U** | **0.050** | 0.003 | **0.048** | 0.002 | **0.047** | 0.002 | **0.047** | 0.002 |

*Supplementary Table S3. Bulk elemental compositions of total process blanks*

| | Total Process Blank (along with SEIWG) | | Total Process Blank (along with SOAWG) | | Total Process Blank (along with SEIWG) | | Total Process Blank (along with SEIWG) | |
|---|---|---|---|---|---|---|---|---|
| **Lab** | **WUSTL** | | **WUSTL** | | **LLNL** | | **UBC** | |
| **Method** | **Q-ICP-MS** | | **Q-ICP-MS** | | **SF-ICP-MS** | | **SF-ICP-MS** | |
| | **μg** | ±SD | **μg** | ±SD | **μg** | ±SD | **μg** | ±SD |
| **Li** | **0.000** | 0.000 | **0.000** | 0.000 | | | **0.000** | 0.000 |
| **Be** | **0.000** | 0.000 | **0.000** | 0.000 | **0.000** | 0.000 | | |
| **Na** | **0.000** | 0.000 | **0.000** | 0.000 | **0.003** | 0.001 | **0.001** | 0.004 |
| **Mg** | **0.000** | 0.000 | **0.000** | 0.000 | **0.002** | 0.000 | **0.037** | 0.015 |
| **Al** | **0.000** | 0.001 | **0.000** | 0.001 | **0.005** | 0.000 | **0.088** | 0.021 |
| **P** | **0.000** | 0.000 | **0.002** | 0.001 | **0.010** | 0.001 | **0.003** | 0.158 |
| **S** | | | | | | | | |
| **K** | **0.004** | 0.003 | **0.001** | 0.002 | **0.000** | 0.001 | **0.013** | 0.008 |
| **Ca** | **0.011** | 0.022 | **0.011** | 0.023 | **0.013** | 0.000 | **0.171** | 0.021 |
| **Sc** | **0.000** | 0.000 | **0.000** | 0.000 | **0.000** | 0.000 | **0.000** | 0.006 |
| **Ti** | **0.000** | 0.000 | **0.000** | 0.000 | **0.000** | 0.000 | **0.006** | 0.000 |
| **V** | **0.000** | 0.000 | **0.000** | 0.000 | **0.000** | 0.000 | **0.000** | 0.002 |
| **Cr** | **0.000** | 0.000 | **0.000** | 0.000 | **0.000** | 0.000 | **0.018** | 0.000 |
| **Mn** | **0.000** | 0.000 | **0.000** | 0.000 | **0.000** | 0.000 | **0.000** | 0.000 |
| **Fe** | **0.000** | 0.000 | **0.000** | 0.000 | **0.001** | 0.000 | **0.009** | 0.000 |
| **Co** | **0.000** | 0.000 | **0.000** | 0.000 | **0.000** | 0.000 | **0.001** | 0.012 |
| **Ni** | **0.000** | 0.000 | **0.000** | 0.000 | **0.000** | 0.000 | **0.061** | 0.000 |
| **Cu** | **0.000** | 0.000 | **0.000** | 0.000 | **0.000** | 0.000 | **0.000** | 0.001 |
| **Zn** | **0.000** | 0.000 | **0.000** | 0.000 | **0.000** | 0.000 | **0.002** | 0.003 |
| **Ga** | **0.000** | 0.000 | **0.000** | 0.000 | **0.000** | 0.000 | **0.000** | 0.001 |
| **As** | **0.000** | 0.000 | **0.000** | 0.000 | | | | |
| **Se** | **0.000** | 0.000 | **0.000** | 0.000 | | | | |
| **Rb** | **0.000** | 0.000 | **0.000** | 0.000 | **0.000** | 0.000 | **0.000** | 0.000 |
| **Sr** | **0.000** | 0.000 | **0.000** | 0.000 | **0.000** | 0.000 | **0.001** | 0.000 |
| **Y** | **0.000** | 0.000 | **0.000** | 0.000 | **0.000** | 0.000 | **0.000** | 0.000 |
| **Zr** | **0.000** | 0.000 | **0.000** | 0.000 | **0.000** | 0.000 | **0.000** | 0.000 |
| **Nb** | **0.000** | 0.000 | **0.000** | 0.000 | **0.000** | 0.000 | **0.000** | 0.000 |
| **Mo** | **0.000** | 0.000 | **0.000** | 0.000 | **0.000** | 0.000 | **0.000** | 0.000 |
| **Ag** | **0.000** | 0.000 | **0.000** | 0.000 | | | **0.000** | 0.001 |
| **Cd** | **0.000** | 0.000 | **0.000** | 0.000 | **0.000** | 0.000 | **0.000** | 0.000 |
| **In** | **0.000** | 0.000 | **0.000** | 0.000 | | | | |
| **Sn** | **0.000** | 0.000 | **0.000** | 0.000 | **0.000** | 0.000 | **0.000** | 0.000 |
| **Sb** | **0.000** | 0.000 | **0.000** | 0.000 | **0.000** | 0.000 | **0.000** | 0.001 |
| **Te** | **0.000** | 0.000 | **0.000** | 0.000 | | | | |
| **Cs** | **0.000** | 0.000 | **0.000** | 0.000 | **0.000** | 0.000 | **0.000** | 0.000 |
| **Ba** | **0.000** | 0.000 | **0.000** | 0.000 | **0.000** | 0.000 | **0.000** | 0.000 |
| **La** | **0.000** | 0.000 | **0.000** | 0.000 | **0.000** | 0.000 | **0.000** | 0.001 |
| **Ce** | **0.000** | 0.000 | **0.000** | 0.000 | **0.000** | 0.000 | **0.000** | 0.000 |
| **Pr** | **0.000** | 0.000 | **0.000** | 0.000 | **0.000** | 0.000 | **0.000** | 0.000 |
| **Nd** | **0.000** | 0.000 | **0.000** | 0.000 | **0.000** | 0.000 | **0.000** | 0.000 |
| **Sm** | **0.000** | 0.000 | **0.000** | 0.000 | **0.000** | 0.000 | **0.000** | 0.000 |
| **Eu** | **0.000** | 0.000 | **0.000** | 0.000 | **0.000** | 0.000 | **0.000** | 0.000 |
| **Gd** | **0.000** | 0.000 | **0.000** | 0.000 | **0.000** | 0.000 | **0.000** | 0.000 |
| **Tb** | **0.000** | 0.000 | **0.000** | 0.000 | **0.000** | 0.000 | **0.000** | 0.000 |
| **Dy** | **0.000** | 0.000 | **0.000** | 0.000 | **0.000** | 0.000 | **0.000** | 0.000 |
| **Ho** | **0.000** | 0.000 | **0.000** | 0.000 | **0.000** | 0.000 | **0.000** | 0.000 |
| **Er** | **0.000** | 0.000 | **0.000** | 0.000 | **0.000** | 0.000 | **0.000** | 0.000 |
| **Tm** | **0.000** | 0.000 | **0.000** | 0.000 | **0.000** | 0.000 | **0.000** | 0.000 |
| **Yb** | **0.000** | 0.000 | **0.000** | 0.000 | **0.000** | 0.000 | **0.000** | 0.000 |
| **Lu** | **0.000** | 0.000 | **0.000** | 0.000 | **0.000** | 0.000 | **0.000** | 0.000 |
| **Hf** | **0.000** | 0.000 | **0.000** | 0.000 | **0.000** | 0.000 | **0.000** | 0.000 |
| **Ta** | **0.000** | 0.000 | **0.000** | 0.000 | **0.000** | 0.000 | **0.000** | 0.000 |
| **W** | **0.000** | 0.000 | **0.000** | 0.000 | **0.000** | 0.000 | **0.000** | 0.000 |
| **Ir** | **0.000** | 0.000 | **0.000** | 0.000 | | | | |
| **Pt** | **0.000** | 0.000 | **0.000** | 0.000 | | | | |
| **Tl** | **0.000** | 0.000 | **0.000** | 0.000 | **0.000** | 0.000 | **0.000** | 0.003 |
| **Pb** | **0.000** | 0.000 | **0.000** | 0.000 | **0.000** | 0.000 | **0.000** | 0.000 |
| **Bi** | **0.000** | 0.000 | **0.000** | 0.000 | | | **0.000** | 0.000 |
| **Th** | **0.000** | 0.000 | **0.000** | 0.000 | **0.000** | 0.000 | **0.000** | 0.000 |
| **U** | **0.000** | 0.000 | **0.000** | 0.000 | **0.000** | 0.000 | **0.000** | 0.000 |

## *Supplementary Table S4. Bulk elemental compositions of geological reference materials*

| | GSP1 | | SDC1 | | BHVO-2 | | JB2 | | JA1 | | JB2 | | JA1 | | BCR-2* | |
|---|---|---|---|---|---|---|---|---|---|---|---|---|---|---|---|---|
| **Lab** | LLNL | | LLNL | | LLNL | | LLNL | | LLNL | | LLNL | | LLNL | | UBC | |
| **Method** | SF-ICP-MS | | SF-ICP-MS | | SF-ICP-MS | | SF-ICP-MS | | SF-ICP-MS | | SF-ICP-MS | | SF-ICP-MS | | SF-ICP-MS | |
| **Date** | Feb-24 | | Feb-24 | | Feb-24 | | Oct-24 | | Oct-24 | | Mar-25 | | Mar-25 | | | |
| | µg/g | ±SD | µg/g | ±SD | µg/g | ±SD | µg/g | ±SD | µg/g | ±SD | µg/g | ±SD | µg/g | ±SD | µg/g | ±SD |
| **Li** | | | | | | | | | | | | | | | 8.60 | 0.47 |
| **Be** | 1.404 | 0.122 | 2.721 | 0.153 | 1.122 | 0.243 | -0.020 | -0.191 | 0.038 | 0.145 | 0.224 | 0.056 | 0.437 | 0.041 | | |
| **Na** | 19805 | 76 | 14559 | 105 | 15299 | 133 | 15507 | 406 | 28209 | 370 | 15104 | 68 | 27896 | 91 | 21673 | 664 |
| **Mg** | 5674 | 23 | 9835 | 63 | 30454 | 191 | 26761 | 305 | 9236 | 136 | 28461 | 162 | 9089 | 115 | 19938 | 1622 |
| **Al** | 77066 | 419 | 79186 | 527 | 42585 | 219 | 79646 | 851 | 83345 | 509 | 78225 | 327 | 77730 | 550 | 64421 | 4101 |
| **P** | 1392 | 46 | 771 | 54 | 1200 | 13 | 919 | 62 | 1246 | 96 | 679 | 38 | 1010 | 48 | 1480 | 88 |
| **S** | | | | | | | | | | | | | | | | |
| **K** | 45021 | 286 | 26692 | 240 | 4274 | 25 | 3499 | 38 | 6029 | 85 | 3391 | 46 | 6191 | 67 | 14017 | 1018 |
| **Ca** | 14371 | 97 | 10118 | 92 | 77999 | 575 | 71556 | 793 | 42052 | 434 | 70377 | 416 | 40218 | 135 | 46362 | 5523 |
| **Sc** | 5.472 | 0.084 | 14.183 | 0.257 | 13.762 | 0.134 | 56.206 | 0.514 | 28.962 | 0.392 | 55.848 | 0.473 | 28.197 | 0.105 | 33.8 | 4.9 |
| **Ti** | 3837 | 38 | 5697 | 44 | 15822 | 104 | 6962 | 94 | 5094 | 32 | 6814 | 39 | 4873 | 38 | 13326 | 202 |
| **V** | 48.82 | 0.42 | 90.82 | 0.78 | 308.52 | 1.94 | 588.42 | 5.03 | 107.73 | 1.00 | 586.53 | 5.34 | 103.93 | 0.72 | 472 | 85 |
| **Cr** | 10.410 | 0.109 | 58.833 | 0.561 | 290.233 | 3.506 | 24.916 | 0.200 | 3.817 | 0.137 | 26.253 | 1.053 | 6.434 | 0.714 | 13.6 | 1.1 |
| **Mn** | 291.03 | 2.21 | 821.84 | 5.38 | 1247.06 | 6.07 | 1729.87 | 9.09 | 1205.75 | 7.19 | 1670.82 | 14.17 | 1166.29 | 4.24 | 1453 | 47 |
| **Fe** | 28442 | 115 | 45550 | 303 | 83354 | 466 | 103330 | 680 | 49462 | 301 | 99812 | 657 | 48493 | 262 | 91080 | 2176 |
| **Co** | 6.429 | 0.037 | 17.150 | 0.175 | 46.033 | 0.199 | 36.619 | 0.426 | 10.940 | 0.197 | 36.927 | 0.174 | 11.211 | 0.089 | 41.6 | 4.4 |
| **Ni** | 7.741 | 0.579 | 33.547 | 0.478 | 119.431 | 0.592 | 13.810 | 0.786 | 1.684 | 0.330 | 22.703 | 0.550 | 11.133 | 0.451 | 13.2 | 1.8 |
| **Cu** | 30.597 | 0.608 | 26.135 | 0.402 | 127.345 | 0.439 | 219.979 | 2.563 | 33.641 | 1.003 | 224.785 | 1.661 | 51.197 | 1.180 | 15.8 | 2.0 |
| **Zn** | 90.907 | 11.964 | 82.141 | 10.540 | 101.562 | 0.205 | 183.181 | 0.804 | 347.127 | 5.702 | 118.345 | 1.318 | 101.521 | 1.761 | 128 | 17 |
| **Ga** | 22.250 | 0.373 | 21.190 | 0.272 | 20.153 | 0.141 | 16.560 | 0.290 | 18.271 | 0.429 | 16.524 | 0.260 | 17.235 | 0.538 | 22.7 | 2.7 |
| **As** | | | | | | | | | | | | | | | | |
| Se | | | | | | | | | | | | | | | | |
| Rb | 243.503 | 1.695 | 117.968 | 0.386 | 4.852 | 0.082 | 5.985 | 0.212 | 9.999 | 0.179 | 5.347 | 0.129 | 9.385 | 0.123 | 47.1 | 3.7 |
| Sr | 229.885 | 1.960 | 172.726 | 0.798 | 385.250 | 1.721 | 180.238 | 1.499 | 256.416 | 2.683 | 172.063 | 0.824 | 250.581 | 1.051 | 331 | 20 |
| Y | 24.817 | 0.244 | 32.600 | 0.102 | 21.941 | 0.166 | 24.464 | 0.187 | 28.498 | 0.242 | 23.788 | 0.120 | 28.515 | 0.187 | 36.4 | 2.1 |
| Zr | 41.638 | 1.186 | 46.271 | 0.388 | 164.105 | 1.151 | 47.407 | 0.500 | 82.440 | 3.833 | 46.042 | 0.480 | 78.407 | 0.414 | 185 | 12 |
| Nb | 27.637 | 0.194 | 19.169 | 0.082 | 18.313 | 0.252 | 0.492 | 0.021 | 1.311 | 0.039 | 0.439 | 0.121 | 1.263 | 0.124 | 13.1 | 0.7 |
| Mo | 0.329 | 0.025 | 0.122 | 0.045 | 4.029 | 0.257 | 1.064 | 0.097 | 1.815 | 0.106 | 1.255 | 0.074 | 2.762 | 0.085 | 269 | 18 |
| Ag | | | | | | | | | | | | | | | 1.17 | 0.11 |
| Cd | 0.060 | 0.017 | 0.035 | 0.025 | 0.199 | 0.002 | 0.188 | 0.081 | 0.112 | 0.049 | 0.163 | 0.249 | 0.066 | 0.196 | 0.959 | 0.064 |
| In | | | | | | | | | | | | | | | | |
| Sn | 7.302 | 0.134 | 3.107 | 0.134 | 1.409 | 0.005 | -2.749 | -1.412 | -2.617 | -1.399 | 1.012 | 0.106 | 1.307 | 0.064 | 2.05 | 0.13 |
| Sb | 2.895 | 0.068 | 0.446 | 0.024 | 0.108 | 0.002 | -0.060 | -0.035 | -0.065 | -0.035 | 0.124 | 0.010 | 0.157 | 0.007 | 0.259 | 0.032 |
| Te | | | | | | | | | | | | | | | | |
| Cs | 0.961 | 0.021 | 3.710 | 0.060 | 0.059 | 0.013 | 0.498 | 0.074 | 0.225 | 0.067 | 0.555 | 0.046 | 0.313 | 0.042 | 1.12 | 0.05 |
| Ba | 1259.200 | 4.872 | 615.286 | 4.418 | 122.763 | 0.610 | 212.388 | 2.365 | 282.831 | 1.415 | 208.565 | 0.921 | 282.133 | 1.516 | 574 | 42 |
| La | 187.538 | 1.715 | 40.394 | 0.142 | 14.049 | 0.078 | 2.240 | 0.058 | 4.778 | 0.039 | 2.241 | 0.015 | 4.804 | 0.029 | 21.8 | 1.6 |
| Ce | 443.034 | 2.108 | 86.504 | 0.342 | 34.787 | 0.172 | 6.624 | 0.066 | 12.806 | 0.069 | 6.385 | 0.048 | 12.717 | 0.078 | 46.0 | 2.5 |
| Pr | 56.128 | 0.250 | 10.407 | 0.071 | 4.834 | 0.023 | 1.110 | 0.019 | 2.055 | 0.032 | 1.130 | 0.005 | 2.054 | 0.035 | 5.62 | 0.39 |
| Nd | 204.779 | 0.818 | 39.225 | 0.214 | 22.373 | 0.117 | 6.169 | 0.132 | 10.503 | 0.245 | 6.193 | 0.046 | 10.515 | 0.051 | 23.9 | 2.2 |
| Sm | 26.404 | 0.188 | 7.805 | 0.068 | 5.400 | 0.072 | 2.256 | 0.065 | 3.265 | 0.122 | 2.197 | 0.032 | 3.308 | 0.055 | 5.53 | 0.43 |
| Eu | 2.299 | 0.039 | 1.552 | 0.019 | 1.828 | 0.007 | 0.789 | 0.009 | 1.055 | 0.024 | 0.789 | 0.027 | 1.043 | 0.014 | 1.61 | 0.16 |
| Gd | 17.119 | 0.203 | 6.817 | 0.104 | 5.610 | 0.030 | 2.805 | 0.082 | 3.752 | 0.145 | 2.691 | 0.022 | 3.645 | 0.026 | 5.55 | 0.43 |
| Tb | 1.575 | 0.022 | 1.031 | 0.010 | 0.870 | 0.005 | 0.572 | 0.015 | 0.716 | 0.023 | 0.537 | 0.007 | 0.692 | 0.008 | 0.821 | 0.085 |
| Dy | 5.666 | 0.034 | 6.055 | 0.038 | 4.831 | 0.023 | 3.945 | 0.057 | 4.729 | 0.079 | 3.943 | 0.070 | 4.709 | 0.056 | 5.26 | 0.51 |
| Ho | 0.905 | 0.017 | 1.189 | 0.027 | 0.893 | 0.007 | 0.883 | 0.037 | 1.004 | 0.039 | 0.868 | 0.020 | 1.010 | 0.011 | 1.03 | 0.13 |
| Er | 2.059 | 0.038 | 3.410 | 0.026 | 2.284 | 0.007 | 2.565 | 0.056 | 2.966 | 0.035 | 2.561 | 0.042 | 2.943 | 0.030 | 2.85 | 0.31 |
| Tm | 0.238 | 0.006 | 0.487 | 0.017 | 0.295 | 0.002 | 0.383 | 0.023 | 0.437 | 0.015 | 0.390 | 0.018 | 0.437 | 0.007 | 0.405 | 0.040 |
| Yb | 1.276 | 0.018 | 3.054 | 0.026 | 1.790 | 0.006 | 2.574 | 0.152 | 2.920 | 0.069 | 2.553 | 0.045 | 2.907 | 0.068 | 2.64 | 0.34 |
| **Lu** | **0.161** | 0.007 | **0.442** | 0.005 | **0.249** | 0.002 | **0.400** | 0.008 | **0.448** | 0.008 | **0.392** | 0.008 | **0.439** | 0.009 | **0.326** | 0.045 |
| **Hf** | **1.089** | 0.031 | **1.294** | 0.037 | **3.979** | 0.022 | **1.439** | 0.035 | **2.419** | 0.039 | **1.606** | 0.222 | **2.441** | 0.012 | **4.59** | 0.57 |
| **Ta** | **0.888** | 0.009 | **1.210** | 0.013 | **1.156** | 0.010 | **0.036** | 0.011 | **0.119** | 0.037 | **0.391** | 0.672 | **0.077** | 0.034 | **0.675** | 0.083 |
| **W** | **0.378** | 0.025 | **0.954** | 0.029 | **0.219** | 0.002 | **0.478** | 0.041 | **0.662** | 0.032 | **0.675** | 0.022 | **1.017** | 0.062 | **0.464** | 0.055 |
| **Ir** | | | | | | | | | | | | | | | | |
| **Pt** | | | | | | | | | | | | | | | | |
| **Tl** | | | | | | | **0.317** | 0.008 | **0.273** | 0.016 | **0.151** | 0.057 | **0.139** | 0.043 | **0.198** | 0.029 |
| **Pb** | **58.741** | 0.665 | **23.326** | 0.577 | **1.340** | 0.037 | **1.877** | 0.172 | **1.939** | 0.160 | **7.578** | 0.087 | **11.307** | 0.102 | **8.05** | 0.92 |
| **Bi** | | | | | | | | | | | | | | | **0.040** | 0.006 |
| **Th** | **112.299** | 0.614 | **11.184** | 0.067 | **0.812** | 0.017 | **0.274** | 0.012 | **0.742** | 0.009 | **0.258** | 0.007 | **0.732** | 0.012 | **4.34** | 1.14 |
| **U** | **2.915** | 0.026 | **2.658** | 0.027 | **0.368** | 0.005 | **0.163** | 0.019 | **0.352** | 0.016 | **0.107** | 0.061 | **0.324** | 0.069 | **1.27** | 0.35 |

* Please note that this BCR-2 analyzed by UBC was NOT matrix-corrected by the Orgueil run but only calibrated by the 5-point synthetic standard solution calibration.

*Supplementary Table S5. The DOIs associated with the data detailed in this paper.*

| DOI | Product Name | Product Type |
|---|---|---|
| **10.60707/bdq3-px39** | 20250128_Q-ICP-MS_WUSTL_multiSample_1_QICPMSProcessedTabular_1.csv | QICPMSProcessedTabular |
| **10.60707/fzjk-kt84** | 20250128_Q-ICP-MS_WUSTL_multiSample_1_QICPMSRawTabular_1.csv | QICPMSRawTabular |
| **10.60707/3qvv-wy22** | 20240125_ICP-OES_UCB_multiSample_1_ICPOESIntermediateTabular_1.csv | ICPOESIntermediateTabular |
| **10.60707/rm08-bg04** | 20240125_ICP-OES_UCB_multiSample_1_ICPOESProcessedTabular_1.csv | ICPOESProcessedTabular |
| **10.60707/fj10-mc96** | 20240125_ICP-OES_UCB_multiSample_1_ICPOESRawTabular_1.csv | ICPOESRawTabular |
| **10.60707/2tde-w141** | 20241202_ICP-OES_UCB_multiSample_1_ICPOESIntermediateTabular_1.csv | ICPOESIntermediateTabular |
| **10.60707/bsne-yq93** | 20241202_ICP-OES_UCB_multiSample_1_ICPOESProcessedTabular_1.csv | ICPOESProcessedTabular |
| **10.60707/03bz-6d45** | 20241202_ICP-OES_UCB_multiSample_1_ICPOESRawTabular_1.csv | ICPOESRawTabular |
| **10.60707/zhe4-0086** | 20250317_HR-ICP-MS_LLNL_multiSample_1_HRICPMSProcessed_1.xlsx | HRICPMSProcessed |
| **10.60707/4h3r-3h19** | 20250317_HR-ICP-MS_LLNL_multiSample_1_HRICPMSRaw_1.csv | HRICPMSRaw |
| **10.60707/q4b0-v203** | 20240206_HR-ICP-MS_LLNL_OREX-803015-101_1_HRICPMSProcessed_1.csv | HRICPMSProcessed |
| **10.60707/d62h-j641** | 20240206_HR-ICP-MS_LLNL_OREX-803015-101_1_HRICPMSRaw_1.csv | HRICPMSRaw |

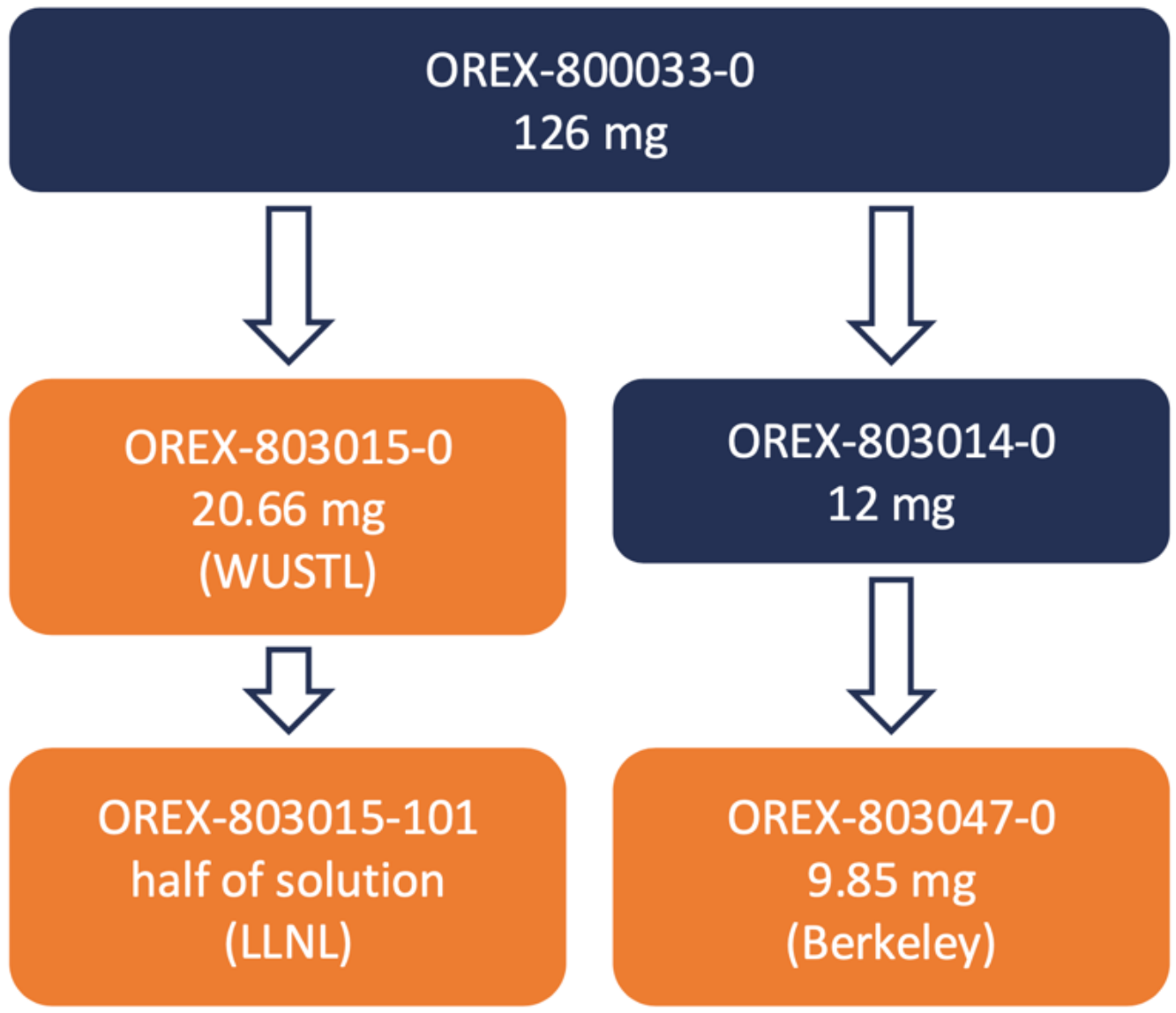

**Supplementary Figure S1.** Flow chart illustrating the division history of the small non-homogenized aggregate sample (OREX-800033-0). Only samples used for bulk elemental analysis in this study are included. Boxes in orange indicate the final samples analyzed.

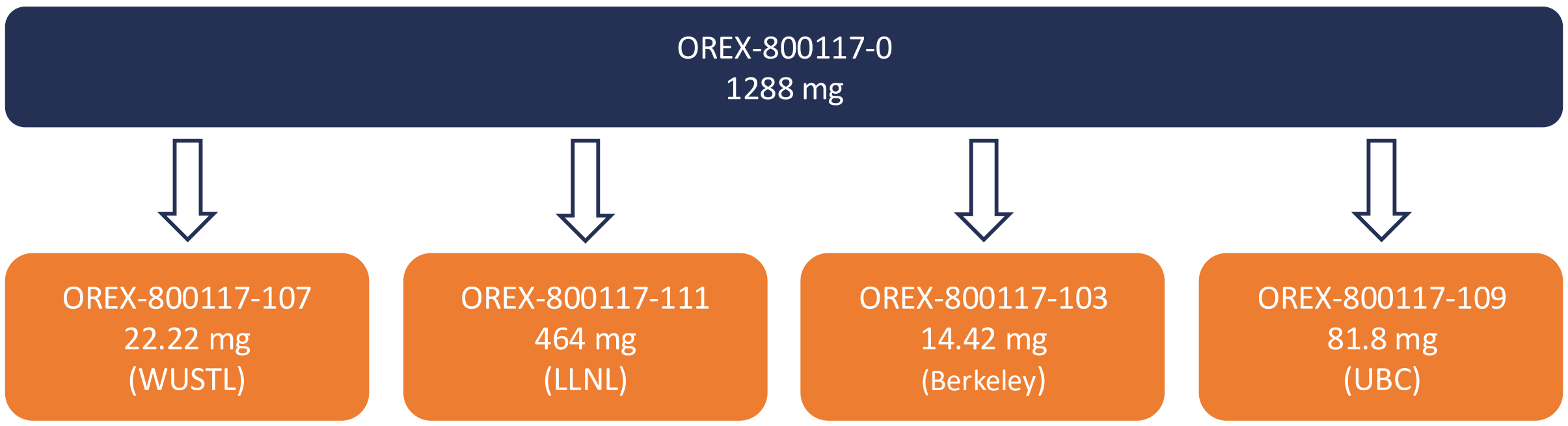

**Supplementary Figure S2.** Flow chart illustrating the division history of the large homogenized aggregate powder for SEIWG (OREX-800117-0). Only samples used for bulk elemental analysis in this study are included. Boxes in orange indicate the final samples analyzed.

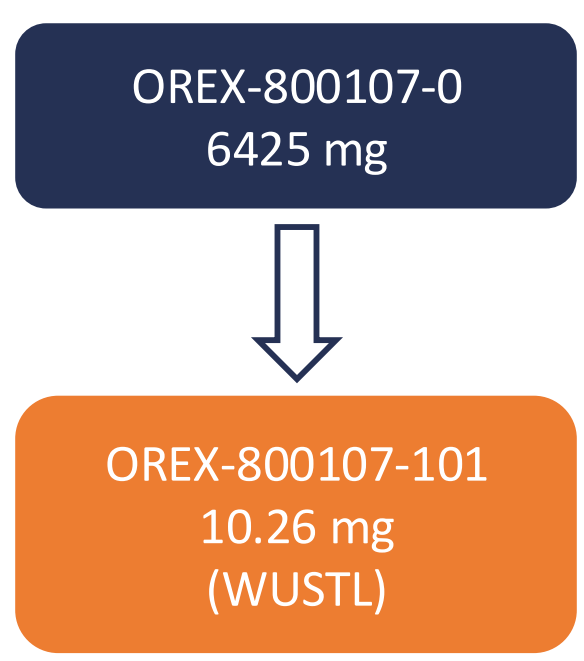

**Supplementary Figure S3.** Flow chart illustrating the division history of the large homogenized aggregate powder for SOAWG (OREX-800107-0). Only samples used for bulk elemental analysis in this study are included. Boxes in orange indicate the final samples analyzed.

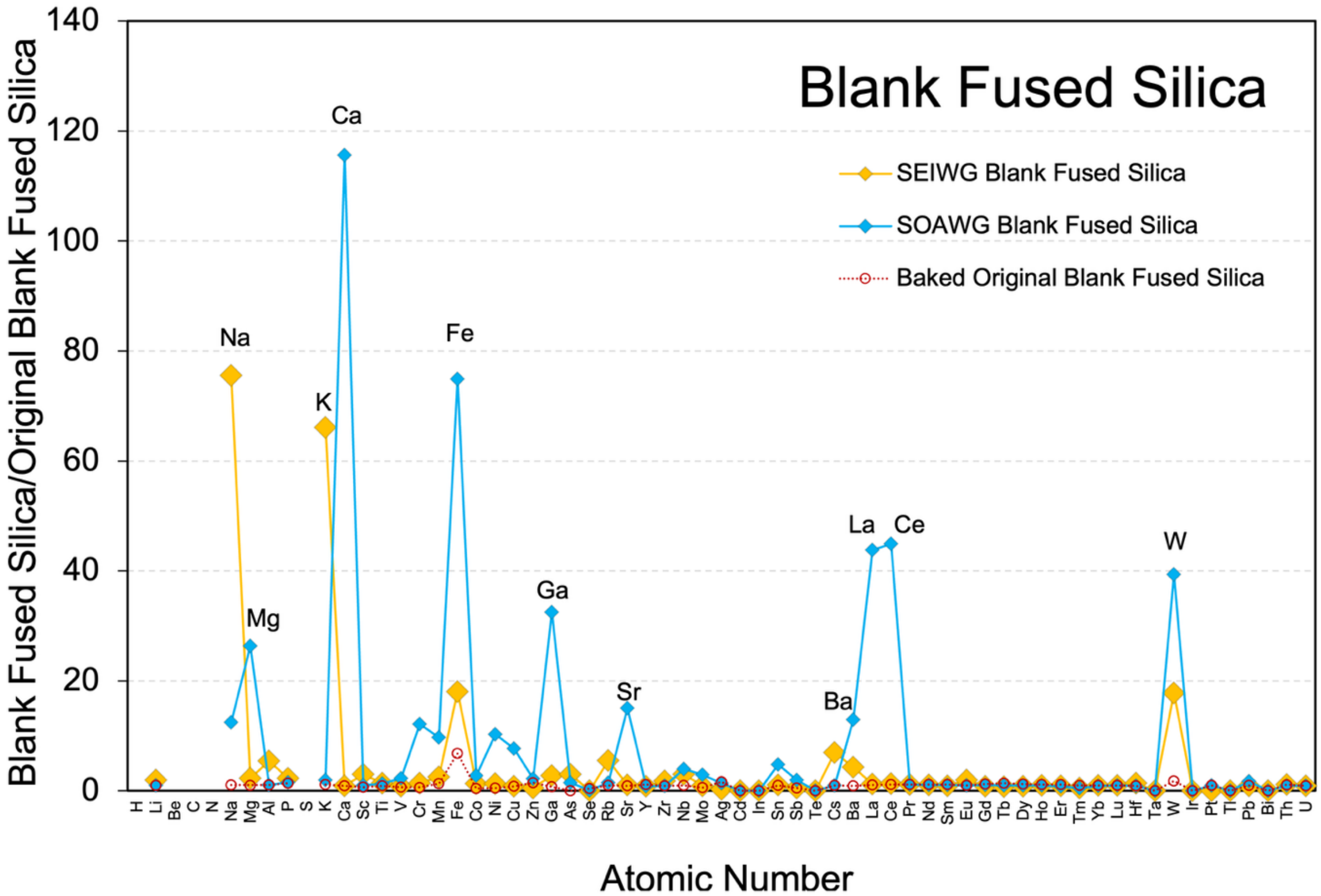

**Supplementary Figure S4.** Bulk elemental compositions of SEIWG, SOAWG, and baked original blank fused silica samples, normalized to the original blank fused silica sample. The baked original blank fused silica sample shows no significant difference from the unbaked original blank fused silica, indicating that baking alone (a step in SOAWG sample homogenization processing) does not introduce measurable contamination. In contrast, the SEIWG and SOAWG blank fused silica samples exhibit clear enrichment in Na, Mg, K, Ca, Fe, Ga, Sr, Ba, La, Ce, and W, which were likely introduced during SEIWG and SOAWG sample processing.

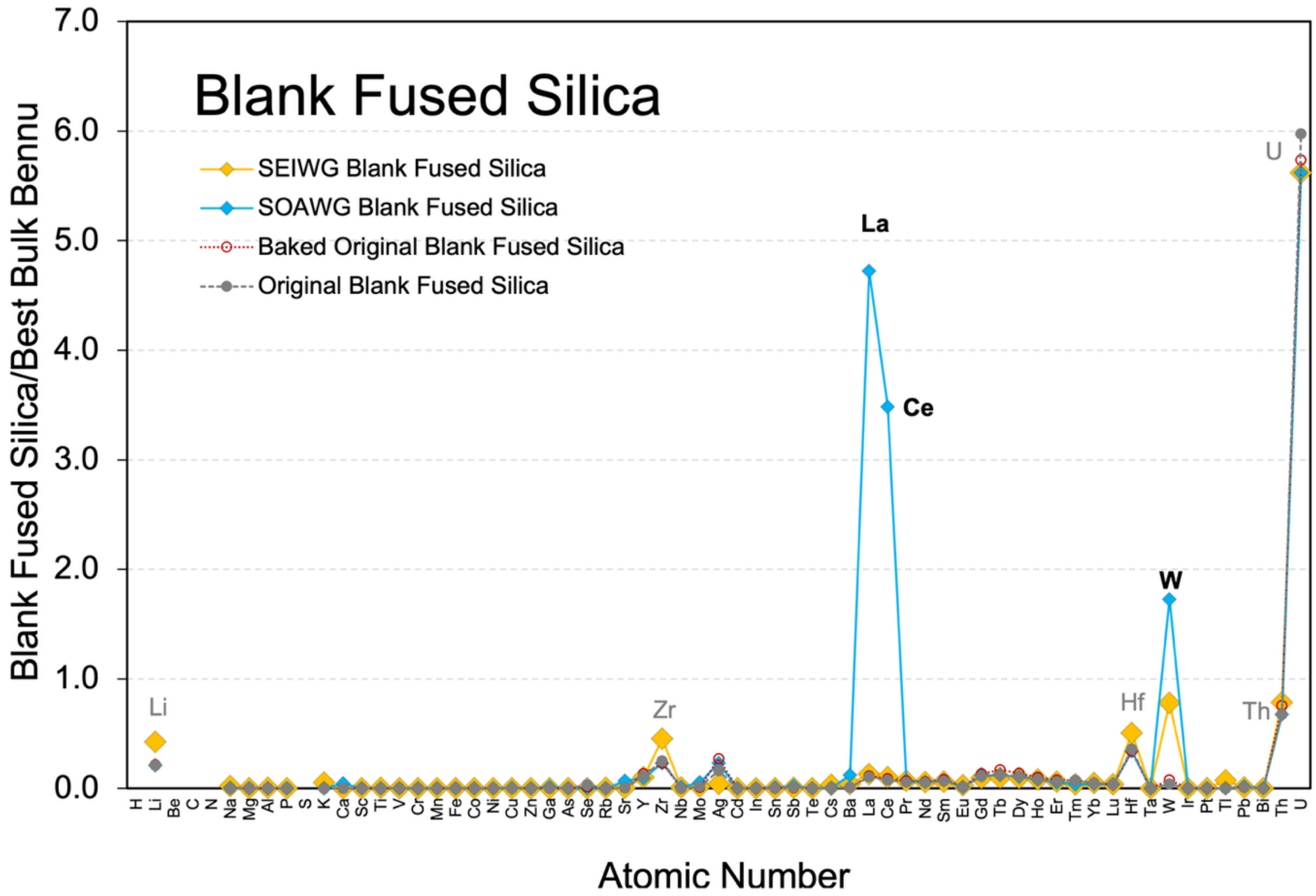

**Supplementary Figure S5.** Bulk elemental compositions of SEIWG, SOAWG, baked and unbaked original blank fused silica samples, normalized to the $B^3$ composition. For most elements, the blank fused silica samples show negligible concentrations (well below analytical uncertainty); however, a few elements, such as U, Light Rare Earth Element (LREE: La and Ce), W, and Th, exhibit significant enrichment, particularly in the SOAWG blank fused silica. Both blank fused silica samples (SEIWG and SOAWG) were sourced from the same supplier and used to clean the mortar and pestle prior to homogenizing the Bennu samples. The fused silica samples were also processed along with Bennu samples as contamination control. The observed spikes, especially in the SOAWG blank fused silica, suggest possible contamination introduced during the powdering process. However, this does not necessarily indicate contamination of the Bennu samples themselves, as the mortar and pestle were thoroughly rinsed with Milli-Q water following the fused silica cleaning step. When calculating the $B^3$ composition, three potential elements (La, Ce and W) in SOAWG Bennu samples were excluded to avoid skewing the results (see Table 7). These three elements are deemed more likely to be contamination due to large difference between SEIWG and SOAWG blank fused silica. Other spikes (e.g., U, Th, Hf, Zr and Li) are interpreted as inherited from the commercial silica itself as all SEIWG, SOAWG, baked and unbaked original blank fused silica samples show almost identical concentrations.

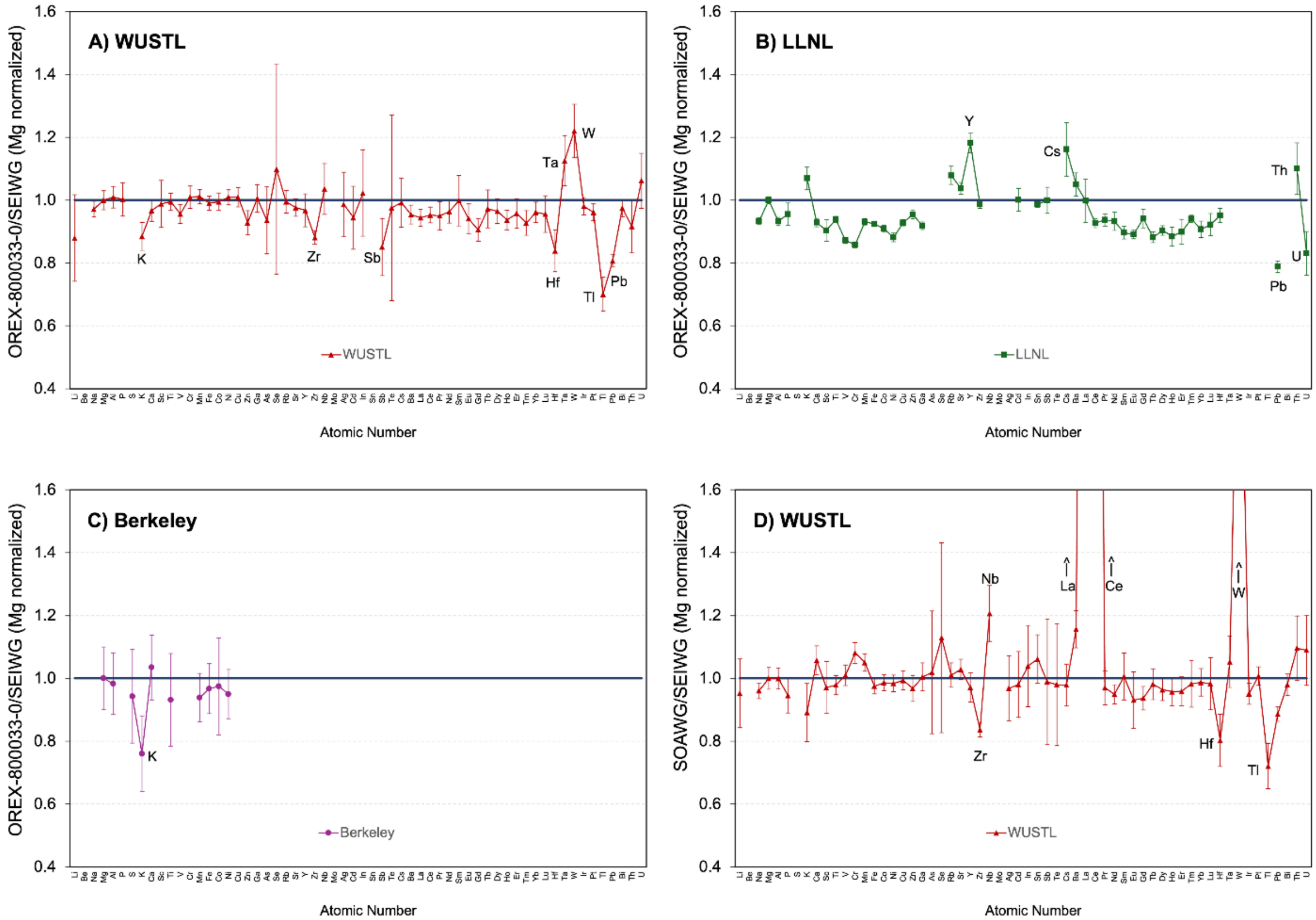

**Supplementary Figure S6. (A–C)** Compositions of the non-homogenized aggregate sample OREX-800033-0 normalized to those of the homogenized SEIWG sample (OREX-800117-0) analyzed in the same laboratory. The differences between these two samples are not consistent across laboratories, indicating that they do not reflect true compositional differences but are instead attributable to analytical uncertainty within individual laboratories. **(D)** Composition of the homogenized SOAWG sample (OREX-800107-0) normalized to that of the homogenized SEIWG sample (OREX-800117-0) analyzed in the same laboratory. As in Figure 4, obvious contamination is observed for La, Ce, and W.

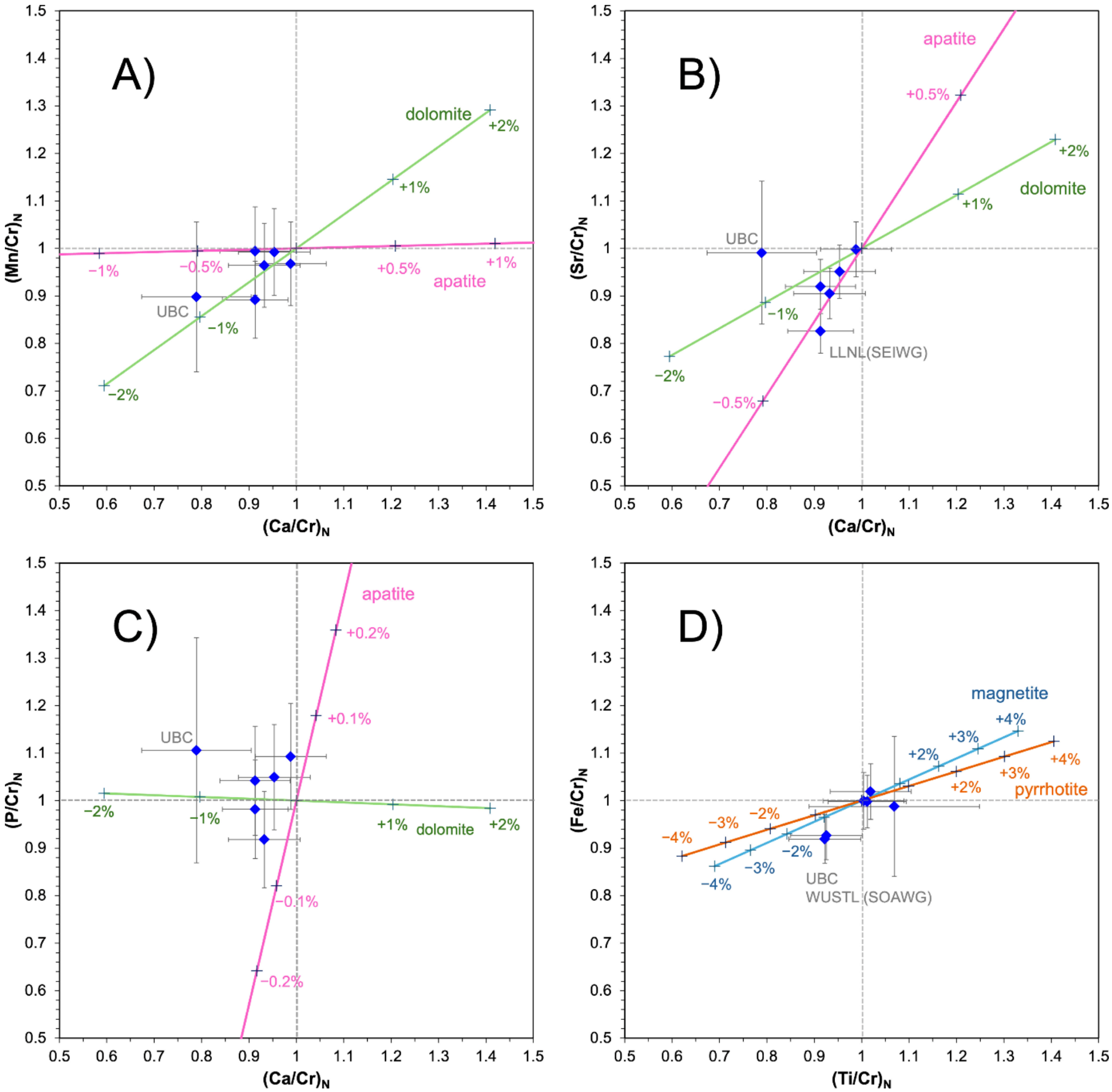

**Supplementary Figure S7.** Correlations between CI-normalized elemental ratios. (A) Mn/Cr vs. Ca/Cr. (B) Sr/Cr vs. Ca/Cr. (C) P/Cr vs. Ca/Cr. (D) Fe/Cr vs. Ti/Cr. CI chondrite concentrations are from Lodders et al. (2025a). Samples plotting outside the CI value are labeled in light grey. The lines represent the addition or removal of dolomite (green), apatite (pink), magnetite (blue), and pyrrhotite (orange) relative to the bulk CI chondrite composition. Mineral compositions are primarily from Lauretta et al. (2026) for Bennu minerals. Where Bennu mineral compositions were unavailable, we used mineral compositions for Ryugu reported by Bazi et al. (2022) and Nakamura et al. (2022). This figure follows the approach of Yokoyama et al. (2025), who applied the same analysis to bulk Ryugu samples.

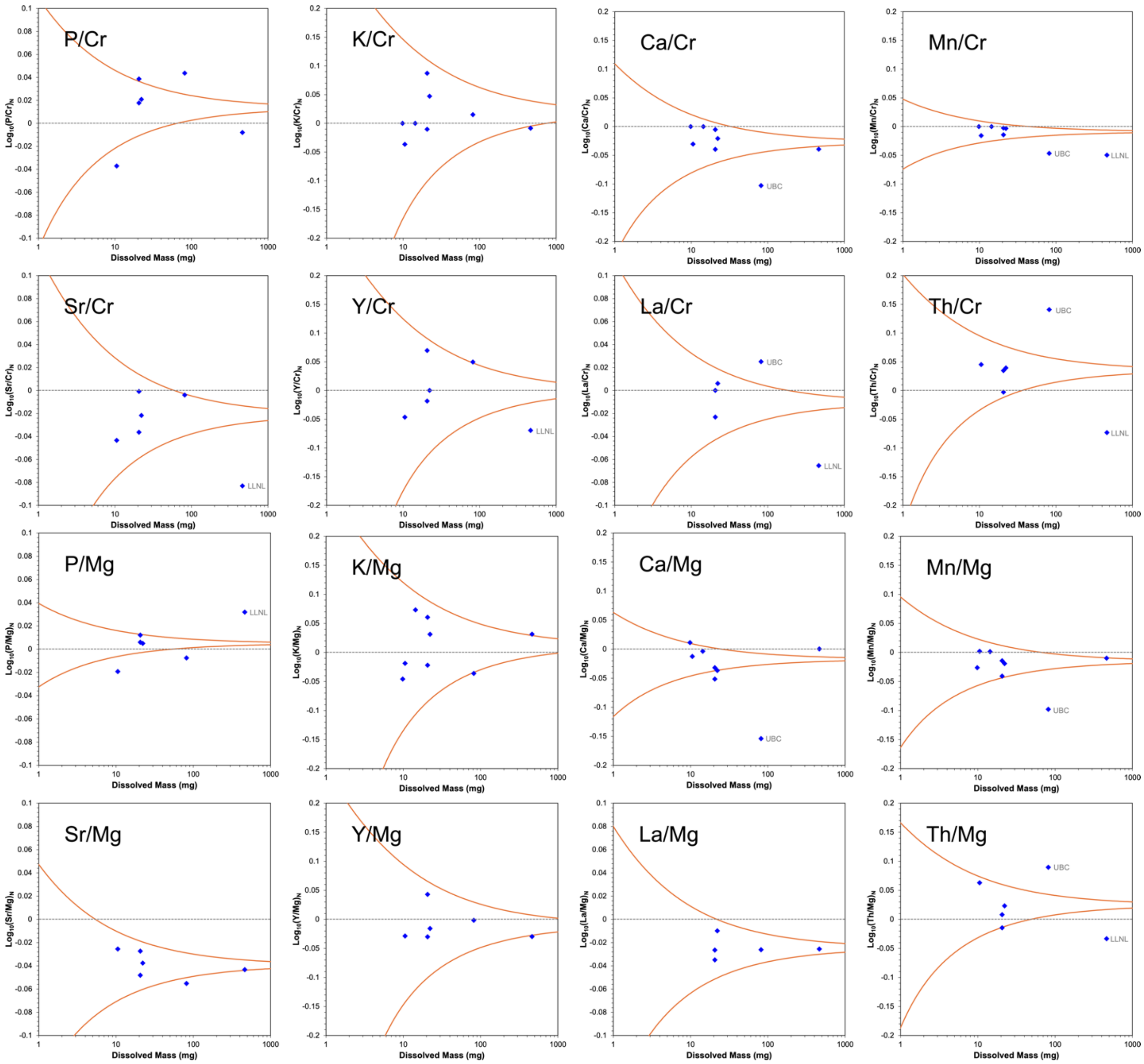

**Supplementary Figure S8.** CI-normalized elemental ratios plotted against **dissolved sample mass**. CI chondrite concentrations are from Lodders et al. (2025a). If nugget effects are responsible for the observed elemental ratio variations, the dispersion of the data (i.e., along the y-axis) would be expected to decrease with increasing sample mass (Dauphas and Pourmand, 2015). The lines represent the estimated data dispersion calculated from the variability observed in smaller samples. Larger samples that fall outside the expected range are labeled in light gray. This figure follows the approach of Yokoyama et al. (2025), who applied the same analysis to bulk Ryugu samples. The lines are calculated using the Equations 5 and 6 in Yokoyama et al. (2025).

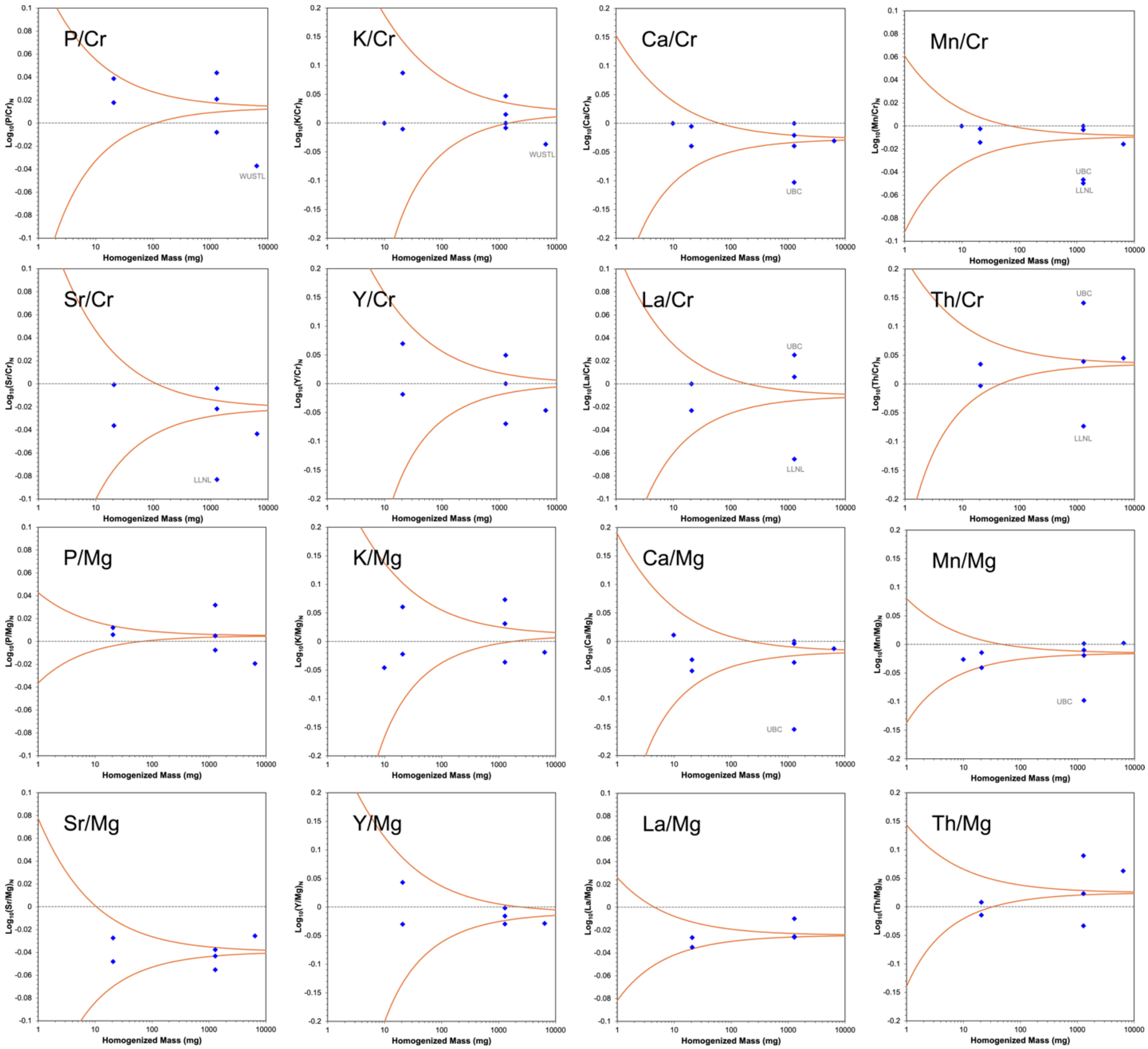

**Supplementary Figure S9.** CI-normalized elemental ratios plotted against **homogenized sample mass**. CI chondrite concentrations are from Lodders et al. (2025a). If nugget effects are responsible for the observed elemental ratio variations, the dispersion of the data (i.e., along the y-axis) would be expected to decrease with increasing sample mass (Dauphas and Pourmand, 2015). The lines represent the estimated data dispersion calculated from the variability observed in smaller samples. Larger samples that fall outside the expected range are labeled in light gray. This figure follows the approach of Yokoyama et al. (2025), who applied the same analysis to bulk Ryugu samples. The lines are calculated using the Equations 5 and 6 in Yokoyama et al. (2025).